\documentclass[fleqn,usenatbib]{mnras}
\usepackage{newtxtext,newtxmath}
\usepackage[T1]{fontenc}
\usepackage{ae,aecompl}
\usepackage{appendix}

\usepackage{graphicx}
\usepackage{amssymb,amsmath,relsize}
\usepackage{natbib}
\usepackage{booktabs,multirow}

\usepackage[usenames, dvipsnames]{color}

\def\lesssim{\mathrel{\hbox{\rlap{\hbox{\lower5pt\hbox{$\sim$}}}\hbox{$<$}}}}
\def\gtrsim{\mathrel{\hbox{\rlap{\hbox{\lower5pt\hbox{$\sim$}}}\hbox{$>$}}}}

\title[Non-thermal Emission from M82]
{Cosmic Rays and Magnetic Fields in the Core and Halo of the Starburst M82: Implications for Galactic Wind Physics}

\author[Buckman et al.]{
Benjamin J.~Buckman,$^{1,2}$\thanks{E-mail: buckman.12@osu.edu} 
Tim Linden,$^{1,2}$ 
and Todd A.~Thompson$^{1,3,4}$
\\
$^1$Center for Cosmology \& Astro-Particle Physics, The Ohio State University, Columbus, Ohio 43210, USA\\
$^2$Department of Physics, The Ohio State University, Columbus, Ohio 43210, USA\\
$^3$Department of Astronomy, The Ohio State University, Columbus, Ohio 43210, USA\\
$^4$Institute for Advanced Study, Princeton, New Jersey 08540, USA
}

\date{Accepted XXX. Received YYY; in original form ZZZ}

\pubyear{2019}

\begin{document}
\label{firstpage}
\pagerange{\pageref{firstpage}--\pageref{lastpage}}
\maketitle

\begin{abstract}
Cosmic rays (CRs) and magnetic fields may be dynamically important in driving large-scale galactic outflows from rapidly star-forming galaxies. We construct two-dimensional axisymmetric models of the local starburst and super-wind galaxy M82 using the CR propagation code {\tt{GALPROP}}. Using prescribed gas density and magnetic field distributions, wind profiles, CR injection rates, and stellar radiation fields, we simultaneously fit both the integrated gamma-ray emission and the spatially-resolved multi-frequency radio emission extended along M82's minor axis. We explore the resulting constraints on the gas density, magnetic field strength, CR energy density, and the assumed CR advection profile. In accord with earlier one-zone studies, we generically find low central CR pressures, strong secondary electron/positron production, and an important role for relativistic bremsstrahlung losses in shaping the synchrotron spectrum. We find that the relatively low central CR density produces CR pressure gradients that are weak compared to gravity, strongly limiting the role of CRs in driving M82's fast and mass-loaded galactic outflow. Our models require strong magnetic fields and advection speeds of order ${\sim}1000$\,km/s on kpc scales along the minor axis in order to reproduce the extended radio emission. Degeneracies between the controlling physical parameters of the model and caveats to these findings are discussed.
\end{abstract}


\section{Introduction}
\label{section:introduction}

Cosmic rays (CRs) are the intermediaries between high-energy astrophysics and the underlying dynamics of the galactic interstellar medium. Accelerated primarily in supernova (SN) remnants, CRs propagate through galaxies via a combination of diffusion, advection, and streaming \citep{2007ARNPS..57..285S}, interacting with the ambient galactic interstellar medium (ISM), interstellar radiation field (ISRF), and magnetic field. As CR protons, anti-protons, and nuclei (collectively CRp) propagate, they collide with the ISM and undergo hadronic interactions producing pions and secondary protons/nuclei. Neutral pions ($\pi^0$) decay into gamma-rays while charged pions ($\pi^\pm$) decay into secondary electrons/positions and neutrinos. Low energy CRp ($E_{\mathrm{kin}} \lesssim 1$\,GeV) also lose energy to the ISM through ionization. Primary and secondary CR electrons/positrons (collectively CRe) interact with the ISM directly through ionization losses and relativistic bremsstrahlung, producing gamma-ray emission. CRe also interact with the ISRF through inverse-Compton (IC) scattering, producing X-rays and gamma-rays, and with the magnetic field, producing synchrotron radiation from the radio to the X-ray bands.

In addition to dominating the non-thermal emission of star-forming galaxies, CRs may also be dynamically important. In particular, in the Milky Way, the CR pressure is comparable to that required to maintain vertical hydrostatic equilibrium of the gas disk \citep{1990ApJ...365..544B}, implying that CRs may contribute significantly to the local vertical pressure support, and that they might drive large-scale mass-loaded winds \citep{1975ApJ...196..107I, 1991A&A...245...79B, 2008ApJ...674..258E}. Galactic winds are essential to the evolution of galaxies and their surrounding circumgalactic media, but the dominant launching mechanisms for the outflowing cool gas remain uncertain, and may vary from normal star-forming galaxies to dense starbursts  \citep{2005ARA&A..43..769V,galactic_winds_review_heckman,Zhang2018}. Analytic arguments and multi-dimensional numerical simulations of star-forming galaxies with increasingly sophisticated CR transport algorithms indicate that CRs could be dynamically important across a wide range of galaxy parameters (e.g., \citealt{2008ApJ...687..202S, 2008A&A...481...33J, 2012MNRAS.423.2374U, 2016ApJ...827L..29S, 2016ApJ...824L..30P, 2017MNRAS.467..906W,  2017ApJ...834..208R}). 

A self-consistent model for CR injection, transport, and cooling is necessary to constrain the effect of CRs on galaxy evolution. Due to their very high gas and radiation densities, strong magnetic fields, and evident outflows, rapidly star-forming galaxies (``starbursts") like the nearby super-wind system M82, serve as ideal laboratories to test physical models in a system significantly different from the Milky Way. In particular, the starburst core of M82 is a disk with a diameter $\sim$500\,pc and gas scale height of order $\sim$50\,pc \citep{Lynds_Sandage_1963,1997MNRAS.291..517W,2002A&A...383...56G}, that is characterized by a large star formation and SN rate of $0.02-0.1$\,SN\,yr$^{-1}$ \citep{2003ApJ...599..193F} and a large gas reservoir of $2.0\times 10^8$\,M$_{\sun}$ \citep{2002A&A...383...56G}, making it an ideal environment to observe the effects of CR interactions.   

Previous analytic and one-zone numerical diffusion/advection models have focused on simultaneously modeling the integrated radio and gamma-ray emission of starburst galaxies, including M82 (e.g., \citealt{Torres,Thompson2007, LTQ, Lacki2011,lacki_thompson, yoast-hull, 2014A&A...567A.101P, 2016ApJ...821...87E, 2019MNRAS.tmp.1118P, 2019MNRAS.484.3665Y}). Overall, these studies highlight the importance of (1) secondary electrons and positrons from hadronic interactions in reproducing the radio emission \citep{Torres,Rengarajan,LTQ}, (2) bremsstrahlung and ionization losses in setting the spectral slope of the integrated radio emission \citep{Thompson2006}, and (3) hadronic interactions with the dense ISM in producing the bright gamma-ray emission --- so called, CRp ``calorimetry" \citep{Loeb_Waxman,Thompson2007}. Importantly, because these models are single-zone and tuned to fit the integrated emission, they did not exploit the spatially extended features of the galaxy seen in the radio bands.

\subsection{A Multi-Dimensional Model}

In this paper, we model M82 using a suite of axisymmetric {\tt GALPROP}\footnote{https://galprop.stanford.edu/} calculations that self-consistently include CR injection, diffusion, advection, and loss processes, with the aim of constraining the dynamical importance of CRs and magnetic fields in the core and halo of this important local starburst wind system. Like previous one-zone models, we match the integrated gamma-ray emission and radio emission from the starburst core. However, we significantly improve previous models by using the resolved radio continuum emission along the minor axis of M82 to directly probe the CR, magnetic field, and wind parameters in this region by connecting the integrated constraints with models for the wind physics. In particular, the radio morphology and spectral index along the minor axis provides a powerful probe that can be utilized to constrain the gas density, magnetic field and CR energy densities, the velocity of CR advection, and the CR diffusion constant. This modeling then allows us to constrain the large-scale gradients in both the CR and magnetic pressure to understand their importance in driving the observed outflow.

We show that under a variety of assumptions, CRs are dynamically weak with respect to gravity in the starburst core and along the minor axis, limiting their ability to accelerate large-scale, heavily mass-loaded winds. Although we substantiate this conclusion here with a suite of {\tt GALPROP} models that demonstrate the resilience of these results to changes in our model assumptions, the basic conclusion that CRs are weak with respect to gravity in the cores of dense, gas-rich, rapidly star-forming galaxies follows from earlier work (see, e.g., \citealt{Thompson_Lacki}). In short, the shallow radio spectral indices of starbursts imply that relativistic bremsstrahlung and ionization losses dominate the cooling of the CRe population, implying that the CRe must interact with gas at nearly the mean density of the ISM. Since relativistic CRe and CRp have similar CR transport properties, CRp will also largely interact with this same gas. One then finds that the equilibrium energy density of CRp is set by the hadronic loss time and that the CRp energy density is small with respect to the corresponding pressure required by hydrostatic equilibrium. 

This result is broadly applicable to star-forming galaxies that have large gas densities, which quickly cool the accelerated CRp population via pion losses, but it does not typically apply to galaxies like the Milky Way, which have smaller gas densities and weaker CRp pion cooling. Quantitatively, \cite{LTQ} find that CRs become dynamically weak with respect to gravity for galaxy-averaged gas surface densities above $\Sigma_g\gtrsim25-50$\,M$_\odot$ pc$^{-2}$ (see their Section 5.6, Fig.\ 15). Specifically for the case of M82, \cite{Lacki2011} (see their Section~6.3) find that the CR pressure is only 2\% of the pressure required for hydrostatic equilibrium if CRs interact with ISM at its mean density. They show that such a scenario is required by both the shallow radio spectral index and the luminous observed gamma-ray emission relative to the star formation rate. 

This paper presents our 2D models of the starburst galaxy M82 and our constraints on the CR population, the magnetic field strength, and other properties as a function of distance along the minor axis. In particular, we produce a first comparison of cosmic ray propagation models with the resolved extended radio halo emission. In Section \ref{section:model}, we present our implementation of {\tt GALPROP} and our model parameters. Section \ref{section:results} presents the results, and compares our models with the integrated and resolved data, including a discussion of the effects of various modeling parameters on the results. We also calculate the cosmic-ray spectrum as a function of position. In Section \ref{section:discussion}, we discuss the implications of our models on scenarios where cosmic rays or magnetic fields drive galactic outflows. Specifically, we show that cosmic rays are dynamically weak with respect to gravity while the gradient of the magnetic field energy density is comparable to gravity. Lastly, we summarize our results, provide additional discussion, and conclude in Section \ref{section:conclusion}.

\section{Numerical Model}
\label{section:model}

\begin{table}
	\begin{center}
		\begin{tabular}{ccc}
		    \toprule
			Distribution & Core Value & Outside Core Spatial Dependence \\
			\midrule \\[-1ex]
			$B(\vec r)$ & $B_0$ & $\max \left\{ B_\mathrm{exp}, B_\mathrm{pow} \right\}$ \\[4ex]
			& & $ \dfrac{ B_\mathrm{exp}}{B_0} = \exp \left(- \dfrac{r-r_\mathrm{core} (\phi)}{r_\mathrm{scale} (\phi)} \right)$ \\[4ex]
			& & $ \dfrac{B_\mathrm{pow}}{B_0} = \left(\dfrac{r -z_\mathrm{core} + z_\mathrm{scale}}{z_\mathrm{scale}}\right)^{{{\mathlarger{\mathlarger{{-\beta}}}}}}$ \\[4ex]
			$n(\vec r)$ & $n_0$ &  $\max \left\{ n_\mathrm{exp}, n_\mathrm{wind} \right\}$  \\[4ex]
			& & $ \dfrac{n_\mathrm{exp}}{n_0} = \exp \left(- \dfrac{r-r_\mathrm{core} (\phi)}{r_\mathrm{scale} (\phi)} \right)$ \\[4ex]
			& & $n_\mathrm{wind} = \dfrac{\dot M}{4\pi r^2 V_n m_p}$ \\[4ex]
			$q_{\mathrm{CR}}(\vec r)$ & $q_{\mathrm{CR},0}$ & $\exp\left(-\dfrac{r - r_\mathrm{core}(\phi)}{r_{\mathrm{q}}(\phi)}\right)$  \\[4ex]
			$q_{\mathrm{ISRF}}(\vec r)$ & $q_{\mathrm{ISRF},0}$ & $\exp\left(-\dfrac{r-r_\mathrm{core}(\phi)}{r_{\mathrm{core}}(\phi)}\right)$  \\ \\[-6pt]
			\bottomrule
		\end{tabular}
	\end{center}
	\caption{ The key spatial distributions utilized in this study. $B$~($\mu$G), $n$~(cm$^{-3}$), $q_{\mathrm{CR}}$~(cm$^{-3}$\,s$^{-1}$\,MeV$^{-1}$), and $q_{\mathrm{ISRF}}$~($\mu$eV\,cm$^{-3}$\,s$^{-1}$\,$\mu$m$^{-1}$) are the magnetic field, gas density, CR source, and ISRF source functions, respectively. The default values for each parameter are given in Table~\ref{parameters}. Each distribution is continuous, having a constant value in an ellipsoidal core with an exponential/power-law suppression in spherical radius, $r=(R^2+z^2)^{1/2}$. For the ellipsoidal shape, we define $ \phi =  \tan^{-1} |z|/R$ and \mbox{$r_i(\phi) = ( (\cos\phi/R_i)^2+(\sin\phi/z_i)^2 )^{-1/2} $} for $i=\{ \mathrm{core, scale, q} \}$. $B$ has a spherical power-law index of $-\beta$ outside the core. $q_{\mathrm{CR}}$ has the same form for both CRp and CRe, with different normalizations but identical rigidity dependencies proportional to $\mathcal{R}^{p_0}$, where $\mathcal{R}$ is the rigidity and $p_0$ is the CR injection spectral index (See Table~\ref{parameters} and Section~\ref{section:model:galprop:injection}). The normalization of $q_{\mathrm{CR},0}$ is determined by fitting the models to data (see Section~\ref{section:model:observations}). $q_{\mathrm{ISRF}}$ has a frequency dependence fit to \citet{1998ApJ...509..103S} and we use it to calculate $U_\mathrm{ISRF}$, the energy density of the ISRF at all positions (see Section~\ref{section:model:galprop:isrf}). We also note that we have an additional gas density component from the observed wind that depends on the ratio of the mass-loss rate to the wind velocity, $\dot M/V_n$ (See Section~\ref{section:model:galprop:mag_gas}). 
	}
	\label{distributions}
\end{table}

\begin{table}
	\begin{center}
		\begin{tabular}{lcccc}
		    \toprule
			Parameter & A & B & B$'$ & Search \\ \midrule
			\multicolumn{5}{|c|}{} \\[-12pt]
			\midrule
			\multicolumn{5}{c}{{\bf Core Scale Lengths} \textsection{\ref{section:model:galprop:resolution}}}  \\
			\midrule
            $R_{\mathrm{core}}$ (kpc) & 0.2 & --- & --- & \\
            $z_{\mathrm{core}}$ (kpc) & 0.05 & --- & --- & \\
            \midrule
            \multicolumn{5}{c}{{\bf CR Source Parameters} \textsection{\ref{section:model:galprop:injection}}} \\
			\midrule
			${Q_{\mathrm{CR},p}}/{Q_{\mathrm{CR},e}}$ & 10 & --- & --- & \\
			$p_0$ & -2.2 & --- & --- & \\
			$R_{\mathrm{q}}$ (kpc) & 0.02 & --- & --- & \\
            $z_{\mathrm{q}}$ (kpc) & 0.005 & --- & --- & \\
			\midrule
			\multicolumn{5}{c}{{\bf Magnetic Field \& Gas Parameters} \textsection{\ref{section:model:galprop:mag_gas}}} \\
			\midrule
			$R_\mathrm{scale}$ (kpc) & 0.2 & --- & --- & \\
			$z_\mathrm{scale}$ (kpc) & 0.2 & --- & --- & \\
			$B_0$ ($\mu$G) & 150 & 325 & 325 & $[10 - 10,000]$ \\
			$\mathlarger{\beta}$ & 1.0 & 1.2 & 0.2 & $\ge 0$ \\
			$n_0$ (cm$^{-3}$) & 150 & 675 & 1000 & $[10 - 10,000]$ \\
			${\dot M}/{V_n}$ $\left(\frac{M_\odot\,\mathrm{yr}^{-1}}{\mathrm{km}\,\mathrm{s}^{-1}} \right)$  & 0.025 & 0.15 & 0.15 & $[0.005 - 0.5]$ \\
            ${n_\mathrm{HI}}/{n_\mathrm{HII}}$ & 19 & --- & --- & \\
            $C_\mathrm{ff}$ & 20.7 & 0.784 & 0.483 & $[10^{-5} - 10^2]$ \\
			\midrule
			\multicolumn{5}{c}{{\bf Propagation Parameters} \textsection{\ref{section:model:galprop:isrf} \& \ref{section:model:galprop:propagation}}} \\
			\midrule
			$D_{xx,0}$  (cm$^2$\,s$^{-1}$) & $5\times 10^{27}$ & $10^{28}$ & $10^{28}$ & $[10^{26} - 10^{29}]$ \\
            $\delta$ & 0.31 & --- & --- & \\
			$V_0$ (km\,s$^{-1}$) & 800 & 1000 & 1000 & $[0 - 2000]$ \\
			$U_{\mathrm{ISRF},0}$ (eV\,cm$^{-3}$) & 1000 & --- & --- & \\ \\[-7.5pt]
			\bottomrule
		\end{tabular}
	\end{center}
	\caption{ Numerical values for the parameters of Models A, B, and B$'$. Column ``Search'' denotes the extremal values of our parameter space scan to find best-fit models. Entries shown as ``---'' imply models~B or B$'$ have the same value as Model~A. Parameters given in Table~\ref{distributions} are defined consistently here. Additionally, ${n_\mathrm{HI}}/{n_\mathrm{HII}}$ is the ratio of neutral to ionized gas, $C_\mathrm{ff}$ is the free-free clumping factor, ${Q_{\mathrm{CR},p}}/{Q_{\mathrm{CR},e}}$ is the ratio of the total energy injection rate between protons and electrons, $D_{xx,0}$ is the diffusion coefficient normalization at a rigidity of 4\,GV/$c$, $\delta$ is the rigidity power-law dependence of the diffusion coefficient, $V_{0}$ is the maximum assumed wind velocity, and $U_{\mathrm{ISRF},0}$ is the central energy density of the ISRF.
	}
	\label{parameters}
\end{table}

In this section, we describe our numerical model, which is based on the CR propagation code {\tt GALPROP}. In Section~\ref{section:model:galprop}, we provide a brief overview of the algorithm along with the modifications necessary to model the M82 galaxy. We describe the  existing observations and then the two-staged process employed to match our models to the observations. Lastly, we provide analytic estimates for the cooling and propagation timescales of CRe and CRp in models with parameters similar to those observed in M82.

\subsection{\tt GALPROP}
\label{section:model:galprop}

We use the CR propagation code {\tt GALPROP} to self-consistently model the distribution of CRs and their diffuse emission in M82. {\tt GALPROP} numerically solves for the steady-state solution of the diffusive transport equation on a fixed spatial grid. Using spatially-dependent models for CR sources, magnetic fields, gas densities, and the ISRF, the code calculates the relevant energy losses during CR propagation. Subsequently, the secondary CR production rates are determined and the secondary CRs are propagated. All species are propagated until the CR distribution comes to a steady-state \citep{1998ApJ...509..212S, 1998ApJ...493..694M}. For protons and heavier nuclei, the code accounts for hadronic interactions, fragmentation, radioactive decays, ionization, and Coulomb losses, while for electrons, it takes into account synchrotron, IC, bremsstrahlung, ionization, and Coulomb losses \citep{2007ARNPS..57..285S}. After finding a steady-state solution, {\tt GALPROP} calculates the emission due to synchrotron, $\pi^0$-decay, IC, bremsstrahlung, and free-free processes at every grid point and integrates this emission along the line of sight toward an observer to make a 2D projected image \citep{2013MNRAS.436.2127O}.

While {\tt GALPROP} has typically been used to model CR propagation and emission within the Milky Way (e.g. \citealt{2017arXiv171209755O, 2010ApJ...722L..58S, 2012ApJ...750....3A, 2016A&A...594A..25P, 2011A&A...534A..54S, 2011ApJ...729..106T, 2016ApJ...831...18C, 2019arXiv190305509J}), we have made several necessary adjustments to allow {\tt GALPROP} to accurately model actively star-forming galaxies. For M82, we produce new spatial distributions for all input parameters, including the magnetic field, gas density, source morphology, wind velocity, and ISRF. We modify {\tt GALPROP} to utilize new analytic functions for the magnetic field distribution, gas density distribution, wind models (including the outflow velocity and density prescriptions), and primary CR source distributions. We produce an analytic model for the ISRF, which is described in detail in Section~\ref{section:model:galprop:isrf}. We make physically motivated changes to propagation parameters and investigate their effects. For simplicity and computational efficiency, we restrict ourselves to two spatial dimensions, and treat the system as axisymmetric. We used the output of {\tt GALPROP} and our own line-of-sight integrator to create 2D projections of the diffuse emission, assuming that M82 is located at a distance of 3.5\,Mpc \citep{2009AJ....138..332J} and an inclination angle of 80$^\circ$ \citep{2014A&A...570A..13M}\footnote{http://leda.univ-lyon1.fr/}. For the radio emission, we solve the radiation transport equation taking into account free-free absorption. 

In the following subsections, we discuss our input model parameters and distributions, including the {\tt GALPROP} spatial and energy resolution, the CR injection properties, the ISRF, the magnetic field \& gas densities, and the propagation parameters (diffusion \& advection). 

We note here that our models explicitly ignore several physical effects that might shape the integrated and resolved non-thermal emission of M82. First, we explicitly seek steady-state solutions and thus ignore the potential time-dependence of the CR energy injection on Myr timescales as is indicated by the star formation rate history and associated SN rate \citep{2003ApJ...599..193F}. We also ignore CR reacceleration in shocks in the M82 outflow, which may affect the underlying CRe energy spectrum and the large-scale synchrotron halo. Finally, we also ignore the possible contribution of ``tertiary'' CRe from $\gamma$-rays interacting with the ISRF. We note that \citet{2005A&A...444..403D} and \citet{2011ApJ...728...11I} calculated the gamma-ray optical depth due to the ISRF in M82 and a similar starburst galaxy, NGC 253, and found absorption was only important for gamma-ray emission above $\sim$10\,TeV. \citet{lacki_thompson} did a similar analysis and calculated the tertiary CRe spectrum and their diffuse emission. They find the tertiary CRe contribution to the the integrated gamma-ray spectrum negligible and the contribution to the integrated synchrotron spectrum only important in the X-ray band, which we do not analyze in this paper. We mention these issues again and point to future avenues of investigation in Section~\ref{section:conclusion}.

\subsubsection{Resolution}
\label{section:model:galprop:resolution}
We model the starburst core of M82 as an oblate-ellipsoid with a fiducial cylindrical-radius, $R_\mathrm{core}=200$\,pc, and half-thickness, $z_\mathrm{core}=50$\,pc (These numbers, along with other scale lengths, are summarized in Table~\ref{parameters}). We note that we will denote cylindrical-radius as $R$ and spherical-radius as $r$. Thus, to resolve the starburst core in space and time, we modify {\tt{GALPROP}} to use smaller grids compared to standard Milky Way values. The spherical-radius of the oblate-ellipsoid as a function of azimuthal angle, $\phi$, is given by \mbox{$r_\mathrm{core} (\phi) = ((\cos \phi/R_\mathrm{core})^2+(\sin \phi/z_\mathrm{core})^2)^{-1/2}$}, which we use to parameterize our spatial distributions.

In this paper, we propagate CRs in a 2-dimensional cylindrical axisymmetric geometry. Our model considers a cylindrically \mbox{radial-$R$} (vertical-$z$) spatial domain of $[0,5]$~($[-4,4]$)\,kpc with an bin size of $0.05$~($0.01$)\,kpc. We set the minimum time resolution to be \mbox{10 years} to account for the rapid energy losses of high-energy CRe. We evaluate our CRs on an energy grid spanning from $[1,10^9]$\,MeV in increments of 6.85 bins per decade.

\subsubsection{CR Injection Properties}
\label{section:model:galprop:injection}

We assume that all CR sources are isotropically distributed throughout the starburst core. Outside the core, we assume that the CR source density rapidly decreases exponentially in $r$ with a scale length of \mbox{$r_\mathrm{q} (\phi) = ((\cos \phi/R_\mathrm{q})^2+(\sin \phi/z_\mathrm{q})^2)^{-1/2}$}. $R_\mathrm{q}$ and $z_\mathrm{q}$ are given in Table~\ref{parameters}. We ignore CR sources in the portion of the galactic disk that fall outside the starburst core, since the diffuse radio emission is dominated by the core. The analytic form of the CR source injection function is shown in Table~\ref{distributions}. The normalization of the CR source injection function is obtained using a fit to the radio and gamma-ray data (See Section~\ref{section:model:observations}). 

All primary CRs are injected with a power-law rigidity spectrum, $\mathcal{R}^{p_0}$ ($\mathcal{R}$ is the rigidity), with index $p_0= -2.2$. We note that we use $p_0$ to refer to the \emph{injected} spectral index while the variable $p$ denotes the \emph{observed/steady-state} spectral index. We normalize the injected spectral model for the primary CRe and CRp such that the total relative energy contribution is
\begin{equation}
Q_{\mathrm{CR},p}/Q_{\mathrm{CR},e} = 10,
\label{CRepratio}
\end{equation} 
which we keep constant throughout this analysis. At energies above 1\,TeV, this corresponds to a proton to electron ratio of ${\sim}38$ due to the power-law injection spectrum and a minimum energy of 1\,MeV for both protons and electrons. We take this as a conservatively high value for electron injection, noting that charge conservation implies \mbox{$Q_{\mathrm{CR},p}/Q_{\mathrm{CR},e} \approx16$} \citep{2002cra..book.....S} while simulations show \mbox{$Q_{\mathrm{CR},p}/Q_{\mathrm{CR},e} \gtrsim26$} (e.g. \citealt{2015PhRvL.114h5003P, 2016MNRAS.463..394V} and references therein). The numerical values of all propagation parameters are provided in Table~\ref{parameters}.

For computational simplicity, we only model the propagation of Helium-4, Helium-3, deuterium, and primary protons, assuming the same relative abundances as in the Milky-Way. We also model primary electrons, secondary electrons/positrons, knock-on electrons, and secondary protons/antiprotons. 

\subsubsection{Interstellar Radiation Field}
\label{section:model:galprop:isrf}

The ISRF energy density (from reprocessed starlight) is calculated under the assumption that the starlight is produced by an optically-thin disk with a stellar density that exponentially decays with $r$ with a scale-length of $r_\mathrm{core}(\phi)$. The ISRF model is normalized to an energy density of $1000$\,eV\,cm$^{-3}$ at the center of the galaxy, which is comparable to \citet{lacki_thompson} and \citet{yoast-hull}. Moreover, this value is consistent with a model where half \citep{2002ApJ...574..709M} of the total infrared luminosity of M82, ${\sim}10^{10.7}$\,L$_\odot$ \citep{2003AJ....126.1607S}, is isotropically emitted from a thin galactic disk (i.e. \mbox{$U\approx L\,/\,2\pi r_\mathrm{core}^2 c$}). We assume the spectral dependence of the ISRF is uniform throughout M82, and utilize the best-fit spectral model from \citet{1998ApJ...509..103S} to model the wavelength dependence. We also include the cosmic microwave background, which is negligible compared to the contribution from stars and dust in our region of interest. 

For simplicity, we keep the ISRF constant throughout our analysis. The ISRF energy density as a function of height above the disk is shown in the right panel Figure~\ref{acceleration} as the dashed black line. Changing the normalization of the ISRF by a factor of 2 does not drastically affect the majority of our qualitative results. For further discussion of the ISRF, see Section~\ref{section:results:emission:isrf} and Appendix~\ref{appendix:spectral}.

\subsubsection{Magnetic Field \& Gas Density}
\label{section:model:galprop:mag_gas}

For simplicity, we assume that the magnetic field ($B$) and gas number density ($n$) are constant throughout the core with values of $B_0$ and $n_0$. Outside the core, the magnetic field and gas density is constructed as the maximum of two components. The first component is an exponentially decreasing $B$ and $n$ with $r$, with a scale length \mbox{$r_\mathrm{scale} (\phi) = ((\cos \phi/R_\mathrm{scale})^2+(\sin \phi/z_\mathrm{scale})^2)^{-1/2}$}. The second component for $B$ is a power-law fall off with index $-\beta$ and scale length $r_\mathrm{scale}(\phi)$. The second component for $n$ assumes spherical mass-outflow with a constant mass loss rate, $\dot M$, and a constant wind velocity, $V_n$. For simplicity, we combine $\dot M$ and $V_n$ into a single parameter, $\dot M/V_n$, that sets the normalization of the gas density. Note that the wind velocity $V_n$ need not be equal to the CR advection velocity $V_0$ discussed in Section \ref{section:model:galprop:propagation}.

For functional forms of magnetic field and gas density, see Table~\ref{distributions}. Numerical values for the parameters are given in Table~\ref{parameters}. See Figure~\ref{acceleration} for the shape of the magnetic energy density, \mbox{$U_B = B^2/8\pi$}. See Appendix~\ref{appendix:gas_wind} and the left panel of Figure~\ref{gas_wind} for the shape of the gas density distribution. 

We further assume that the magnetic field is randomly oriented (consistent with recent results from \citet{2017A&A...608A..29A}) and that the gas is locally homogeneous. We add an additional isotropic magnetic field with a strength of $0.1$\,$\mu$G throughout the computational volume. This magnetic field is included primarily to remove artifacts from the {\tt GALPROP} CR diffusion calculation outside the core of the galaxy in cases where we set the wind velocity to zero and the wind component of the gas density to be small. This small background field has no significant effect on the properties of any of our solutions. 

For simplicity, we assume that 5\% of the gas is ionized (i.e. $n_\mathrm{HI}/n_\mathrm{HII} = 19$). This does not have a large effect on our results. The ionized gas is also responsible for free-free absorption and emission in the radio band. To change the amount of free-free processes in fitting the data, we allow the free-free clumping factor, $C_\mathrm{ff}$, to vary. See Section~\ref{section:results:emission:mag_gas} for a more indepth discussion on the effects of $C_\mathrm{ff}$.

\subsubsection{CR Diffusion \& Advection}
\label{section:model:galprop:propagation}

The diffusion coefficient is kept constant throughout the galactic volume but it is rigidity dependent. We use the standard rigidity dependence of {\tt GALPROP}: \mbox{$D_{xx} = D_{xx.0}\,(\mathcal{R}/4\,\mathrm{GeV}c^{-1} )^\delta$} where $D_{xx,0}$ and $\delta$ are given in Table~\ref{parameters}, and $\mathcal{R}$ is the CR rigidity. In reality, the diffusion coefficient may differ inside and outside the core, but we find that outside the core, the wind strongly dominates diffusion unless the diffusion coefficient increases by more than an order of magnitude. For simplicity, we avoid changing the diffusion coefficient in different regions. Changing the diffusion coefficient by a factor of 2 does not affect our solutions, but for larger changes in the diffusion coefficient, we would need to change the wind velocity accordingly to fit the degeneracy discussed in Section~\ref{section:results:extended:diff_wind}.

We incorporate a spherical radial CR advection velocity profile, where the  normalization of the wind density profile, $\dot{M}/V_n$, and the asymptotic CR advection speed, $V_0$, are allowed to vary. The observed wind itself has been extensively studied and has been shown to have multiple components with different velocities, ranging from somewhat slower cold/cool molecular and neutral gas \citep{2015ApJ...814...83L, 2018ApJ...856...61M}), to high velocity warm ionized gas of ${\sim}600$\,km\,s$^{-1}$ \citep{1998ApJ...493..129S}. Additionally, X-ray emission indicates the presence of a hot component that would have a large asymptotic velocity of ${\sim}1000-2000$\,km\,s$^{-1}$ \citep{2009ApJ...697.2030S}. The question of which gas the CRs sample is critical, as this directly affects the advection speed and the density distribution, which in turn affects the relative importance of secondary CRe production in the outflow, CRe bremsstrahlung and ionization losses, and the relative importance of advection versus diffusion. Moreover, self-consistent dynamical CR transport models generally incorporate both diffusion and streaming at a multiple of the local Alfv\'en velocity (e.g., \citealt{1975ApJ...196..107I,2017MNRAS.467..906W,Chan2018,Jiang2018}). Thus, the CR advection velocity $V_0$ need not be identical to the hydrodynamical wind velocity $V_n$ of any given observed wind component. For these reasons, the wind density profiles employed are not directly connected to the CR advection velocity through the hydrodynamical mass outflow rate, as would be the case if the CRs were only advected and did not stream or diffuse. As a consequence, we give ourselves the freedom to vary the CR advection velocity (set by $V_0$) and the wind density profile (set by $\dot{M}/V_n$) independently.

The profile of CR advection velocity we employ is parameterized, but is based on the velocity structure of the energy-driven supersonic wind model of \cite{Chevalier_Clegg}. The advection speed is 0\,km\,s$^{-1}$ at the disk midplane and reaches half of its maximum velocity at a scale length of 200\,pc. The maximum velocity, $V_0$, is specified for each given model (see Table~\ref{parameters}). We consider values of $V_0$ from $0-2000$\,km\,s$^{-1}$. We assume that within the disk, the wind is only perpendicular to the disk. For this, we multiply the $R$ component of the wind by $1-\exp (-|z|/z_\mathrm{core})$. See Appendix~\ref{appendix:gas_wind} for more discussion and the right panel of Figure~\ref{gas_wind} for the profile of the magnitude of the CR advection speed along the minor axis. We explore the importance of the wind velocity structure  and diffusion for our best-fit models in Sections~\ref{section:results:extended:diff_wind} \& \ref{section:results:extended:beta_wind}.

With this modified version of {\tt{GALPROP}}, we obtain steady state solutions for the CR distribution and emission for M82 that are consistent with observations. 

\subsection{Comparison with Observations}
\label{section:model:observations}

A very broad range of input parameters -- $n_0$, $B_0$, etc. (see right column of Table \ref{parameters}) -- are explored. For each choice of {\tt GALPROP} inputs, we evolve the simulation until the system reaches a steady-state solution, and compare the resulting non-thermal emission with observations. We compare the emergent $\gamma$-ray emission from each model with the spatially unresolved $\gamma$-ray emission observed by the Fermi-LAT (first observed by \citet{2010ApJ...709L.152A}) and VERITAS \citep{2009Natur.462..770V} telescopes. To provide an improved spectral constraint on the M82 GeV $\gamma$-ray emission, we re-analyzed Fermi-LAT data taken from August~4,~2008 to July~26,~2018. We produced a binned analysis utilizing eight energy bins per decade spanning the range \mbox{100\,MeV---100\,GeV}, following standard procedures for data selection, likelihood fitting and diffuse modeling. We obtained results that are statistically consistent with the intensity and spectral parameters of M82 obtained in the 3FGL catalog. 

In addition, we use the spatially integrated radio emission from \citet{2010ApJ...710.1462W, 2013A&A...555A..23A, 2015A&A...574A.114V}; and \citet{1988A&A...190...41K}. For data on the spatial extension of the radio halo, we quantitatively compare our models to \citet{2013A&A...555A..23A}, who analyzed the halo at 92\,cm~(326\,MHz), 22\,cm~(1.36\,GHz), 6\,cm~(5\,GHz), and 3\,cm~(10\,GHz). In Section~\ref{section:results:extended}, we convolve our models with the appropriate beam size, then integrate the radio emission within 30"~(0.5\,kpc) of the minor axis and compare the output to the data within a distance of 210"~(3.5\,kpc) above/below the major axis. We provide an additional comparison of our models to \citet{2015A&A...574A.114V} in Appendix~\ref{appendix:emission_maps}.

We set the overall normalization for each {\tt GALPROP} model by allowing the normalization of the CR energy injection rate to float to best-fit the available data. While we could use the star-formation rate data to normalize the cosmic-ray injection parameters as an input, we note that significant uncertainties remain in the expected SN rate and the efficiency of cosmic-ray injection per unit star formation. However, we find that our best-fitting models have implied CR injection rates and SN rates that are consistent with the star formation rate implied by the global far-infrared spectral energy distribution.

\subsubsection{Fitting Procedures}
\label{section:model:observations:fitting}

Our attempts to fit the {\tt GALPROP} models to the data immediately reveal a complicated $\chi^2$ space. In an effort to understand the interplay between the large number of parameters controlling the models, we have opted to fit the data in several stages. In the first stage, we make a preliminary scan through a restricted parameter space by varying $B_0$, $n_0$, and $C_\mathrm{ff}$, while holding  $\beta=1$, $\dot M/V_n=0.005$, $D_{xx,0}=5\times 10^{28}$\,cm$^{2}$\,s$^{-1}$, and $V_0=500$\,km\,s$^{-1}$ fixed (see Table~\ref{parameters}). We call this ``Model~I''. With this constrained model, we focus exclusively on the fit to the integrated emission in order to understand how $B_0$, $n_0$, and $C_\mathrm{ff}$ connect when the CR injection rate is left to freely vary (Section~\ref{section:results:emission:mag_gas}). We find a strong degeneracy between $B_0$ and $n_0$, manifest as a minimum in $\chi^2$ shown in Figure~\ref{chi_SNR_plot}. As we discuss in more detail in Section~\ref{section:results:emission:mag_gas} and derive in Appendix~\ref{appendix:spectral}, this degeneracy arises from the interplay between $B_0$-dependent synchrotron cooling and both $n_0$-dependent losses for both CRp and CRe. $B_0$ and $n_0$ affect both the detailed shape of the radio spectrum and the ratio of the integrated synchrotron and $\gamma$-ray luminosities with the result being the positive degeneracy shown in Figure~\ref{chi_SNR_plot}.

With the global space of the \emph{integrated} emission from Model~I mapped, in the second stage of our comparison to the data, we select models that reproduce the \emph{spatially-resolved} extended radio emission along the minor axis, while simultaneously fitting the observed $\gamma$-ray spectrum. Best-fitting models were selected on the basis of comparison with the radio maps of \citet{2013A&A...555A..23A} across 4 radio wavelengths. For a chosen magnetic field strength ($B_0$), we varied the normalization of the gas number density ($n_0$), the spatial power-law drop-off for magnetic field ($\beta$), the wind component of the gas density (whose normalization depends on $\dot M/V_n$), the free-free clumping factor ($C_\mathrm{ff}$), the diffusion coefficient ($D_{xx}$), and the maximum wind velocity ($V_0$).

Due to the $B_0{-}n_0$ degeneracy of Figure~\ref{chi_SNR_plot}, there is a complicated locus of parameters that can be made to fit the spatially-resolved data. In order to make the discussion manageable, we choose two representative points in the $B_0{-}n_0$ plane near the $\chi^2$ minimum picked out by the search with Model~I. We call these Model~A and Model~B. However, we find that the magnetic power-law drop off, $\beta$, also has a large impact on our models (See Section~\ref{section:results:extended:beta_wind}), thus we choose a third model, Model~B$'$, that has the same magnetic field as Model~B with a different $\beta$ that also causes a slight change in best-fitting $n_0$. 

These three models form the basis of much of our discussion in the rest of the paper. Model~A has a relatively low gas density and low magnetic field strength, with a higher proportion of free-free emission at higher radio frequencies, whereas Model~B and B$'$ have higher $B_0$ and $n_0$, and a lower contribution from free-free emission. These models have some qualitatively different features that roughly bracket the available parameter space in the $B_0{-}n_0$ plane. All three are chosen to fit the data well. All three models fall somewhat off the best-fit locus identified in Figure~\ref{chi_SNR_plot} with Model~I because models A, B, and B$'$ include the spatially-resolved radio information.

The final parameters for Models~A, B, and B$'$ are given in Table~\ref{parameters}. We provide a detailed discussion of how variations in the physical parameters affect the model fits in Section~\ref{section:results}.

\begin{figure*}
	\centering
	\makebox[0.49\linewidth][c]{ \includegraphics[width=0.51\linewidth]{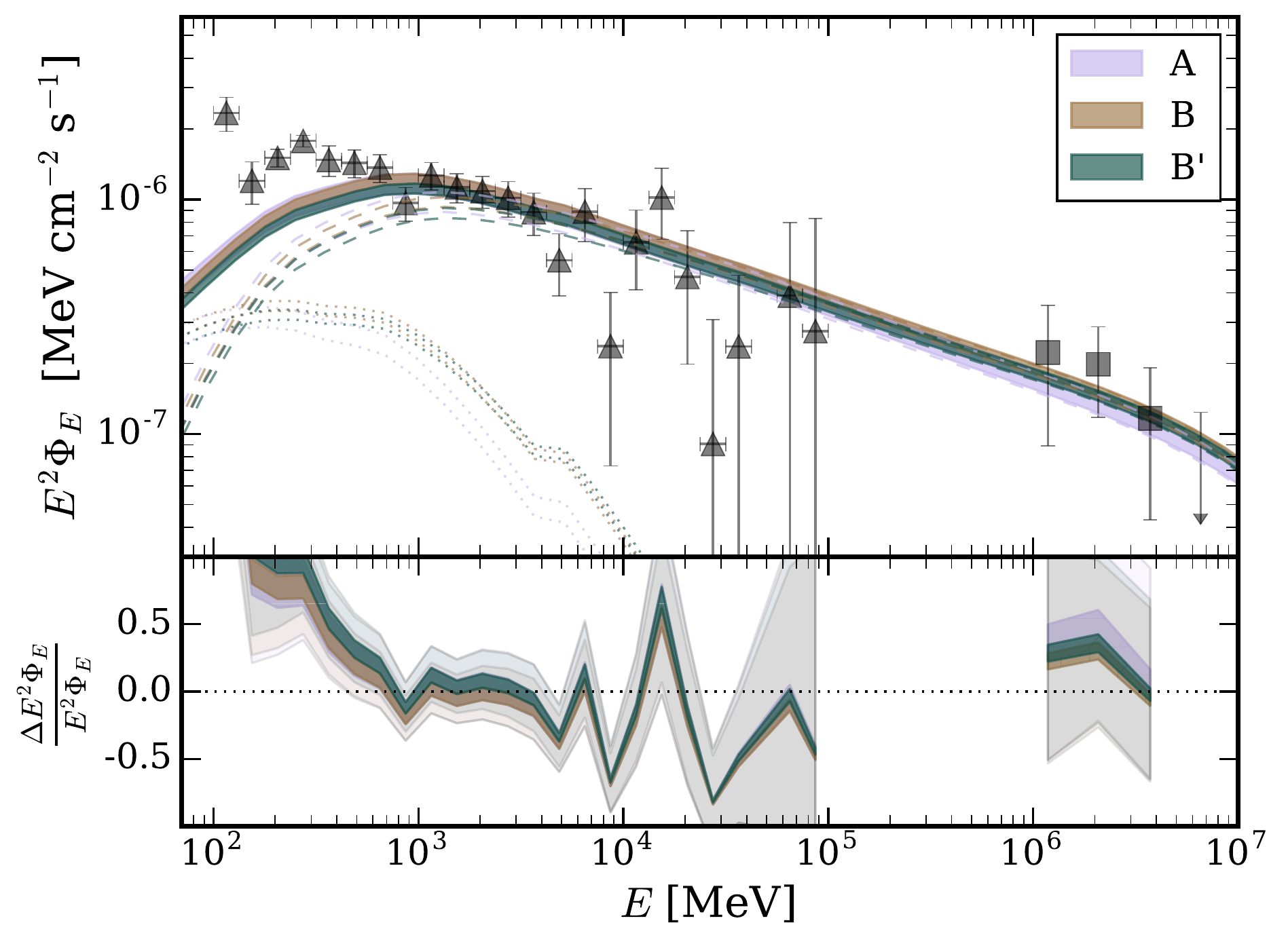} }
	\hspace{0pt}
	\makebox[0.49\linewidth][c]{ \includegraphics[width=0.51\linewidth]{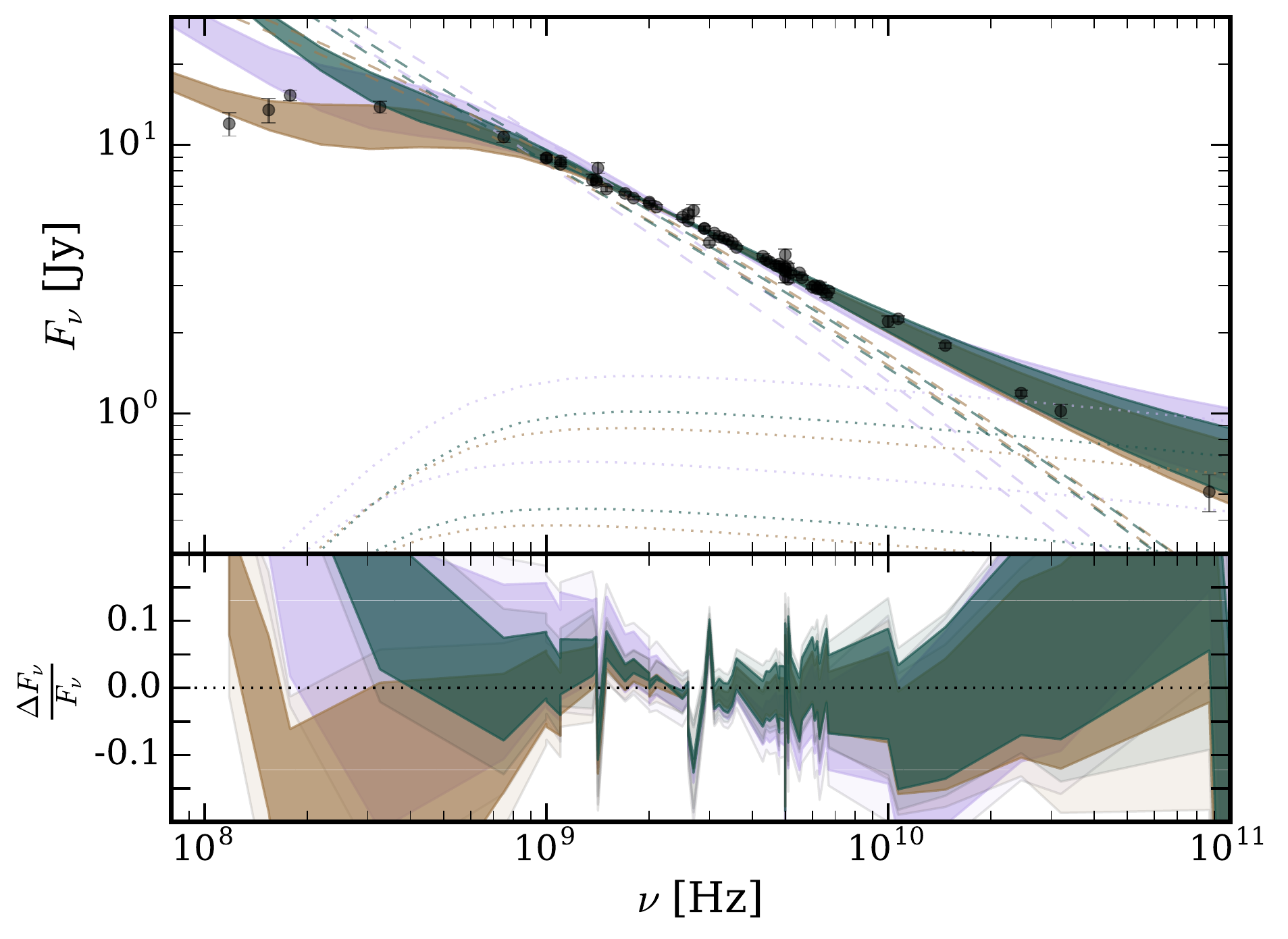} }
	\caption{ The integrated gamma-ray (left) and radio (right) fluxes from Model~A (lavender), Model~B (brown), and Model~B$'$ (green). The width of the band indicates the effect of increasing/decreasing the free-free clumping parameter by a factor of $\sqrt{2}$ compared to the default value given in Table~\ref{parameters}. Increasing the free-free clumping factor increases the radio emission and decreases the gamma-ray emission. We plot gamma-ray data from the Fermi-LAT (triangles) (See Section~\ref{section:model:observations}) and VERITAS (squares) \citep{2009Natur.462..770V}. The gamma-ray emission components stemming from $\pi^0$-decay (dashed) and bremmstrahlung (dotted) are also shown, while the emission from IC falls below the horizontal axis. We compare radio observations with data from multiple sources~\citep{2010ApJ...710.1462W,2013A&A...555A..23A,2015A&A...574A.114V,1988A&A...190...41K} and, additionally, we plot the free-free (dotted) and the unabsorbed synchrotron (dashed) spectra. At the bottom of each plot, we show the difference between our models and the data points normalized by our models. The lighter lavender, brown, and green regions in the error plot show the 1-$\sigma$ error bars for all three models. }
	\label{AB_integrated}
\end{figure*}

\subsection{Analytic Estimates of Key Propagation Timescales}

To better interpret the results of subsequent sections, we record the CR cooling, diffusion, and advection timescales for parameters appropriate to M82 for easy reference.

In the case of CRe, our models are primarily constrained by the resolved radio observations of the starburst core and radio halo. Thus, it is useful to report the electron cooling time as a function of the critical synchrotron emission frequency for a given magnetic field strength rather than as a direct function of the CRe energy \citep{rybicki,ginzburg}. The characteristic emission frequency is related to CRe energy and $B$ by:
\begin{equation}
\nu_{1} \approx\, 0.36 \ E_1^2 \ B_{100}, \label{e_nu_b}
\end{equation}

\noindent where \mbox{$\nu_1~=~\nu\,/\,1\,\mathrm{GHz}$}, \mbox{$E_1~=~E\,/\,1\,\mathrm{GeV}$} is the CRe energy, and \mbox{$B_{100}~=~B\,/\,100\, \mu\mathrm{G}$} is the magnetic field strength. Using this relation, we can write the relativistic bremsstrahlung, synchrotron, inverse Compton, and ionization CRe cooling timescales, respectively, as:
\begin{align}
\tau_{\text{bremss}} \simeq\,& 3.2 \times 10^5 \ n_{100}^{-1} \quad \mathrm{yr} \label{tau_bremss} \\
\tau_\text{synch} \simeq\,& 7.9 \times 10^5 \ \nu_1^{-\frac12} \ B_{100}^{-\frac32} \quad \mathrm{yr} \label{tau_synch} \\
\tau_\text{IC} \simeq\,& 1.9 \times 10^5 \ \nu_1^{-\frac12} \ B_{100}^{\frac12} \ U_{1000}^{-1} \quad \mathrm{yr} \label{tau_ic} \\
\tau_{\text{ion}} \simeq\,& 3.7 \times 10^6 \left( 1 - 0.75 f_\mathrm{ion} \right) \ \nu_1^{\frac12} \ B_{100}^{-\frac12} \ n_{100}^{-1} \quad \mathrm{yr}, \label{tau_ion}
\end{align}

\noindent where \mbox{$n_{100}~=~n\,/\,100\,\mathrm{cm}^{-3}$} is the gas density and \mbox{$U_{1000}~=~U_{\mathrm{ISRF}}\,/\,1000\,\mathrm{eV}$} is the ISRF energy density. For \mbox{$\tau_\mathrm{ion}$}, we approximated a range of timescales based on the ionization fraction of the gas, \mbox{$f_\mathrm{ion}$}, and have ignored terms that are logarithmically dependent on energy.

We note that the synchrotron spectral index, $\alpha$, is determined by the steady-state electron+positron spectral index, $p$, following \mbox{$\alpha~=~(p+1)/2$}. The above equations then imply that if the CRe spectrum is dominantly cooled via synchrotron or IC, the radio spectral index will be $-1/2$ smaller than if there were no cooling. If CRe cooling is instead dominated by bremsstrahlung, then the radio spectral index is identical to that expected for an uncooled population. Finally, if ionization dominates CRe cooling, then the radio spectral index is $1/2$ larger than the radio spectral index of an uncooled or bremsstrahlung cooled CRe spectrum. Thus, the observed synchrotron spectral index provides a powerful diagnostic on the ratio of the gas density to the ISRF and magnetic field energy densities \cite{Thompson2006}.

We also note the magnetic field dependence of these timescales. As the magnetic field strength increases, \mbox{$\tau_\mathrm{IC}$} gets larger while \mbox{$\tau_\mathrm{ion}$} decreases. This is due to the fact that we are examining emission at a particular synchrotron frequency, the value of which depends on the energy of the CRe \emph{and} the magnetic field strength.

CRp cooling is dominated by hadronic losses, following a timescale calculated by~\citet{2015ApJ...802..114K}:
\begin{equation}
\tau_\mathrm{had} \simeq 8.2\times 10^5\ n_{100}^{-1}\ E_1^{-0.28}\ \left( E_1+200 \right)^{0.2} \quad \mathrm{yr}.
\label{tau_hadronic}
\end{equation}
Ionization losses of CRp are not important above a few hundred MeV. 

In addition to radiative cooling, CRs can be lost due to advection or diffusion. Thus, other important timescales include the wind and diffusive propagation times for CRs to leave the starburst core. These values are identical for both CRe and CRp.
\begin{align}
\tau_\mathrm{wind,core} \simeq\,& \frac{z_\mathrm{core}}{V_\mathrm{wind}} \simeq 4.9\times 10^5 \ z_{50} \ V_{100}^{-1} \quad \mathrm{yr} \label{tau_wind} \\
\tau_\mathrm{diff,core} \simeq\,& \frac{z_\mathrm{core}^2}{D_{xx}} \simeq 7.6\times 10^5 \ z_{50}^2 \ D_{10^{27}}^{-1} \quad \mathrm{yr} \label{tau_diff}
\end{align}
where \mbox{$z_{50}~=~{z}\,/\,{50\,\mathrm{pc}}$}, \mbox{$V_{100}~=~{V}\,/\,{100\,\mathrm{km\,s}^{-1}}$}, and \mbox{$D_{10^{27}}~=~{D_{xx}}\,/\,{10^{27}\,\mathrm{cm}^2\,\mathrm{s}^{-1}}$} where $z$ is the vertical height of the core and $V$ is the wind velocity. In our models, the diffusion coefficient has the rigidity dependence defined in Section~\ref{section:model:galprop:propagation}. Additionally, we note that in the center of the starburst core \mbox{$V=0$\,km\,s$^{-1}$}, implying that diffusive processes dominate in this region. Finally, although {\tt GALPROP} does not include streaming losses and instead treats the CR transport in the diffusive regime, we note that the characteristic Alfv\'en time is of order
\begin{equation}
    \tau_A\simeq\frac{z_{\rm core}}{V_A}\simeq 2\times10^6\ z_{50}\ B_{100}^{-1}\ n_{100}^{1/2} \quad \mathrm{yr}.
\end{equation}

\begin{figure}
	\centering
	\makebox[\linewidth][c]{ \includegraphics[width=1.1\linewidth]{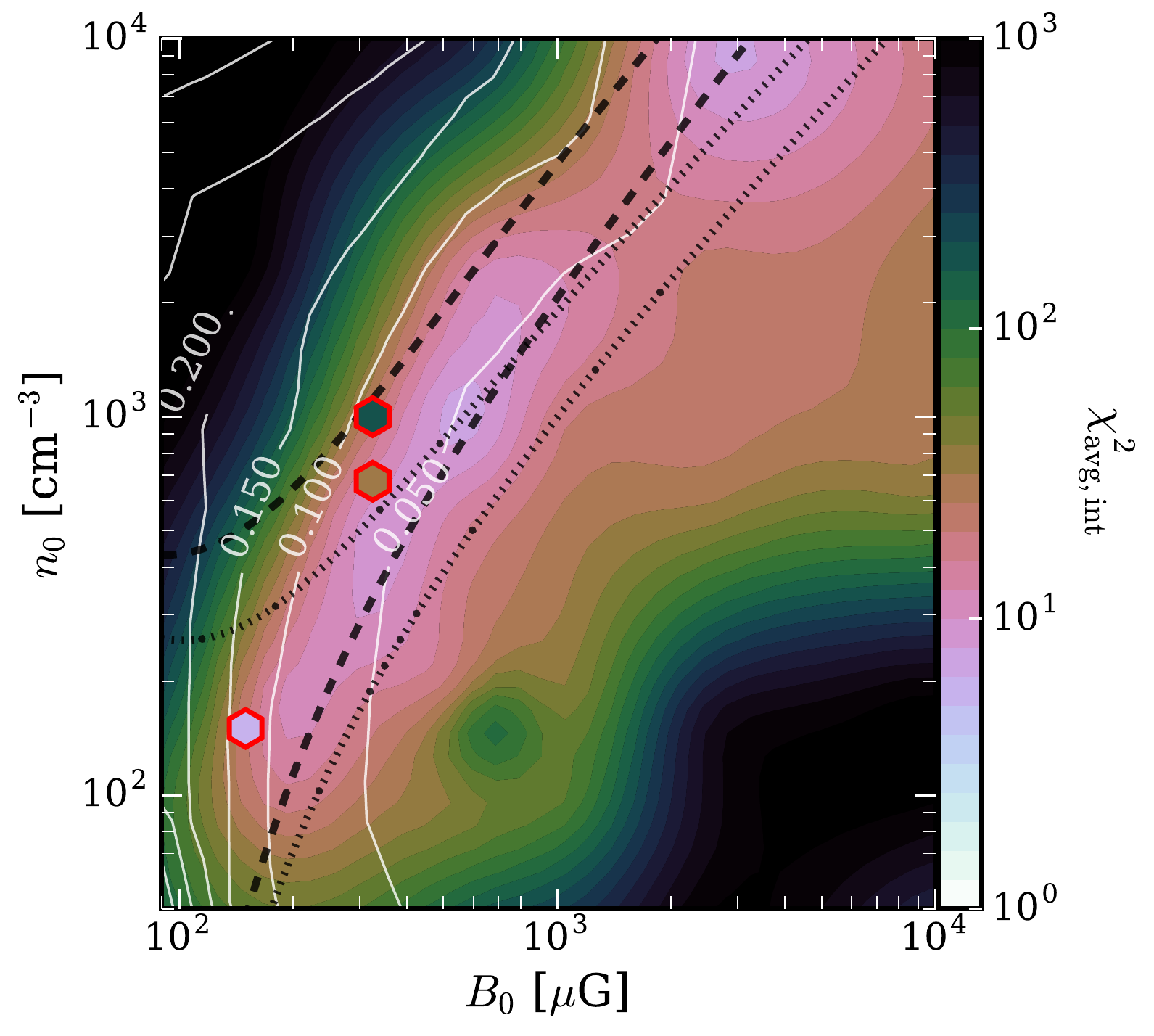} }%
	\caption{ Magnetic field${-}$gas density relation. The filled color contours are the $\chi_{\mathrm{avg,int}}^2$ (see text for definition) of our models. We iterate over the starburst core values of magnetic field and gas density while minimizing over the free-free clumping factor. The solid white lines give the contours of the SN rate (SN\,yr$^{-1}$), assuming a SN kinetic energy of $10^{51}$ ergs with 10\% of the energy going into CRs. The dashed and dotted black lines are a range of best fitting parameters from our analytic estimates (see Appendix~\ref{appendix:spectral}). The dashed (dotted) black lines assume the gas is neutral (ionized). The upper (lower) dashed/dotted line has an CRe spectral index $p=-2.4$ ($-2.6$), respectively. The colored hexagons denote models A (lavender), B (brown) and B$'$ (green). }
	\label{chi_SNR_plot}
\end{figure}

\section{Results}
\label{section:results} 

Here, we present our {\tt GALPROP} models for M82. We divide our analysis into two sections. Section~\ref{section:results:emission} examines the integrated gamma-ray and radio fluxes from M82, which are dominated by the dynamics of the starburst core. This section builds on previous one-zone models by \citet{lacki_thompson} and \citet{yoast-hull}. Section~\ref{section:results:extended} takes advantage of our 2D {\tt GALPROP} models to provide the first comparison between the modeled radio morphology and radio observations above and below the M82 galactic plane. We utilize these observations to strongly constrain the underlying CR population and magnetic field strength along the minor axis of M82. These comparisons imply that the large scale CR pressure gradient is dynamically weak with respect to gravity, while the gradient in the magnetic energy density may be dynamically strong with respect to gravity (See Section~\ref{section:discussion}).

\subsection{Integrated Emission from M82}
\label{section:results:emission}

In Figure~\ref{AB_integrated}, we show the integrated emission from the starburst-galaxy M82 in gamma-rays (left) and radio (right). We show gamma-ray data from both the Fermi-LAT and VERITAS \citep{2009Natur.462..770V}, as well as integrated radio data from \citet{2010ApJ...710.1462W,2013A&A...555A..23A,2015A&A...574A.114V}; and \citealt{1988A&A...190...41K}. Using these datasets, we constrain our CR propagation models in the parameter plane of the core magnetic field strength ($B_0$) and core gas density ($n_0$), and choose three candidate models which provide excellent fits to the combined data and illustrate unique regions of the full parameter space.

\subsubsection{Magnetic Field$-$Gas Density Relation}
\label{section:results:emission:mag_gas}

The characteristics of the integrated gamma-ray and radio spectra (i.e. normalizations and spectral indices) are largely determined by $B_0$ and $n_0$. Roughly speaking, in the non-calorimetric (i.e. losses are dominated by escape) limit for CRs, a larger $B_0$ would increase the total synchrotron power (proportional to $B_0^2$) while a larger $n_0$ would increase the amount of emission from bremsstrahlung and \mbox{$\pi^0$-decay} (both proportional to $n_0$). In this case, just the normalizations of the gamma-ray and radio emission would be enough to directly constrain our models.

However, as CRs approach calorimetry, the spectral normalizations lose their constraining power and a degeneracy between $B_0$ and $n_0$ appears due to the complicated dynamics between the CRe energy losses and the creation of CRe secondaries. Perhaps non-intuitively, a larger value of $n_0$ can increase the total power in radio CRe emission because secondary CRe are created from CRp hadronic interactions with the gas. As an extreme example, if we assume we have a purely \emph{secondary} CRe population that is dominantly cooled by synchrotron, then the CRe energy density, and thus the synchrotron flux, is proportional to the secondary CRe injection rate, which is proportional to the gas density. At the other extreme, if we assume we have a dominantly bremsstrahlung cooled \emph{primary} CRe population, then the primary CRe energy density is proportional to the bremsstrahlung cooling timescale, implying the resulting synchrotron flux is inversely proportional to the gas density.  

In general, as CRe become calorimetric, a larger $B_0$ increases the total synchrotron power, but also decreases the amount of power into bremsstrahlung and IC. Similarly, a larger $n_0$ increases the total bremsstrahlung power, but also decreases the amount of power into synchrotron and IC. However, there is a special case when synchrotron losses dominate the CRe cooling. A larger $B_0$ will \emph{not} increase the power into synchrotron since the radiated synchrotron emission would proportional to the CRe injection rate, which is independent of $B$. Similarly, there is a special case for hadronic-loss dominated CRp in the calorimetric limit, thus implying there would be no $n_0$-dependence in the \mbox{$\pi^0$-emissivity}. As a consequence, there is no way to simultaneously constrain $B_0$ and $n_0$ if synchrotron losses dominate for CRe \emph{and} hadronic losses dominate for CRp. Otherwise, the magnetic field${-}$gas density ($B_0{-}n_0$) degeneracy should exist.

The true degeneracy becomes even more complicated because the degree of calorimetry can change significantly as a function of the CR energy. Both $B_0$ and $n_0$ affect the shape of the radio spectrum as illustrated by the CRe energy-loss timescales (Equations~\ref{tau_bremss}--\ref{tau_ion}). All else fixed, increasing $B_0$ steepens/softens the radio spectrum (especially at higher frequencies) due to synchrotron cooling, while increasing $n_0$ flattens/hardens the radio spectrum (especially at lower frequencies) as a result of bremsstrahlung and ionization cooling of the underlying CRe population.

In addition to complexities stemming from the incomplete CRp calorimetry of M82, the $B_0{-}n_0$ degeneracy is also complicated by the fact that the radio flux is not purely produced through synchrotron processes. The radio spectrum also depends on the density of ionized gas and its ``clumpiness'' through free-free emission at high frequencies (\mbox{$\nu\gtrsim 10^{10}$\,GHz}) and absorption at low frequencies (\mbox{$\nu \lesssim 1$\,GHz}). Keeping the free-free clumping factor ($C_\mathrm{ff}$) constant while increasing the ionized gas density increases the amount of free-free absorption and emission, which is frequency-dependent. For simplicity, we account for these processes by using a fixed fraction of ionized gas, \mbox{$n_\mathrm{HII} = 0.05\,(n_\mathrm{HI}+n_\mathrm{HII})$}, and allow only $C_\mathrm{ff}$ to vary. We note that there is a degeneracy between the ionization fraction of the gas and the clumping factor, as both the absorption coefficient and emissivity of free-free processes are proportional to $C_\mathrm{ff}$ and $n_\mathrm{HII}^2$. Since ionization losses are stronger for ionized gas compared to neutral gas (See Equation~\ref{tau_ion}), changing the ionization fraction would also have an effect on the final CR spectra. However, since ionization only dominates for CRs $\lesssim{1}$\,GeV, a different ionization fraction would have a negligible effect on the final radio and $\gamma$-ray spectra, thus keeping a constant ionization fraction is an appropriate simplification for our analysis.

Figure \ref{chi_SNR_plot} presents the magnetic field${-}$gas density ($B_0{-}n_0$) relation for M82. Specifically, we vary a fiducial model, Model~I (see Section~\ref{section:model:observations:fitting}), over a 3D logarithmic grid of $B_0$, $n_0$, and $C_\mathrm{ff}$. We then minimize over $C_\mathrm{ff}$ by fitting to the observed radio spectral shape for each value of $B_0$ and $n_0$ to account for both free-free emission and absorption. This is done by interpolating our models over $C_\mathrm{ff}$ and minimizing $\chi^2_{\mathrm{radio,int}}$ as defined below. Then, each model is normalized by minimizing the ``averaged'' $\chi^2$ value of all spatially integrated data points defined as:
\begin{equation}
\chi^2_{\mathrm{avg,int}} = \frac{\chi^2_{\mathrm{radio,int}}}{N_{\mathrm{radio,int}}} + \frac{\chi^2_{\mathrm{gamma,int}}}{N_{\mathrm{gamma,int}}}
\end{equation}
where $\chi^2_{\mathrm{radio,int}}$ ($\chi^2_{\mathrm{gamma,int}}$) is the $\chi^2$ of the integrated radio (gamma-ray) data, and $N_{\mathrm{radio,int}}$ ($N_{\mathrm{gamma,int}}$) is the number of integrated radio (gamma-ray) data points. This process provides approximately equal weight to the gamma-ray and radio observations. This approach is warranted both because there are unaccounted systematic errors in the radio observations and non-Gaussianities in the gamma-ray flux uncertainties. Thus, we find that using a standard $\chi^2$ analysis would incorrectly weight our results towards the radio surveys with their smaller reported uncertainties. The $\chi_{\mathrm{avg,int}}^2$ values are plotted as filled color contours in Figure~\ref{chi_SNR_plot}.

In defining Model~I and in constructing Figure~\ref{chi_SNR_plot}, we have allowed the CR injection normalization to vary in our analysis while keeping the primary CRp to primary CRe ratio (Equation~\ref{CRepratio}) fixed. The supernova rate is shown as the solid white contours, in units of SN\,yr$^{-1}$, where we assume an average SN kinetic energy injection of $10^{51}$ ergs per SN with $10$\% of the energy going into CRs. The observed SN rate in M82, as inferred by \citet{2003ApJ...599..193F}, ranges from \mbox{0.02---0.1\,SN\,yr$^{-1}$}. All of our best fit {\tt GALPROP} models have SN rates in the correct range, thus the observed SN rate does not strongly constrain our models in this parameter space.

To the upper left (lower right) of our best-fit $\chi_{\mathrm{avg,int}}^2$ curve, gamma-ray emission is over- (under-) produced and radio emission is under- (over-) produced. We note that changing the diffusion coefficient and maximum wind speed has a minor effect on these results, which we discuss in Section~\ref{section:results:extended:diff_wind}. 

The degeneracy between these two parameters roughly follows a broken power-law divided into two regimes with a break around $\sim$500~$\mu$G, with a steeper slope below the break. In regions above the break, bremsstrahlung/ionization cooling is sufficient to give the synchrotron spectrum the correct shape. In regions below the break, free-free emission and absorption are required to flatten/harden the radio spectrum. These characteristics, above and below the break, are readily apparent in the chosen models from the transition region discussed in Section~\ref{section:results:emission:spectra} and presented in Figure~\ref{AB_integrated}. 

We compare our numerical results from {\tt GALPROP} with an analytic model derived in Appendix~\ref{appendix:spectral} by solving the position-independent energy-loss equation for CRe. We use this solution to constrain the relationship between $B_0$ and $n_0$ assuming we know the steady-state CRe spectral index, $p$. In Figure~\ref{chi_SNR_plot}, we show a range of analytic fits as dashed (dotted) outlined regions. The dashed (dotted) lines denote that the gas is completely neutral (ionized). The upper dashed/dotted line uses $p=-2.4$ and the lower dashed/dotted line uses $p=-2.6$, both of which are plausible values for $p$. For more information, see Appendix~\ref{appendix:spectral}. We find that our analytic results coincide with our numerical results, especially at large gas densities and magnetic fields. We note that for the analytic model, we only took into account the integrated radio spectral measurement and not the relative normalization between the gamma-ray and radio emission. We further note that setting the bremsstrahlung cooling timescale (eq.~\ref{tau_bremss}) equal to the synchrotron cooling timescale (eq.~\ref{tau_synch}) for CRe emitting at GHz frequencies -- the criterion for flattening the GHz spectrum, as suggested by \cite{Thompson2006} and models for the FIR-radio correlation \citep{LTQ} -- one estimates a correlation of the form $B_{100}\sim n_{100}^{2/3}$, which roughly tracks the slope and magnitude of the correlation we find in Figure \ref{chi_SNR_plot} above the break.

Our models independently constrain the magnetic field strength, but we have previous magnetic field estimates that utilize either the minimum energy assumption (where the total energy of CRs+magnetic field is minimized) or the equipartition assumption (where the total energy of CRs is set to equal some factor of the energy of the magnetic field). For example, using the equipartition arguments from \citet{2005AN....326..414B}, \citet{2017A&A...608A..29A} used the total and polarized synchrotron power to determine the turbulent magnetic field magnitude of the core to be 140\,$\mu$G while showing the regular magnetic field component to be $\sim$1\,$\mu$G. \citet{Thompson2006} showed that the minimum energy magnetic field likely  underestimates the true magnetic field in dense starbursts due to strong energy losses, which had not previously been taken into account. For this reason, \citet{2013MNRAS.430.3171L} revised the previous equipartition and minimum energy arguments for starburst galaxies and obtained magnetic field strengths of $220~(240)$\,$\mu$G for the minimum energy (equipartition) magnetic field in M82. Similarly, \citet{2014A&A...567A.101P} determined the magnetic field strength to be 100\,$\mu$G and also calculated the CRp energy density to be 250\,eV\,cm$^{-3}$. \citet{2019MNRAS.tmp.1118P} found a magnetic field of 165\,$\mu$G and a CR energy density of 425\,eV\,cm$^{-3}$. \citet{Thompson2006} estimated a maximum allowable magnetic field strength in the core of M82 of 1.6\,mG, by balancing the magnetic energy density with that required for hydrostatic equilibrium, the total hydrostatic ISM pressure.

Previous models have also found a $B_0{-}n_0$ degeneracy. For example, our results are qualitatively consistent with \citet{2012ApJ...755..106P}. However, our degeneracy has higher best-fit gas densities than those found in \citet{2016ApJ...821...87E}. Our $B_0{-}n_0$ degeneracy shows a range of magnetic field strengths that span the range of previous estimates, although we show that the magnetic field cannot be smaller than $\sim$150\,$\mu$G for our assumed parameters and physical setup. \citet{yoast-hull} also found the magnetic field should be larger than 150\,$\mu$G, but did their analysis in the magnetic field$-$wind velocity plane. In scenarios with a smaller magnetic field strength, IC losses begin to dominate synchrotron losses for the ISRF energy density we use, thus a higher supernova rate is needed to account for the observed synchrotron flux which would then increase the gamma-ray emission above observations.

For reference, molecular tracers indicate that the dense molecular clouds in the core of M82 have densities spanning from \mbox{${\sim}10^3-10^{5}$\,cm$^{-3}$} \citep{1992A&A...265..447W, 2000A&A...358..433M, 2007ApJ...671.1579M, 2008A&A...492..675F, 2010ApJ...722..668N}. However, this dense gas may not fill the entire region. \cite{Kennicutt1998} finds an average total gas surface density for M82 of $\simeq0.7$\,g\,cm$^{-2}$. Assuming a gas scale height of $\simeq50$\,pc, this implies an average density of \mbox{$\langle n \rangle \simeq10^3$\,cm$^{-3}$}. However, because the gas is highly supersonically turbulent, the volume-averaged probability distribution function (PDF) of density will be broad, with a peak significantly below the mean density of the medium \citep{Ostriker2001}. Because CRs may preferentially interact with the gas above or below the mean density of the ISM, in Figure~\ref{chi_SNR_plot} we consider a wide range of densities for the models from $\langle n \rangle/50$ to ${\sim}10\langle n \rangle$.

\subsubsection{Integrated Spectra of Selected Models}
\label{section:results:emission:spectra}

We choose three models, Model~A, Model~B, and Model~B$'$, to best represent two regimes of the $B_0{-}n_0$ degeneracy: Model~A represents a region where free-free processes are essential to fit the radio flux, and both models B and B$'$ represent a region of parameter space with a flatter synchrotron spectra that does not require as much free-free emission. In Figure~\ref{chi_SNR_plot}, we denote models~A, B, and B$'$ as the lavender, brown, and green filled, red outlined hexagons, respectively. 

Figure~\ref{AB_integrated} presents the integrated gamma-ray and radio spectra for models A, B, and B$'$ (see Table~\ref{parameters}). The primary difference between Model~A and the two higher-density Models B and B$'$ are the magnetic field strength ($B_0$) and gas density ($n_0$) in the starburst core. Models~B and B' have the same $B_0$, but different halo magnetic fields, which we discuss in more detail in Section~\ref{section:results:extended:beta_wind}. 

At gamma-ray energies (left panel of Figure~\ref{AB_integrated}), the full spectrum (solid lines) is dominated by $\pi^0$-decay (dashed lines) and all three models have nearly identical fits with assumed CR injection spectral indices of $p_0=-2.2$, although Model~A has a slightly softer spectrum. Emission from bremsstrahlung (dotted lines) is sub-dominant for all three models at high energies, but contributes significantly to the spectrum at $\sim$1\,GeV and dominates below 200\,MeV. Inverse-Compton emission is negligible at all energies in all models. We note that none of our models accurately reproduce the data below 400\,MeV. Increasing the bremsstrahlung emission to account for the low-energy gamma-ray excess with respect to the models would require a much larger electron-to-proton injection ratio (Equation~\ref{CRepratio}); See Section~\ref{section:model:galprop:injection}). We note that observational uncertainties which could contribute to this difference, including the relatively small ROI of our M82 analysis region, which when combined with the relatively poor angular resolution of Fermi observations at MeV energies, could systematically impact the gamma-ray fit. Notably, the 4FGL catalog fits the M82 galaxy with a simple power-law of spectral index -2.2 across the Fermi energy band~\citep{2019arXiv190210045T}. 

However, if this turnover is verified, several theoretical considerations could also affect the low-energy gamma-ray emission without significantly affecting the remainder of our modeling. The low-energy CRe population could potentially be increased through reacceleration, which we do not explore in this paper. Other potential solutions to the discrepancy are from a different injection spectrum for CRs below 1\,GeV or an additional low-energy emission component in the starburst core.

We note that there is a very slight absolute normalization difference between the gamma-ray emission predicted by the models (Figure~\ref{AB_integrated}; approximately the line thicknesses) that can readily be changed by making a very small fractional change in the gas density of the model ($n_0$).

At radio frequencies (right panel of Figure~\ref{AB_integrated}), all models match the observed data, especially in the range $1-10$\,GHz. However, Models A and B+B$'$ produce this emission through a differing combination of synchrotron (dashed lines) and free-free (dotted lines). We demonstrate the effect of increasing/decreasing the free-free clumping factor, $C_{\mathrm{ff}}$, by a factor of $\sqrt{2}$ with the thickness of the solid line for the total integrated emission (or thickness of the outlined regions for individual synchrotron and free-free components). Increasing/decreasing $C_{\mathrm{ff}}$ by a factor $\sqrt{2}$ gives us the upper/lower edge of the region. We note that we have plotted the unabsorbed synchrotron spectrum as the dashed lines. 

For Model~A (lavender), which has lower average gas density, CRe cooling is not completely dominated by bremsstrahlung and ionization losses, allowing synchrotron and IC cooling to steepen the CRe spectrum, making the resulting synchrotron radiation alone too soft to explain the observed emission above $\sim$1\,GHz. Thus, to flatten the radio spectrum, Model~A requires a significant flux of free-free emission (dotted lines) above $\sim$1\,GHz and free-free absorption below 1\,GHz. Model~A does not have enough free-free absorption to match the two lowest frequency data points from LOFAR at ${\sim}100-200$\,MHz \citep{2015A&A...574A.114V}, but the calculation of free-free absorption is complicated by the geometry of the ionized gas.

Models~B and B$'$ (brown and green), on the other hand, have significantly larger gas densities, making bremsstrahlung and ionization cooling dominate the electron spectrum. This produces a harder/flatter synchrotron spectrum that more closely traces the radio spectrum up to ${\sim}5-10$\,GHz, and requires less free-free emission and absorption to fit the data. However, Model~B has enough free-free absorption to reach the lowest frequency points while Model~B$'$ does not. This is because Model~B$'$ has more halo emission from outside the core than Model~B. Specifically, Model~B$'$ has a larger halo magnetic field due to a smaller value for the magnetic field drop-off, $\beta$ (see Section~\ref{section:results:extended:beta_wind}), thus a larger amount of emission comes from just outside the core where there is not as much free-free absorption. This larger halo magnetic field also causes the integrated synchrotron spectrum to be slightly steeper as seen in the difference between the synchrotron spectra of Models~B and B$'$. We note that Model~B$'$ has more free-free emission and absorption inside the core as can be seen by the free-free emission lines (dotted lines), which also take into account absorption.

While all models produce reasonable fits to the observed radio data, we consider models similar to Model~B+B$'$ to be somewhat less fine-tuned, as the observed radio spectrum is less-sensitive to the distribution of ionized gas that causes free-free emission and absorption. 

\begin{table}
	\begin{center}
		\begin{tabular}{llllll}
		    \toprule
			 & A & B & B$'$ & A/B & B/B$'$ \\ \midrule
			\multicolumn{6}{|c|}{} \\[-12pt]
			\midrule
            \multicolumn{6}{c}{\bf Injection Luminosities ($\times 10^{40}$\,ergs s$^{-1}$)} \\
			\midrule
            $^4$He        & 1.34 & 0.95 & 0.91 & 1.41 & 1.04 \\
			Primary $p^+$ & 16.9 & 12.0 & 11.4 & ---  & --- \\
			Primary $e^-$ & 1.69 & 1.20 & 1.14 & ---  & --- \\
            \midrule
            \multicolumn{6}{c}{\bf Secondary Production Rates ($\times 10^{40}$\,ergs s$^{-1}$)} \\
			\midrule
            Secondary $p^+$ & 5.03  & 4.36  & 4.28  & 1.15 & 1.02 \\
			Secondary $e^-$ & 0.235 & 0.221 & 0.223 & 1.06 & 0.99 \\
			Secondary $e^+$ & 0.524 & 0.479 & 0.477 & 1.09 & 1.04 \\
			\midrule
            \multicolumn{6}{c}{\bf Emission Luminosities ($\times 10^{40}$\,ergs s$^{-1}$)} \\
			\midrule
            Synchrotron     & 0.281 & 0.393 & 0.463 & 0.72 & 0.85 \\
            Free-Free       & 22.6  & 19.2  & 9.38  & 1.18 & 2.05 \\
            $\pi^0$-decay   & 1.41  & 1.31  & 1.32  & 1.08 & 0.99 \\
            Bremsstrahlung  & 0.354 & 0.370 & 0.380 & 0.97 & 0.92 \\
            Inverse Compton & 0.612 & 0.215 & 0.139 & 2.85 & 1.55 \\
            \midrule
			\multicolumn{6}{c}{\bf Core Energy Density (eV\,cm$^{-3}$)} \\
			\midrule
            $^4$He            & 721     & 115    & 73.4   & 6.27 & 1.57 \\
			Primary $p^+$     & 1232    & 234    & 151    & 5.26 & 1.55 \\
			Secondary $p^+$   & 302     & 82.0   & 56.9   & 3.68 & 1.44 \\
            Secondary $p^-$   & 1.86    & 0.628  & 0.476  & 2.96 & 1.32 \\
            Primary e$^{-}$   & 16.1    & 3.11   & 2.09   & 5.18 & 1.49 \\
			Secondary e$^{-}$ & 3.78    & 1.24   & 0.828  & 3.05 & 1.50 \\
			Secondary $e^{+}$ & 8.92    & 2.69   & 1.77   & 3.32 & 1.52 \\
            Knock-on e$^{-}$  & 0.0623  & 0.0160 & 0.0105 & 3.89 & 1.52 \\
			\midrule
			\multicolumn{6}{c}{\bf Total Energy ($\times 10^{53}$ ergs)} \\
			\midrule
            $^4$He          & 65.6   & 7.25    & 3.68    & 9.05 & 1.97 \\
			Primary $p^+$   & 584    & 88.9    & 46.7    & 6.57 & 1.90 \\
			Secondary $p^+$ & 251    & 58.2    & 34.3    & 4.31 & 1.70 \\
            Secondary $p^-$ & 2.86   & 0.816   & 0.540   & 3.50 & 1.51 \\
            Primary $e^-$   & 3.15   & 0.360   & 0.0530  & 8.75 & 6.79 \\
			Secondary $e^-$ & 3.34   & 0.944   & 0.136   & 3.54 & 6.94 \\
			Secondary $e^+$ & 3.26   & 1.90    & 0.301   & 1.72 & 6.31 \\
            Knock-on $e^-$  & 0.0558 & 0.0116  & 0.00370 & 4.81 & 3.14 \\
            \midrule
			\multicolumn{6}{c}{\bf Calorimetric Fractions} \\
			\midrule
			Primary CRp       & 0.51 & 0.65 & 0.68 & 0.78 & 0.96 \\
            Primary $e^-$     & 0.88 & 0.98 & 1.00 & 0.90 & 0.98 \\
			Secondary $e^\pm$ & 0.19 & 0.71 & 0.94 & 0.27 & 0.75 \\ \\[-7.5pt]
			\bottomrule
		\end{tabular}
	\end{center}
	\caption{The CR injection luminosities, secondary production rates, emission luminosities, volume-averaged CR energy densities in the core, as well as the total steady-state energies and calorimetric fractions for all three models. In the two right-most columns, we list the ratio between the quantities of Model~A to Model~B and Model~B to Model~B$'$. Our results are consistent with the SN rate of M82, and indicate a CR proton escape timescale of approximately 2\,Myr, consistent with expectations. We note that the calorimetric fractions for CRp and CRe are calculated differently due to boundary conditions. }
	\label{outputs}
\end{table}

\subsubsection{Model CR \& Emission Luminosities}
\label{section:results:emission:lumin}

Table~\ref{outputs} displays the energetics of each model, including the CR injection and emission luminosities, the volume-averaged energy densities in the starburst core, the total CR energies contained within the simulation volume, as well as the calorimetric fractions for important CR species. The CR injection rate for all models are similar, corresponding to an injection rate of approximately $1.5\times10^{41}$\,ergs\,s$^{-1}$. Assuming a typical SN energy injection rate of 10$^{50}$\,ergs into CRs, this corresponds to a rate of 0.047\,SN\,yr$^{-1}$, consistent with the expectations given the star-formation rate of $\sim$20\,$M_\odot$~yr$^{-1}$ \citep{2003ApJ...599..193F}. 

The ratio between the Helium-4 and proton energy injection is set by the {\tt GALPROP} Milky-Way models. We set the total energy injection of electrons to be ${\sim}0.1$ that of protons (eq.~\ref{CRepratio}). The difference between the injection rate of the models are mostly due to the slightly different normalizations of the gamma-ray spectrum but also in small part due to CRp escape in Model~A from the somewhat lower gas density core. For secondary production rates, the total energy injected in positrons exceeds that of electrons by a factor ${\sim}2$. The production of secondary protons is also significant, exceeding 30\% of the primary proton injection rate for all models. The secondary-to-primary injection ratio is similar in each model, which is expected as all models are fit to the gamma-ray data and secondaries are produced at a rate proportional to the $\pi^0$-production/decay rate.

The CR emission luminosities in the models are similar, which is expected because all these models are fit to the same gamma-ray and radio data. Since $\pi^0$-decay dominates the gamma-ray spectra, we expect all models to have the same $\pi^0$-decay power. The total power emitted by CRe does not change between the models since the electrons are nearly completely calorimetric (i.e. all CRe lose their energy). Some minor differences exist. Models~B and B$'$ (more-so for B$'$) have a slightly harder synchrotron spectrum (i.e. more power into synchrotron) due to increased bremsstrahlung cooling while requiring the radio normalization to remain the same. Model~A has more IC emission because the core magnetic energy density is below the fiducial ISRF energy density. Meanwhile Model~B has more IC emission than Model~B$'$ due to the large halo magnetic field of B$'$. We note that Model~A requires the most free-free emission, implying that if $C_{\mathrm{ff}}$ was held constant, Model~A would require the largest ionized gas fraction.  

\begin{figure*}
	\makebox[0.49\linewidth][c]{ \includegraphics[width=.50\linewidth]{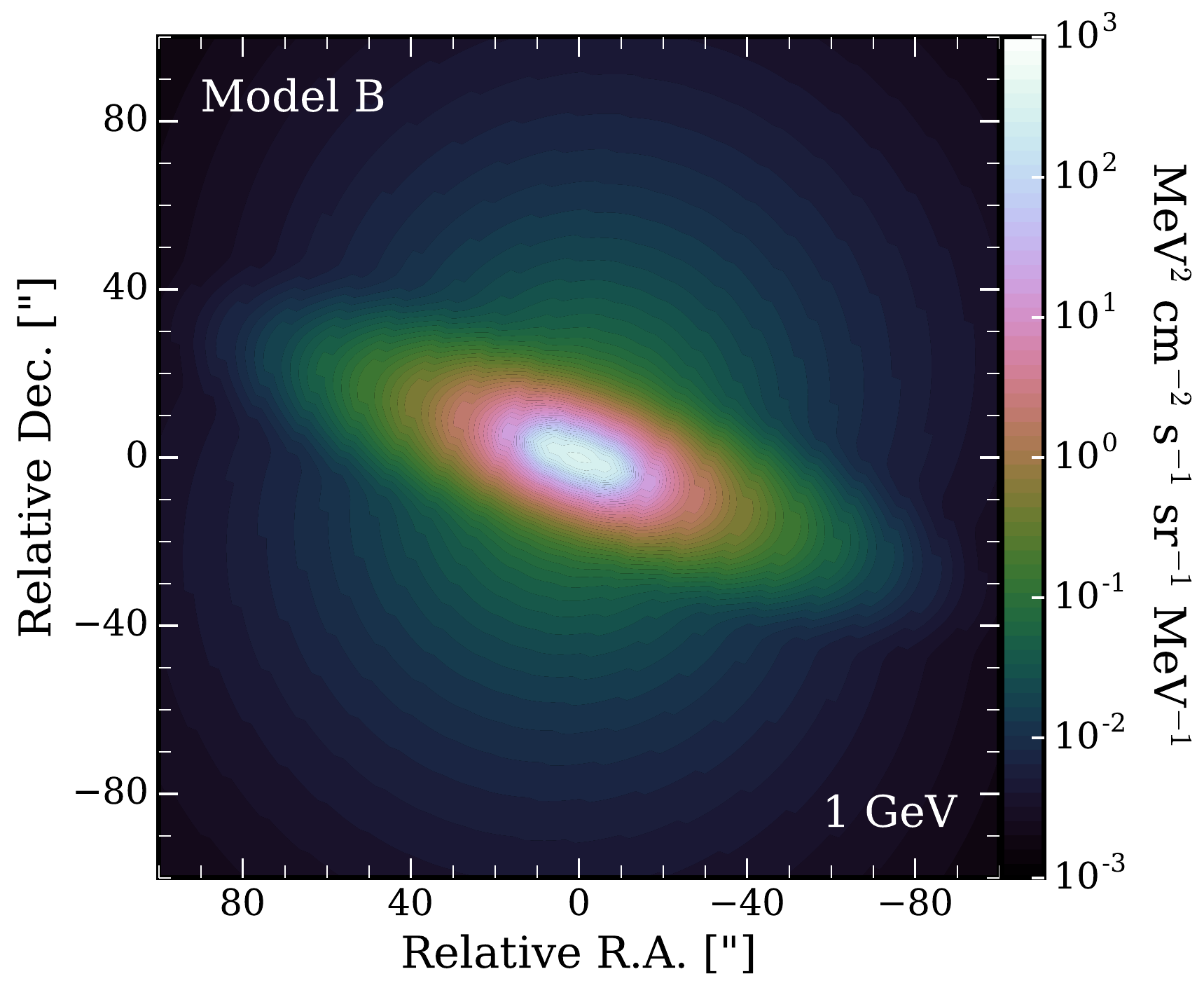} }
	\hspace{0pt}
	\makebox[0.49\linewidth][c]{ \includegraphics[width=.50\linewidth]{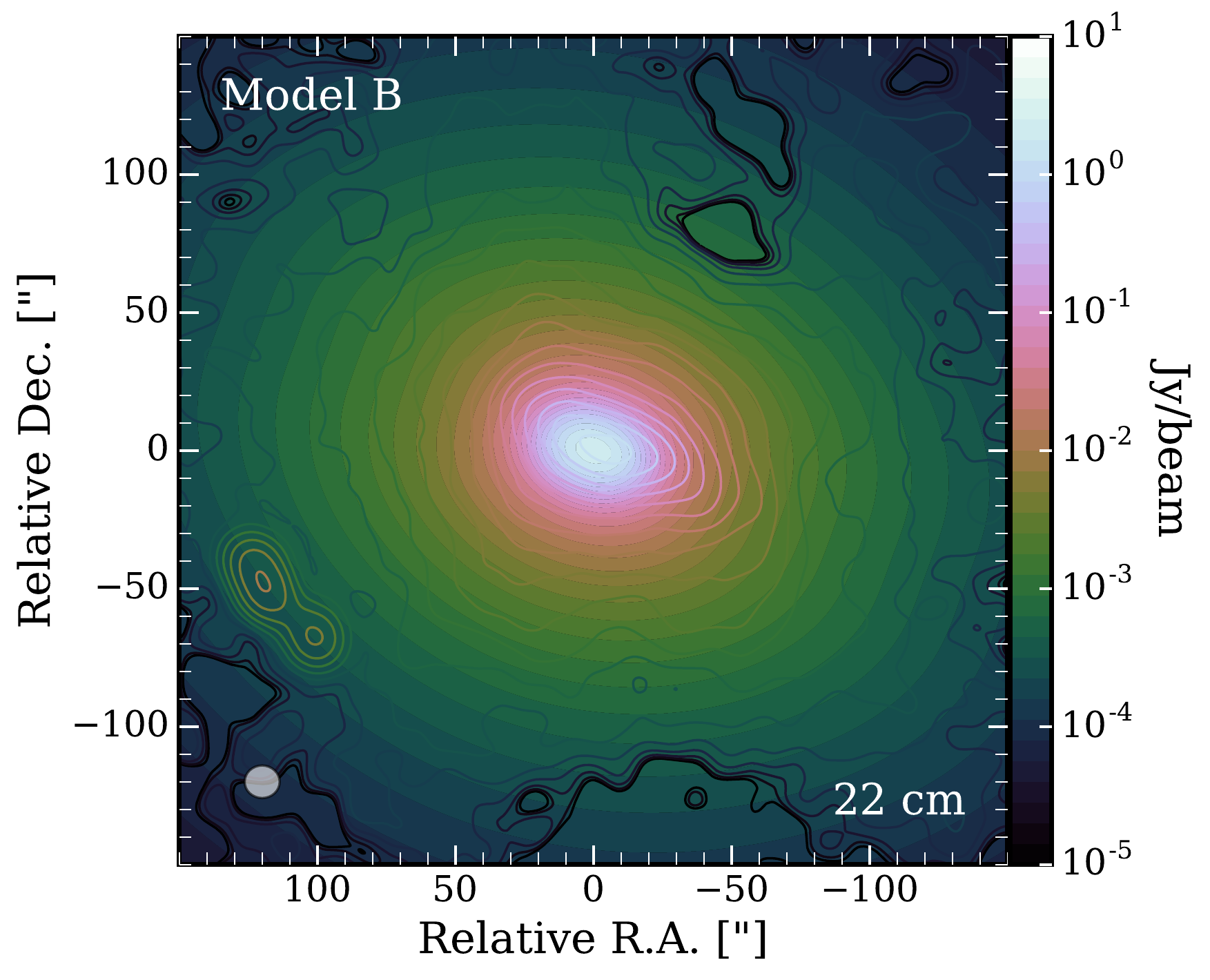} }
	\caption{ Projected images of Model B in gamma-rays (left, 1\,GeV) and radio (right, 22\,cm). On the left, we see the bright gamma-ray core along with disk emission. Above and below the disk, we see a dim gamma-ray halo. On the right is the 22\,cm radio map compared to the colored contours from \citet{2013A&A...555A..23A}. For the radio image, the beam size of 12.7"$\times$11.8" is shown in the bottom left corner. For Model~A, the gamma-ray image has a brighter halo while the radio images are nearly identical. For Model~B$'$, the gamma-ray and radio images are essentially identical. }
	\label{model_image}
\end{figure*}

\subsubsection{CR Energetics of Models}
\label{section:results:emission:energetics}

For Model~A~(B)~[B$'$], we find the total volume-averaged CR energy density of 2436~(468)~[305]~eV\,cm$^{-3}$ in the starburst core\footnote{This includes all CRs, including Helium-3 and deuterium which are not shown in Table~\ref{outputs}}. The magnetic energy density of 560~(2620)~[2620]~eV\,cm$^{-3}$ corresponds to magnetic fields of 150~(325)~[325]~$\mu$G for Model~A~(B)~[B$'$] which spans the range of previous estimates (discussed at the end of Section~\ref{section:results:emission:mag_gas}). None of our models are in equipartition as the ratio of magnetic energy density to CR energy density is 0.23~(5.6)~[8.6] for Model~A~(B)~[B$'$]. But there exists a model in the parameter space between Models~A and B that does have equipartition between the CRs and the magnetic field. As we increase magnetic field strength and gas density from Models~B and B$'$, our models move further away from equipartition but are still able to replicate the data.

As we noted in Section~\ref{section:model:galprop:propagation}, our wind is spherically radial, except in the galactic disk where we multiply the cylindrically radial component of the wind by \mbox{$1-\exp (-|z|/z_\mathrm{core})$}. If we do not take this factor into account, our CR energy densities decrease by a small amount, of order $\sim$10\%. We discuss the wind profile more in Appendix~\ref{appendix:gas_wind}.

Table~\ref{outputs} shows the volume-averaged energy densities of individual CR species within the core. Overall, there is a factor ${\sim}4$ difference between models~A and B and a factor of 1.5 difference between models~B and B$'$. In most previous modeling, secondary protons have been ignored. The ratio of volume-averaged energy densities of secondary protons to primary protons is 0.25~(0.35)~[0.38] and the ratio of the volume-averaged energy densities of secondary CRe to primary CRe is 0.79~(1.26)~[1.24] for Model~A~(B)~[B$'$]. Thus, we show that primary CRe dominate in the starburst core for Model~A, but not for Model~B and B$'$. We find that secondary CRe dominate the total synchrotron emission by a factor of 1.10~(1.36)~[1.46] for Model~A~(B)~[B$'$]. These ratios are larger than the ratio of secondary CRe to primary CRe in the core because secondary CRe dominate primary CRe in the large volume of the wind dominated region (which has a relatively large magnetic field as we discuss in Sections~\ref{section:results:extended:cr} \& \ref{section:results:extended:beta_wind}).

The steady-state CR energy density in each model differs significantly (especially between models~A and B+B$'$) as a result of the difference in CR cooling. Between models~A and B, there is a factor of $3.1-6.2$ difference in the energy density for each CR species. The large range of factors is caused by the different energy losses (e.g.\ protons vs.\ electrons), different sources (e.g.\ primary vs.\ secondaries), different diffusive behavior from different species' rigidities, and different cross-sections with the ISM (e.g.\ $^4$He to primary protons). Between Models~B and B', there is a relatively uniform factor ${\sim}1.5$ difference in the energy densities from the change in gas density. The magnetic field in the core is the same between both of these models. 

For all three models, we also provide the total CR energy per species (integrated over the entirety of M82). Between models~A and B, there is a large range of ratios between different CR species due to different calorimetric fractions. Between models~B and B$'$, the energy ratios are relatively small for CRp, however, B has a much larger CRe total energy because of the very large halo magnetic field in B$'$ which we discuss in Section~\ref{section:results:extended:beta_wind}.

Table~\ref{outputs} shows that the ratio of the total energy in secondary protons to primary protons is 0.43~(0.65)~[0.73] for Model~A~(B)~[B$'$]. We see that primary electrons do not contribute significantly to the total CRe energy at $\sim$0.10 the total energy of secondary CRe for Models~B and B$'$. The factor increases to 0.48 for Model A.

We also present the calorimetric fractions, the fraction of CR energy that does not escape the galaxy, for the CRp and primary and secondary CRe. For CRp, we calculate the calorimetric fraction by summing secondary CRe production rates with the $\pi^0$-decay luminosity and multiplying by $2$ to take into account the emission of neutrinos, then adding the secondary proton production rate and divide the summed value by the sum of primary $p^+$ and $^4$He injection luminosities. We obtain a primary CRp calorimetric fraction of 0.51 (0.65)~[0.68] for Model~A~(B)~[B$'$]. These values are not too far removed from the values previously inferred by \citet{Lacki2011} and \citet{lacki_thompson} (${\sim}0.2-0.4$), \citet{yoast-hull} ($\sim$0.5), and \citet{2018MNRAS.474.4073W} ($\sim$0.35). The difference is not necessarily a physical difference, but could result from our new methodology in calculating the calorimetric fraction. Our method is similar to that of \citet{yoast-hull}, who calculated the calorimetric fraction from the ratios of timescales (i.e. assuming CRs either escape or die in hadronic interactions, \mbox{$\mathrm{cal} = \tau_\mathrm{proton}/\tau_\mathrm{had}$}, where $\tau_\mathrm{proton}$ is the lifetime of a proton).

For primary and secondary CRe, the calorimetric fraction is calculated from \mbox{$\mathrm{cal} = 1-F_\mathrm{NET}/Q$} where $F_\mathrm{NET}$ is the net energy flux of CRe through the surface defined by $R=3$\,kpc and $z=\pm 3$\,kpc and $Q$ is the total injection energy rate. We choose to define ``escape'' at a radius of 3\,kpc in order to try to avoid edge effects in our numeric modeling. We are unable to reliably calculate the calorimetric fraction in this manner for CRp due to these edge effects. We find that primary CRe are essentially calorimetric and that a small fraction of secondary CRe are able to escape. These CRe are important in powering the large radio halo. Summing the total power emitted by CRe and dividing by the total CRe injection rate (both primaries and secondaries), we find a fraction of 0.51~(0.52)~[0.53] for Model~A~(B)~[B$'$]. Ionization cooling dominates at the lowest energies below ${\sim}100$\,MeV, above which bremsstrahlung dominates. Integrating a rigidity power law spectrum ($\mathcal{R}^{-2.2}$) from \mbox{1\,MeV--100\,MeV} and dividing by the same integral from \mbox{1\,MeV--1\,PeV}, we find ionization is the dominant cooling mechanism for ${\sim}55$\% of the total CRe power. This value, along with the ratios of total CRe emission to CRe injection are consistent with CRe being calorimetric.

\subsubsection{Effects of the ISRF}
\label{section:results:emission:isrf}

For our analysis, we keep the energy density of the ISRF, $U_{\mathrm{ISRF},0}$, constant. Because the CRe are strongly calorimetric, there is only a large effect on the CRe population if the IC cooling timescale (Equation~\ref{tau_ic}) becomes comparable to or shorter than the combined cooling timescale from the other processes. If synchrotron cooling (Equation~\ref{tau_synch}) dominates, a magnetic field energy density of 1000\,eV\,cm$^{-3}$ (the energy density of the ISRF we use) corresponds to a magnetic field of 200\,$\mu$G. Increasing the energy density by a factor of 2 corresponds to an increase in the magnetic field to 283\,$\mu$G. Thus, if the magnetic field is $\gtrsim$250\,$\mu$G, then a change in the ISRF energy density by a factor of 2 does not drastically affect our qualitative results. However, for magnetic fields $\lesssim250$\,$\mu$G, changes in the ISRF may affect CRe energy densities if there are no other dominating cooling timescales. 

Similarly, we can compare the ISRF energy density to the gas density. Comparing Equation~\ref{tau_bremss} with Equation~\ref{tau_ic}, we see that for the timescales to be comparable for CRe emitting at a frequency of ${\sim}1$\,GHz, the gas density has to be $\sim$200\,cm$^{-3}$ if the ISRF energy density is 1000\,eV\,cm$^{-3}$. If the ISRF is increased by a factor of 2, the gas density must also increase by a factor of 2. Thus, if the gas density is $\gtrsim$300\,cm$^{-3}$, our qualitative results do not change drastically for $\sim$1\,GeV CRe. However, this is energy dependent since $\tau_\mathrm{bremss}$ and $\tau_\mathrm{IC}$ have different energy dependencies.

For a more quantitative analysis of our models, we find the overall lifetime of GHz-emitting CRe by inversely adding Equations~\ref{tau_bremss}-\ref{tau_ion}. Taking into account all losses except for diffusion and advection, we find that to change the overall lifetime for GHz emitting CRe in the core by $20$\%, the ISRF must increase or decrease by $500$\,eV\,cm$^{-3}$ for Model~A. For Model~B, we find that removing the ISRF entirely only increases the overall lifetime by ${\sim}10$\% and that we must increase the ISRF by a factor of ${\sim}3$ to get a decrease in the overall lifetime of GHz-emitting CRe by $20$\%. Reasonable changes ($<50$\%) in the ISRF energy density do not affect Model~B, but may slightly affect the Model~A CRe normalizations. For more discussion of the effects of the ISRF, see the discussion of our analytic model in Appendix~\ref{appendix:spectral}.

\begin{figure*}
	\makebox[\linewidth][c]{ \includegraphics[width=1.0\linewidth]{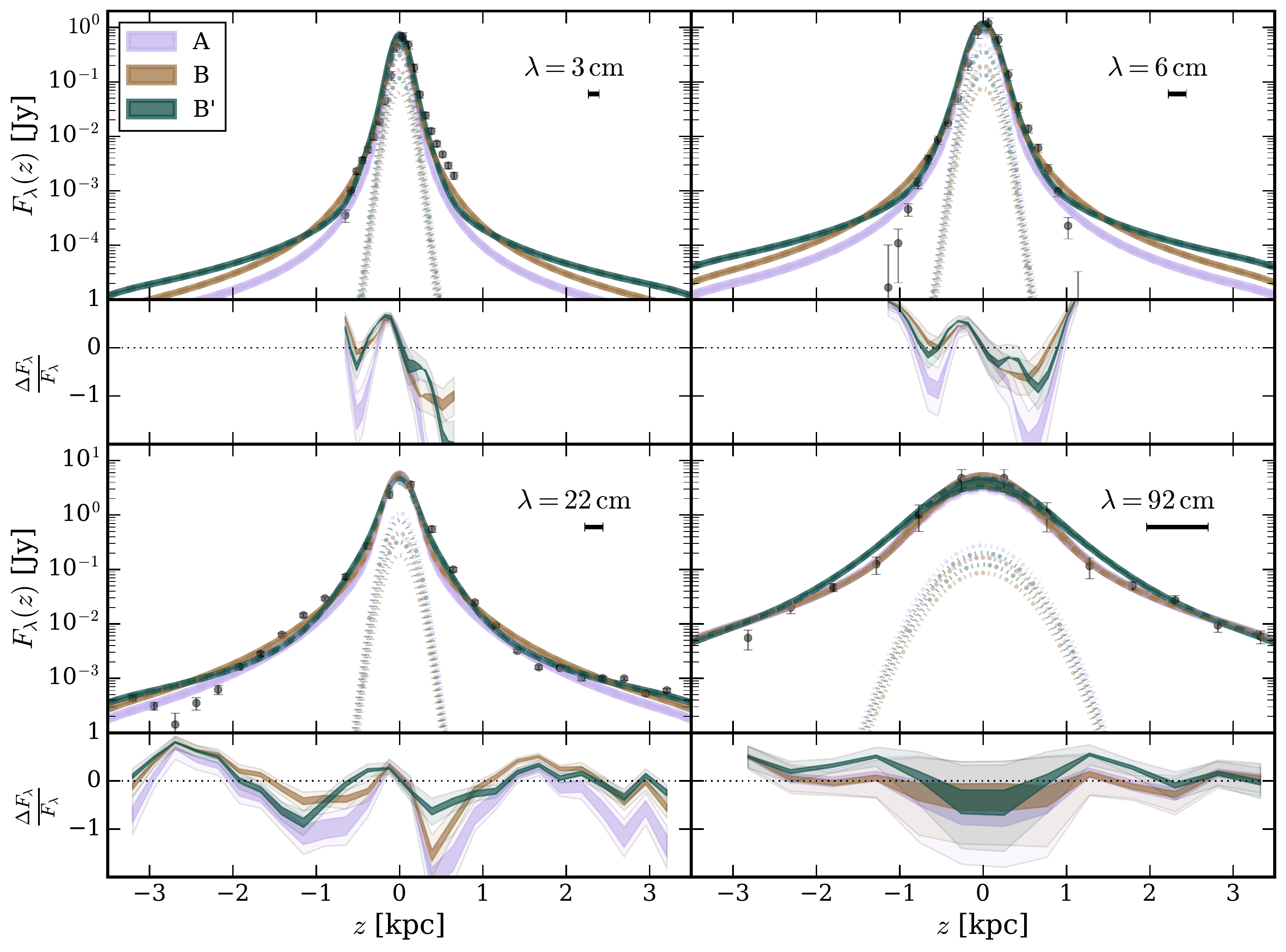} }
	\caption{Extended radio emission along the minor axis of M82 at 3\,cm~(top left), 6\,cm~(top right), 22\,cm~(bottom left), and 92\,cm~(bottom right). Both the data and modeled emission are integrated over regions of 60"$\times$4.5", 60"$\times$7", 60"$\times$15", and 60"$\times$30" for 3, 6, 22, and 92\,cm, respectively. Results are shown for Model~A~(lavender), Model~B~(brown), and Model~B$'$~(green). For each model we show the total emission~(solid), as well as the individual components stemming from synchrotron~(dashed) and free-free~(dot-dashed) emission. The black bar in the upper left of each plot shows the beam size for each observation. The width of the bands indicates the effect of increasing/decreasing the free-free clumping parameter by a factor of $\sqrt{2}$ compared to the default value given in Table~\ref{parameters}.} 
	\label{radio_ext}
\end{figure*}

\subsection{Resolved Maps \& Extra-Planar Emission}
\label{section:results:extended}

To better understand the differences between our CR propagation models, we produce resolved images of M82 at both gamma-ray and radio energies. In Figure~\ref{model_image}, we show the simulated emission morphology for Model~B at a gamma-ray energy of 1\,GeV (left) and at a radio wavelength of 22\,cm (right). In Appendix~\ref{appendix:emission_maps}, we also show the emission morphology for Model~B at 3, 6, and 92\,cm along with a 1\,TeV gamma-ray image. We also show comparisons with LOFAR data at 195 and 254\,cm for models A and B to provide a qualitative comparison.

While gamma-ray observations have not yet resolved M82, the gamma-ray emission map shows the expected features. We observe bright emission from the starburst core and along the galactic plane. Most importantly for our understanding of CR driven winds, we find a dim (6 orders of magnitude dimmer than the core), extended emission component that stretches perpendicular to the galactic plane out to $\sim$1\,kpc in the halo.

\subsubsection{Radio Halo Flux \& Spectral Index}
\label{section:results:extended:radio_halo}

\begin{figure}
    \makebox[\linewidth][c]{ \includegraphics[width=1.02\linewidth]{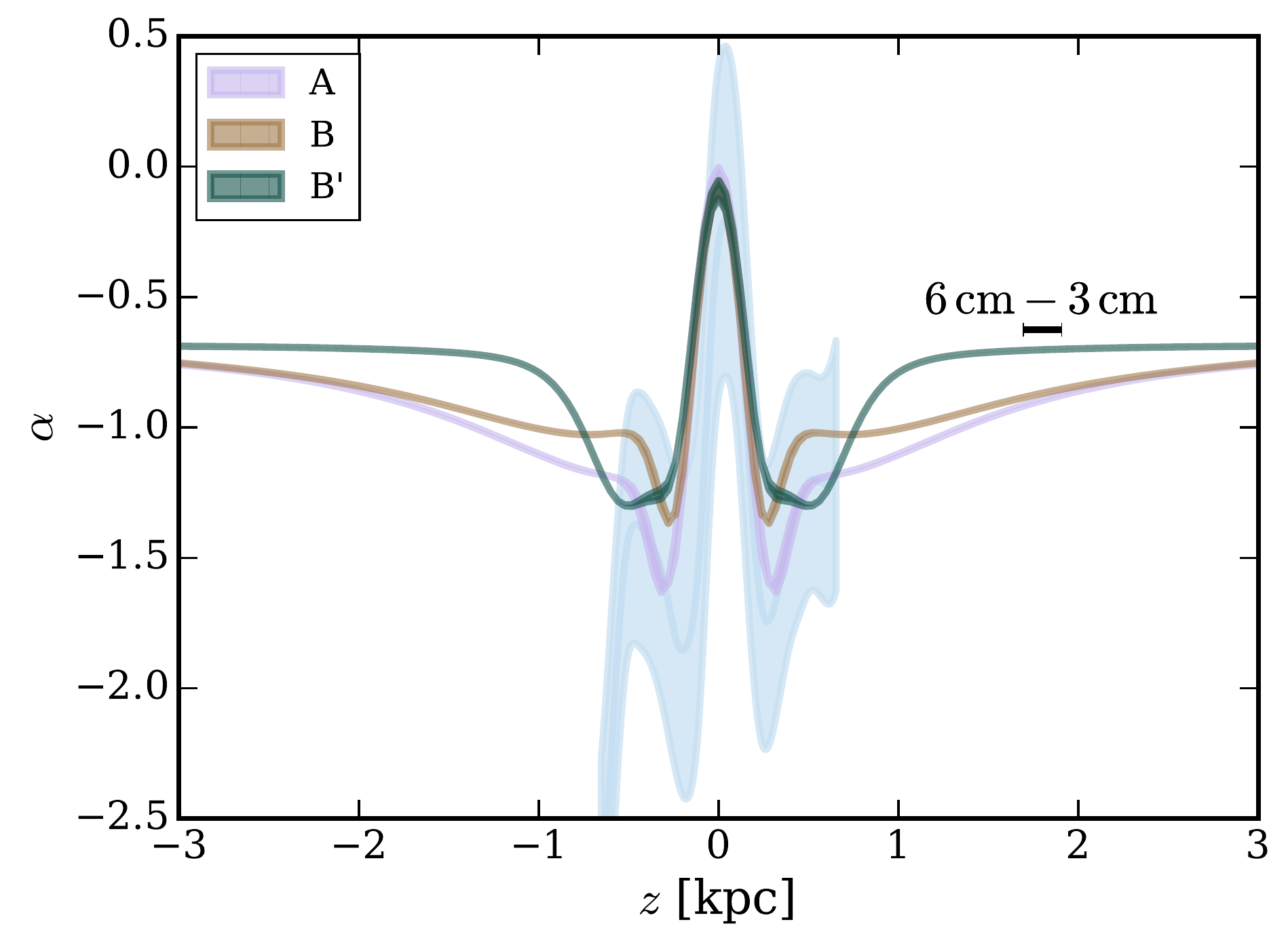} }
	\makebox[\linewidth][c]{ \includegraphics[width=1.02\linewidth]{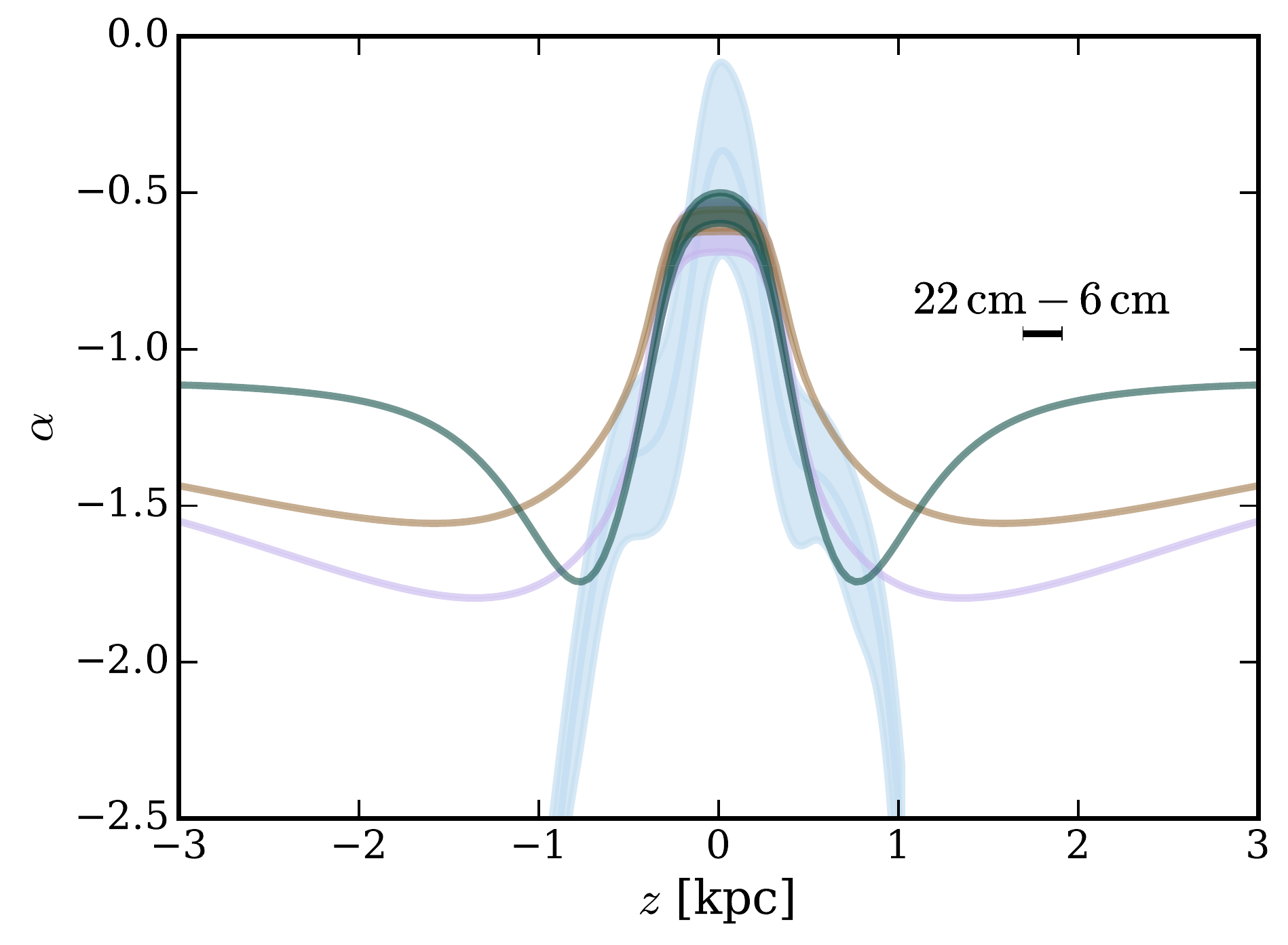} }
    \makebox[\linewidth][c]{ \includegraphics[width=1.02\linewidth]{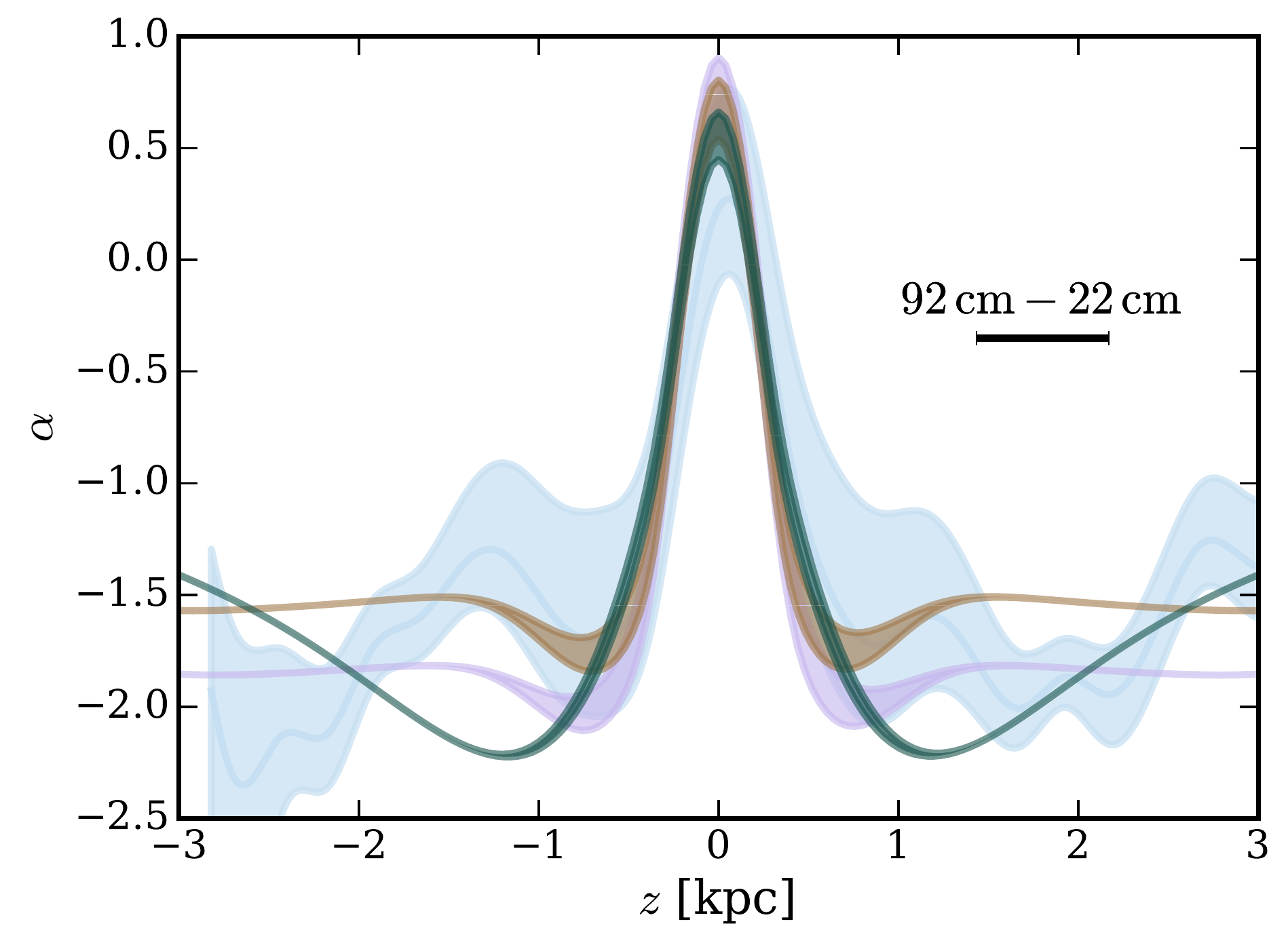} }
	\caption{ The radio spectral index as a function of $z$. Results are averaged over lines of sight that lie within 1\,kpc along the galactic plane. Results are shown for Models~A (lavender), B (brown), and B$'$ (green). The blue region is the measured spectral index between 6\,cm--3\,cm (top) 22\,cm--6\,cm (middle) or 92\,cm--22\,cm (bottom). The uncertainty in the model predictions is derived by increasing/decreasing $C_\mathrm{ff}$ by a factor of $\sqrt{2}$ from our default value. The spectral index is calculated by interpolating the data in Figure \ref{radio_ext}. The black bar denotes the larger beam size of the two wavelengths. }
	\label{b_spectral}
\end{figure}

In the radio band, we can compare our models to several high-resolution observations that resolve the emission along the minor axis of M82. In Figure~\ref{model_image} (right panel), we compare our radio map with \citet{2013A&A...555A..23A} at 22\,cm, by convolving our image with a Gaussian beam of 12.7"$\times$11.8". We find that the emission morphology predicted by Model~B reasonably matches observations along both the major and minor axes. Note that, aside from specifying the basic configuration of sources, density, and magnetic field motivated by observations of M82, we have not made an explicit attempt to fit the data along the major axis. Note also that M82 is naturally not symmetric about its major or minor axes, unlike our {\tt{GALPROP}} models, which are constrained to be axisymmetric. Thus our models generally over- (under-) predict emission above (below) the disk. We note that at frequencies below 1\,GHz, there is an asymmetry above/below the disk due to free-free absorption. 

In Figure~\ref{radio_ext}, we quantitatively evaluate the extended radio emission along the minor axis of M82 at wavelengths spanning from \mbox{$3-92$\,cm} for models~A (lavender), B (brown), and B$'$ (green). The solid, dashed, and dotted lines indicate total, synchrotron, and free-free emission, respectively. In each case, our modeled emission is smoothed by a Gaussian beam corresponding to the angular resolution of the observations and binned according to the analysis of \citet{2013A&A...555A..23A}.

These observations strongly constrain the propagation of CRe in the region where the M82 wind is observed. We note two important trends in the data and our models. The first is the high luminosity of the starburst core relative to the extended low-surface brightness emission along the minor axis. The second is the increasing spatial extension at low frequencies. In our models, this is caused by the longer lifetime of the wind-driven, low-energy CRe in the predominately synchrotron+IC cooled wind region outside of the high-density, bremsstrahlung-cooled core. Wind transport is energy-independent and diffusion is weakly energy-dependent, making propagation effects, alone, unable to cause the increasing spatial extension at low frequencies.

At 92\,cm, models A and B match the data while Model~B$'$ slightly overshoots the data at distances larger than 1\,kpc from the disk. At 22\,cm, our models match the overall behavior of the halo shape, but undershoot the data at $z=-1$\,kpc and $z=0.5$\,kpc. Figure~\ref{radio_ext} shows that the largest mismatch is the overestimation of all our models in the 6\,cm band on scales larger than $\pm1$\,kpc. Indeed, \citet{2013A&A...555A..23A} report a very sudden drop in the 6\,cm flux at $\pm1$\,kpc with essentially zero flux at larger distances from the core. We are unable to match such a steep decrease with our steady-state models. \citet{2013A&A...555A..23A} comment that the abrupt drop in flux at 3\,cm and 6\,cm may be due to their lack of short spacing data, which may be required to fully recover extended, low surface-brightness regions. The fact that the total flux \citet{2013A&A...555A..23A} measure at 6\,cm is less than that reported by previous lower resolution observations provides some evidence that their observations may indeed miss a low surface brightness halo. Taking the {\tt GALPROP} models at face value, we predict an extended low-surface brightness halo at 3\,cm and 6\,cm with the characteristics shown in the top panels of Figure~\ref{radio_ext}. Emission maps for each wavelength are shown in Appendix~\ref{appendix:emission_maps} in Figure~\ref{all_map}.

In Figure~\ref{b_spectral}, we use the modeled emission and data from Figure~\ref{radio_ext} to calculate the radio spectral index, \mbox{$\alpha= d\log F_\nu / d \log \nu$}, as a function of distance away from the M82 disk. The thickness of the model lines, Model~A~(lavender), Model~B~(brown), and Model~B$'$~(green), indicate the effect of increasing/decreasing the free-free clumping factor, $C_\mathrm{ff}$, by a factor of $\sqrt{2}$. The light-blue line denotes the value of $\alpha$ derived from the data by the interpolation of the fiducial flux at each frequency (see Figure~\ref{radio_ext}). The blue-filled region denotes $\alpha$ as derived from the interpolation of the 1-$\sigma$ flux at each $z$. All models produce reasonable matches to the observed data, but tend to overestimate the data at distances exceeding $\sim$1\,kpc from the galactic plane, especially in the 22\,cm--6\,cm band. The steep spectral indices, \mbox{$\alpha < -1.5$}, observed far from the galactic plane are particularly important and physically constraining, as the overall change in the spectral index from the galactic plane to $\sim$1\,kpc, is approximately 1.5 between $92-22$\,cm. 

\subsubsection{Cosmic-Ray Spectra $\implies$ Radio Halo Index}
\label{section:results:extended:cr}

\begin{figure*}
	\makebox[0.49\linewidth][c]{ \includegraphics[width=0.5\linewidth]{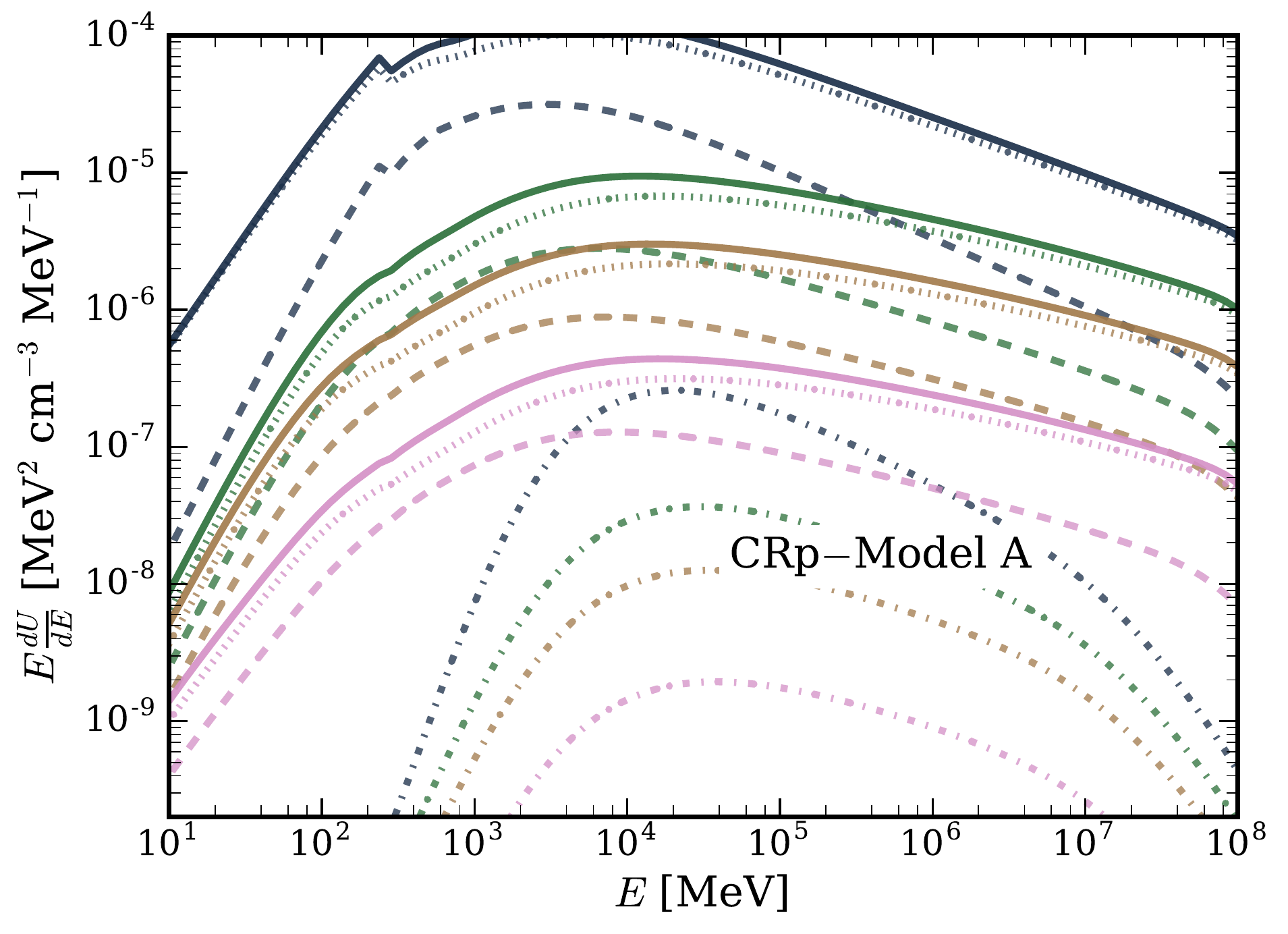} }
	\makebox[0.49\linewidth][c]{ \includegraphics[width=0.5\linewidth]{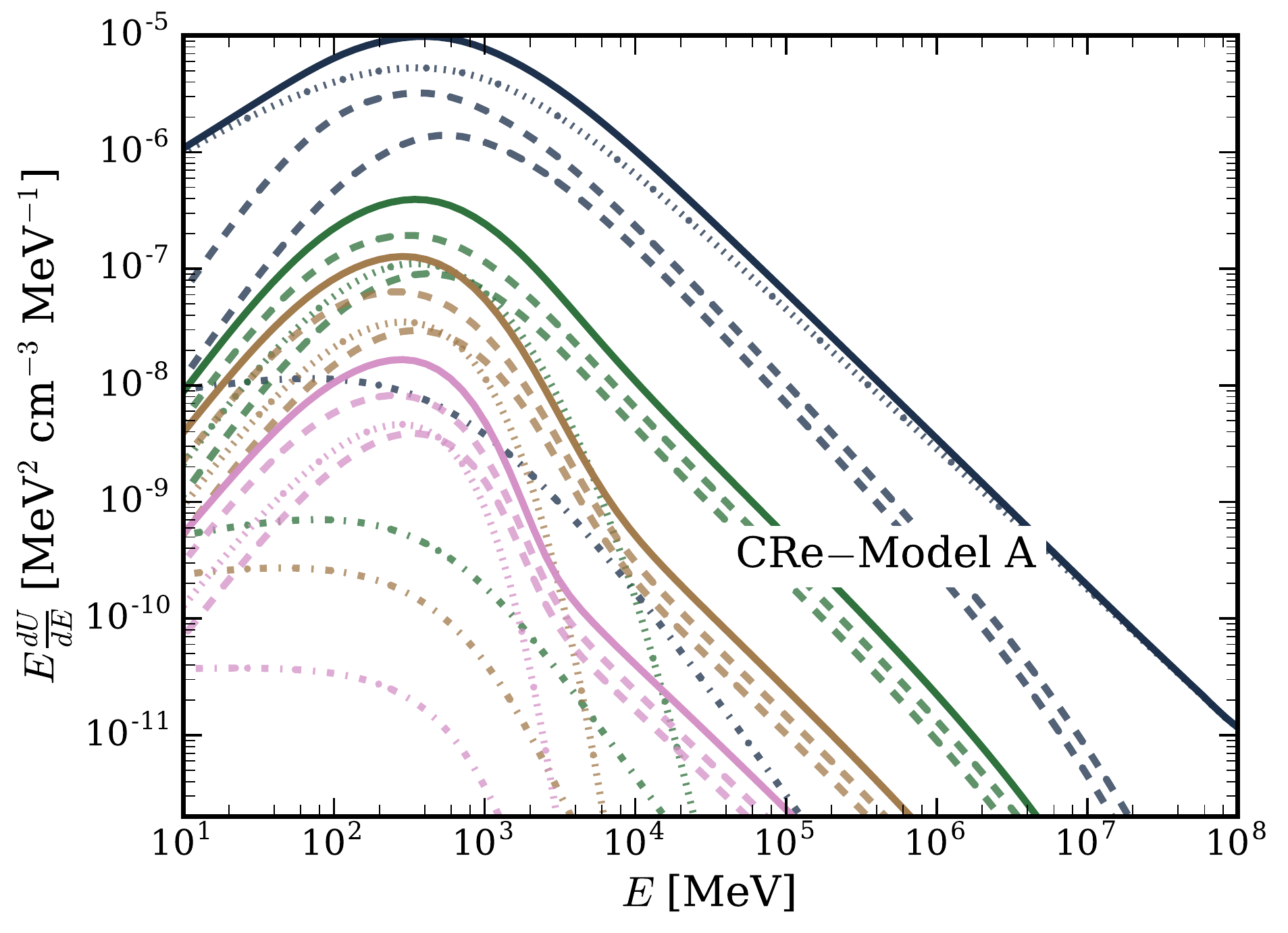} } \\
	\makebox[0.49\linewidth][c]{ \includegraphics[width=0.5\linewidth]{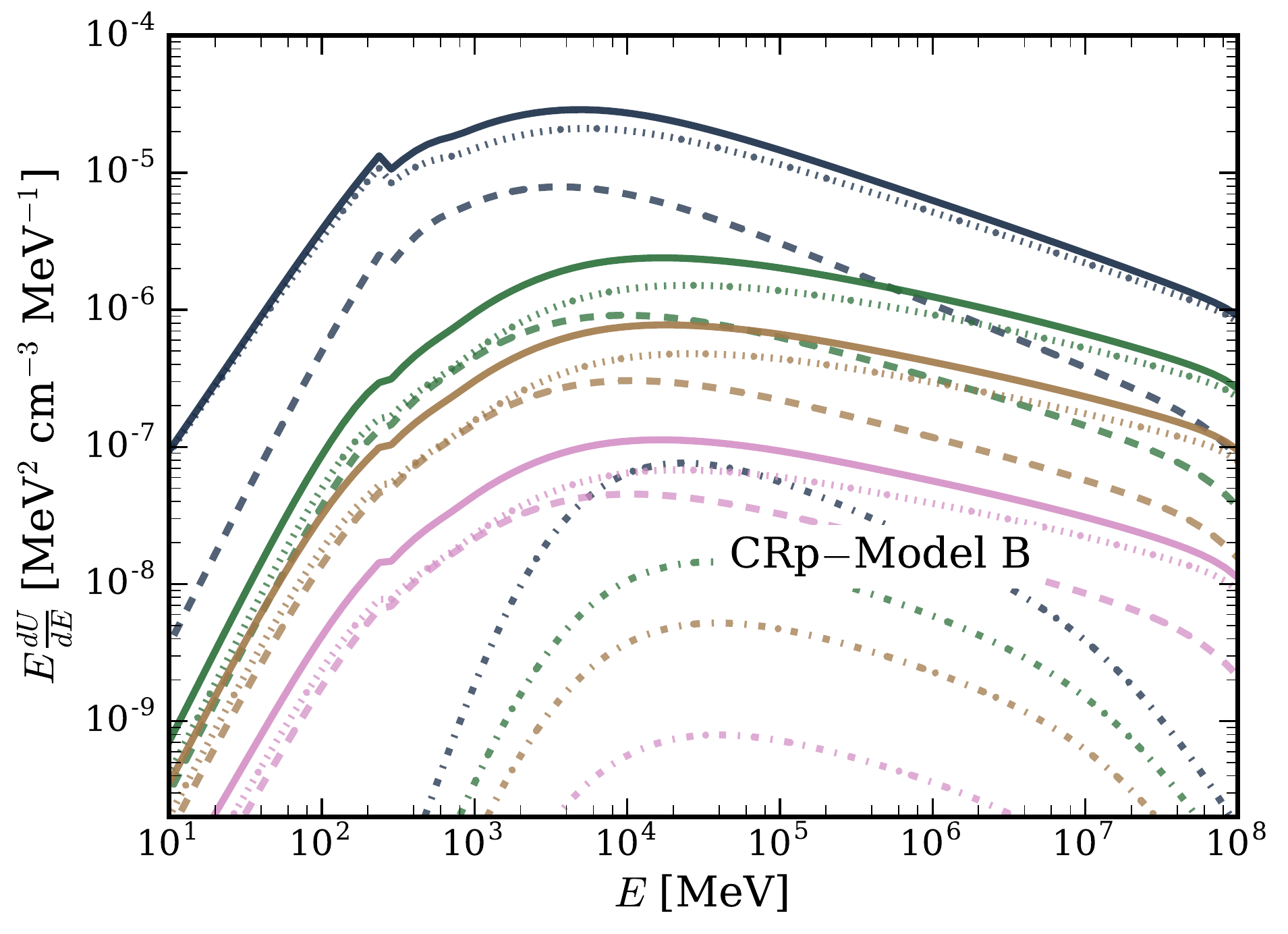} }
	\makebox[0.49\linewidth][c]{ \includegraphics[width=0.5\linewidth]{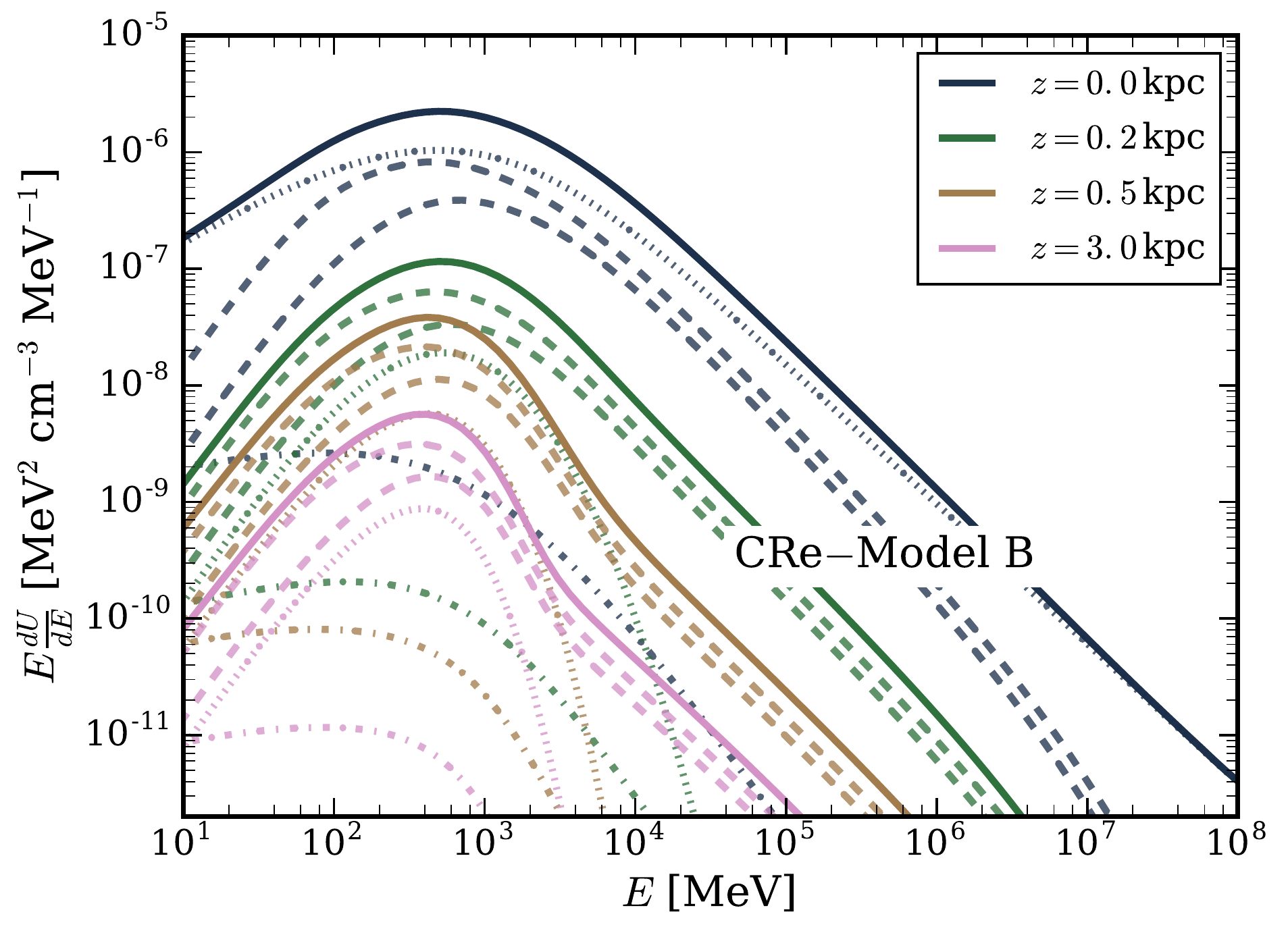} } \\
	\makebox[0.49\linewidth][c]{
	\includegraphics[width=0.5\linewidth]{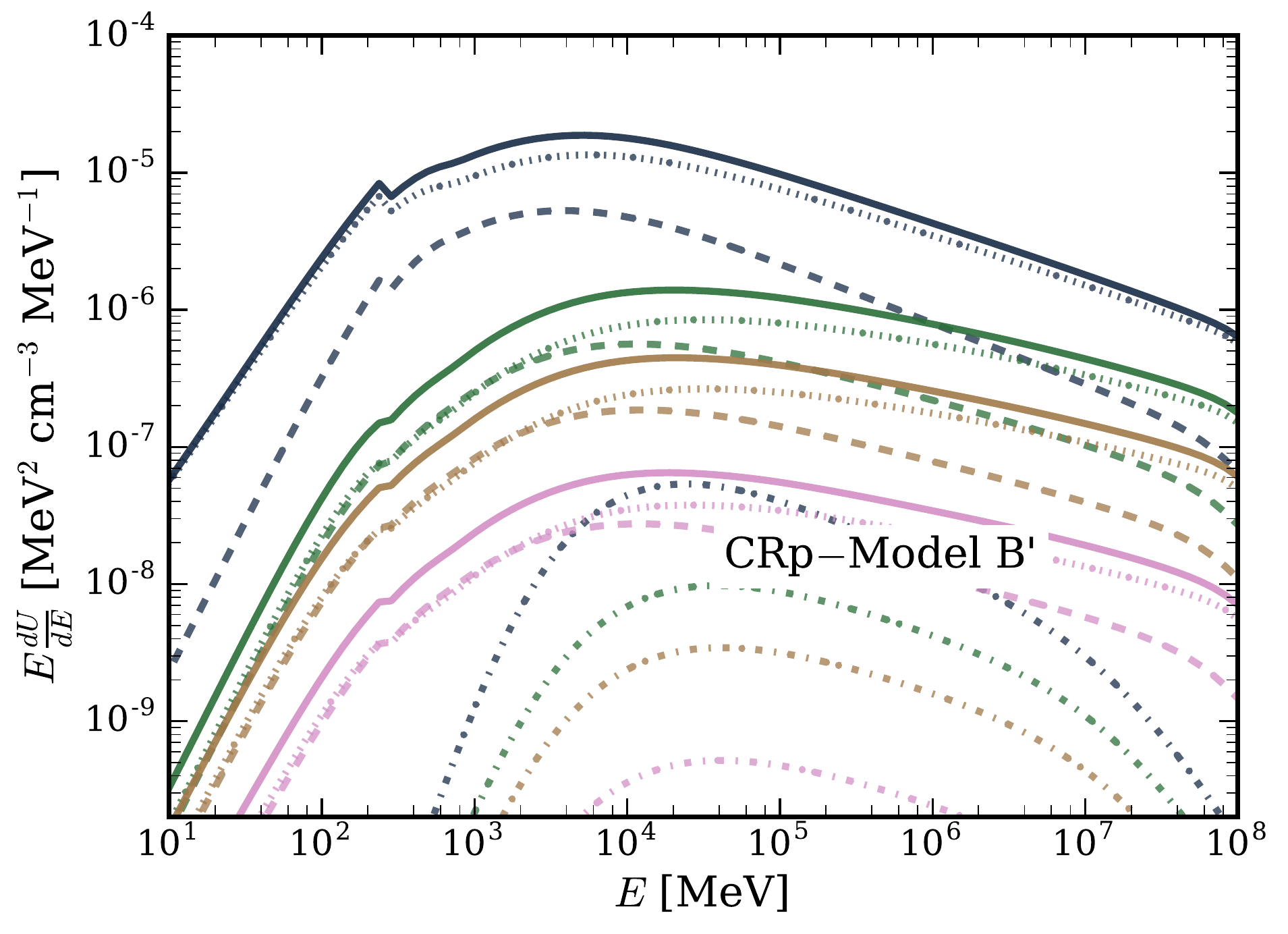} }
	\makebox[0.49\linewidth][c]{ \includegraphics[width=0.5\linewidth]{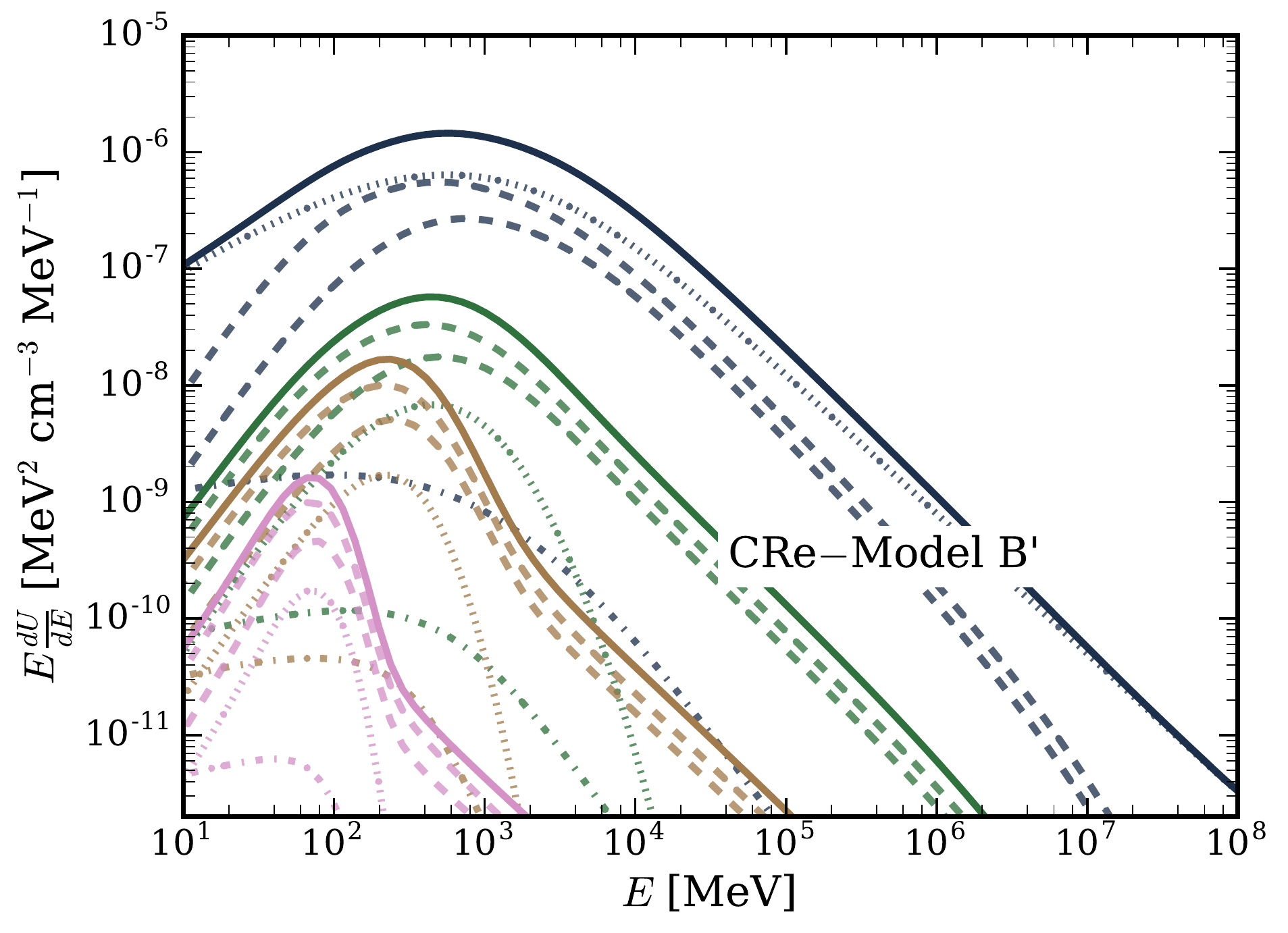} }
	\caption{ The particle spectra at 4 different heights above the M82 galactic disk. Colors denote distances along the minor axis, blue, green, brown, and pink for 0, 0.2, 0.5, and 3\,kpc, respectively. Line types denote particle type. The left panels show the proton spectra for models A (top), B (mid), and B$'$ (bottom), including all protons (solid), primary protons (dotted), secondary protons (dot-dashed), and secondary anti-protons (dashed). The right panels shows leptonic spectra for models A (top), B (mid), and B$'$ (bottom), including all electrons+positrons (solid), primary electrons (dotted), secondary electrons and positrons (dashed), and knock-on electrons (dot-dashed). }
	\label{b_particle}
\end{figure*}

\begin{figure}
	\makebox[\linewidth][c]{ \includegraphics[width=1.03\linewidth]{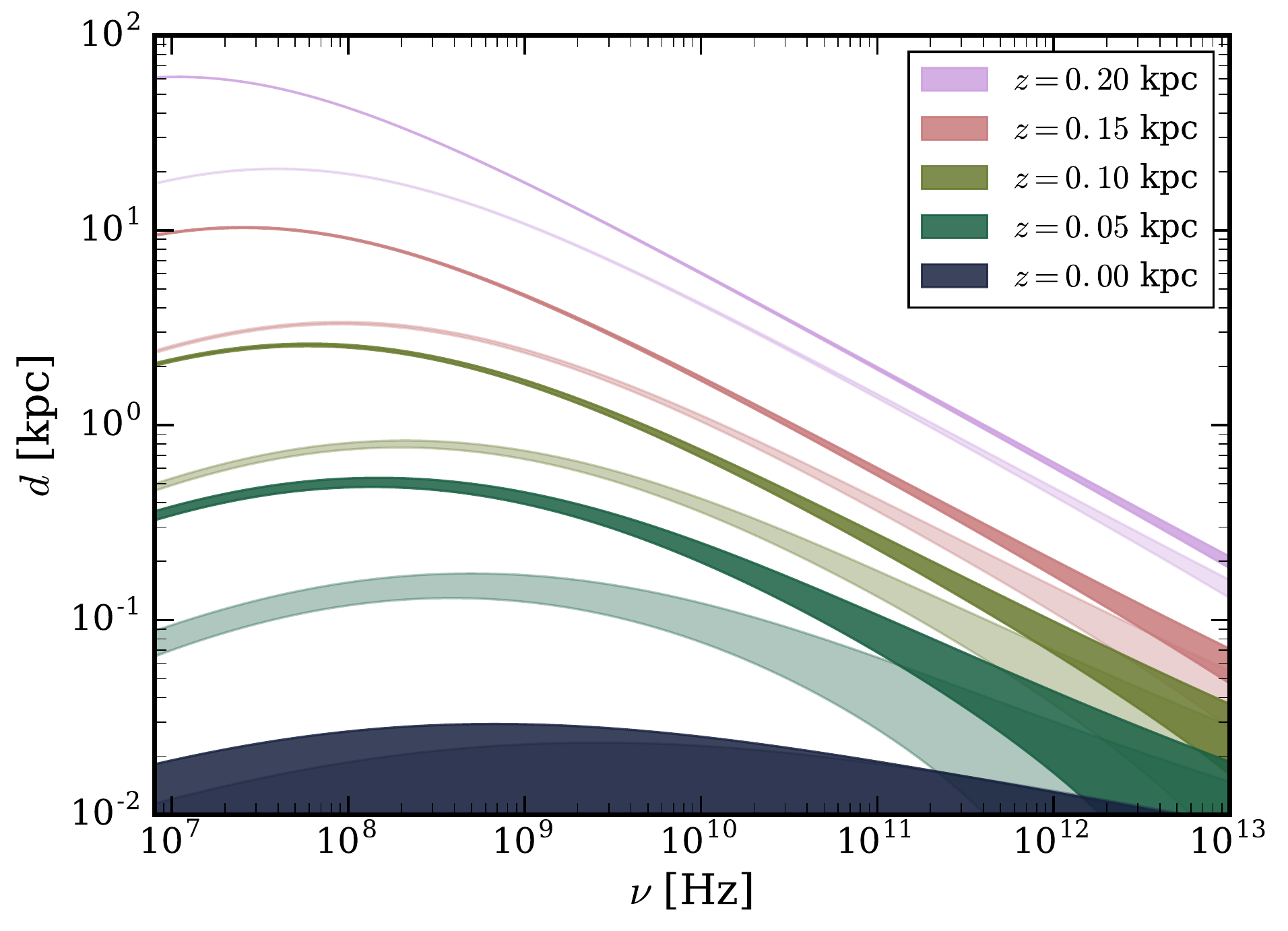} }%
	\caption{ The distance a CRe can travel before it loses all of its energy as a function of the critical synchrotron emission frequency. Results are shown for Models A (dark-shaded) and B (light-shaded). We plot \mbox{$d= V_\mathrm{wind} t_\mathrm{death} \pm \sqrt{D_{xx} t_\mathrm{death}}$} along the minor axis for each value of $z$ given in the legend. We shade the area between the $\pm$ lines for each model and distance. $t_\mathrm{death}$ is the inverse sum of all energy-loss timescales. We assume that the CR travels in a homogeneous environment defined at that value of $z$. }
	\label{plot_distance}
\end{figure}

One physical mechanism that is capable of changing the spectral index as a function of height is the transition from one cooling regime to another. Specifically, a transition from pure ionization cooling in the dense core to pure synchrotron+IC cooling in the halo can accommodate at most a change in the radio spectral index of 1 (as seen in the frequency dependence of the CRe cooling timescales, \mbox{Equations~\ref{tau_synch}--\ref{tau_ion}}). Thus, while a large fraction of the change in the spectral index seen in Figure~\ref{b_spectral} may be attributed to transitions in the dominant cooling process, additional factors are required if the observations are taken at face value. In particular, our models show that additional spectral steepening is due to a combination of wind advection and a rapid decrease in the gas density distribution outside the core. 

To understand the change in the spectral index, we examine the spectra of CRs as a function of height above the core. In Figure~\ref{b_particle}, we show the spectrum of the steady-state CR energy distributions for models A (top panels), B (mid panels), and B$'$ (bottom panels), dividing our analysis into four slices situated at 0, 0.2, 0.5, and 3.0\,kpc above the M82 galactic plane, denoted by the colors blue, green, brown, and pink, respectively. For CRp (left panels), we present the spectrum for all protons (solid), primary protons (dotted), secondary protons (dashed), and secondary antiprotons (dot-dashed). For CRe (right panels), we present the spectrum of all electrons+positrons (solid), primary electrons (dotted), secondary electrons or positrons (dashed), and knock-on electrons (dot-dashed). 

Near the core of M82, the energy density is dominated by $\sim$1\,GeV primary protons. As a result of the energy-dependent diffusion in the core, protons with energies $\lesssim$10\,GeV are less likely to escape than higher energy protons before hadronically interacting, causing the slight hardening in the spectral index from 2.37~(2.35)~[2.33] in the core to 2.16 (2.16) [2.15] at 0.5\,kpc at an energy of 100\,GeV for Model~A~(B)~[B$'$]. As $z$ increases, the dominant energy of all protons increases to $\sim$10\,GeV, although the spectrum is flat. Outside the core, secondary and primary CRp provide similar contributions to the energy density below $\sim$10\,GeV. The proton spectral shape does not change as protons propagate in the halo since propagation is dominated by the energy-independent wind and the energetic losses are minimal. The secondary antiproton density is highly subdominant. We note that the spectral ``spike'' between 200---300\,MeV is due to the $\pi^0$-production theshold. The spectral shapes are nearly identical between the models with only a difference in normalization as discussed in Section~\ref{section:results:emission:energetics}.

For electrons in the starburst core at $z=0$\,kpc, the energy density is dominated by primaries. However, as we move away from the dense core along the minor axis, secondary CRe quickly begin to dominate (top-dashed line is positrons, lower-dashed line is electrons) outside the core. Primary CRe are calorimetric in the core. Interestingly, we see a ``bump'' feature appear in the CRe spectra between $0.2-0.5$\,kpc. This is the feature required to obtain the spectral steeping in the radio halo seen in Figure~\ref{b_spectral}. Specifically, we need the synchrotron spectrum to steepen at wavelengths smaller than 22\,cm, which corresponds to CRe energies greater than $\sim$1\,GeV for a magnetic field $\sim$100\,$\mu$G (See Equation~\ref{e_nu_b}). Since the magnetic field drops off very slowly in Model~B$'$ (bottom-right panel), we see the ``bump'' feature appear at lower energies, especially at 3\,kpc (pink lines). This steepening is seen at 0.5\,kpc (brown) and continues further into the halo (as seen at 3\,kpc in pink) and is due to several factors: (1) secondaries are no longer produced by hadronic CRp interactions because of the low gas density, (2) CRe of all energies are driven from the core by the strong wind, and (3) CRe experience large synchrotron and IC losses outside the core. We also note that since the magnetic field decreases as we go further into the halo, the energy of the CRe needed to emit a certain synchrotron frequency increases, thus the synchrotron spectrum is expected to flatten as a function of distance due to the somewhat harder secondary CRe spectra at higher energies (between $10^3-10^4$\,MeV in the CRe panels for Models A and B at 3\,kpc (pink)). 

The ``bump'' feature in the CRe spectrum at GeV energies is due to two effects. The first is that CRe production turns off rapidly outside the core because there is no source of primary CRe and secondary CRe production by CRp hadronic interactions decrease rapidly at the edge of the core where the density drops precipitously. Second, (mostly secondary) high-energy CRe quickly cool outside the core as a result of still-strong synchrotron and IC losses, while low-energy CRe are allowed to propagate large distances. The ``bump'' cannot be produced by a new injection of low-energy secondary CRe, since there is no corresponding feature in the secondary CRp spectrum. Thus, our models require the efficient elimination (no sources, rapid cooling) of CRe with energies above a few GeV. The sources of primary CRe are constrained to be created inside the core while secondary CRe are created in the core and in the decreasing gas density outside of the core. There is a small population of secondary CRe that is created at large distances by the small wind component of the gas density which is seen at $>$100\,GeV in the CRe spectrum at 0.5\,kpc (brown) and 3\,kpc (pink). If the wind component of the gas density could be decreased, the CRe spectrum would be made even steeper at these higher CRe energies and then we could better fit the spectral steepening seen in the data between 22 and 6\,cm. However, we would still require the lower energy CRe secondaries created at distances outside the core to replicate the extended observed 22\,cm halo. 

Overall, we find a predominantly bremsstrahlung-cooled primary and secondary CRe population in the starburst core, which is then driven into a dominantly synchrotron- and IC-cooled halo by the wind. As seen in the right panel of Figures~\ref{b_particle}, without a large, new secondary population above 0.2\,kpc to replenish high energy CRe, all the previously high-energy CRe rapidly lose their energy, steepening the spectrum, and becoming $\sim$GeV CRe that then live long enough to propagate large distances into the halo. 

In Figure~\ref{plot_distance}, we illustrate this point by showing the displacement a CRe can travel \mbox{($d= V_\mathrm{wind} t_\mathrm{death} \pm \sqrt{D_{xx} t_\mathrm{death}}$}, where $t_\mathrm{death}$ is the lifetime of a CRe) before losing all of its energy as a function of the emitted synchrotron frequency and height along the minor axis. The different colors denote the height at which the CRe are injected and we assume it is always traveling through a homogeneous medium that is identical to its origin. Model~A is denoted by the dark-shaded regions and Model~B is denoted by the light-shaded regions. Model~B$'$ has smaller displacements than Model~B because of its slightly larger gas density and its larger halo magnetic field. The thickness of the shaded regions denote the effect of random motions of CRe due to diffusion. Wind advection begins to dominate diffusion just outside the core at 0.05\,kpc for CRe emitting $\lesssim$100\,GHz synchrotron emission. 

The majority of the CRe are created in the middle of the core, where the wind has little effect. We see that all CRe at $z=0$\,kpc travel less than 0.02\,kpc, well within the core, implying the majority of the CRe are calorimetric. For $z\approx 0.1$\,kpc, we see that high frequency ($\nu \gtrsim 10^{11}$\,Hz) CRe are unable to escape, while lower frequency CRe are free to travel $\sim$\,kpc distances. For $z>0.05$\,kpc, CRe are no longer produced at a rate comparable to the core and the CRe enter a synchrotron cooled region, which steepens the spectrum. These processes thus combine to produce the ``bump" feature and steep CRe spectrum seen in the right hand panel of  Figure~\ref{b_particle}.

We note that we have used a simple parameterized gas density distribution. As we discuss in Appendix \ref{appendix:gas_wind}, changes in the shape of the density profile and wind profile immediately outside the core region can quantitatively affect the strength of the spectral steepening found in the models.

\subsubsection{Effects of Changes to Wind \& Diffusion}
\label{section:results:extended:diff_wind}

The advective wind and diffusion constant have a direct effect on the resulting large-scale radio halo.

Unlike one-zone models, we use a spatially-dependent wind velocity profile motivated by models of hot thermally-driven galactic winds \citep{Chevalier_Clegg}. These profiles have zero radial velocity at the very center of the starburst and increase to half the maximum velocity at a spherical radius of 0.2\,kpc \citep{2009ApJ...697.2030S}. In general, one-zone models have neglected diffusion and focused on advective wind losses in computing steady-state cosmic-ray spectra. With a spatially varying wind profile, we find that our models require a diffusion coefficient large enough to transport CRs out of the zero-velocity core and closer to the wind-dominated region so that secondary CRe produced by hadronic interactions can escape into the halo, to produce the extended radio emission seen along the minor axis. Conversely, the diffusion coefficient cannot be too large, as it would then dominate the wind and overproduce the halo emission, especially at 3 and 6\,cm.

We have explored the interaction between diffusion and advection in some detail, and the related degeneracy between the wind velocity and magnetic field strength of the halo (Section \ref{section:results:extended:beta_wind}).  For Model~A, the degeneracy in the \mbox{$(\log_{10} {D_{xx}}\,/\,{\mathrm{cm}^2\,\mathrm{s}^{-1}},\ {V_0}\,/\,{\mathrm{km\,s}^{-1}})$} plane spans from $(26, 1000)-(28, 200)$. For Models~B+B$'$, which have a larger diffusion coefficent than Model A, the degeneracy spans from $(26, 2000)-(28.5, 500)$. At smaller diffusion coefficients, CRp cannot propagate far enough to create secondary CRe in the wind dominated region that, in turn, creates the large radio halo. At smaller wind velocities, the CRe population cannot be advected away fast enough, thus producing a halo that is too small to fit observations. For all models, at larger diffusion coefficients, $D_{xx}\gtrsim 10^{28.5}$~cm$^2$~s$^{-1}$, the halo is over produced, especially at 3 and 6\,cm. Compared to Model~A, models~B+B$'$ require a larger overall diffusion constant because the larger gas density and stronger magnetic field produce faster CR losses.

The assumed wind also affects the details of the CRe spectra. For example, compared to the CRe spectra in Figure~\ref{b_particle} at 0.5\,kpc (brown), as the $V_0$ wind speed is increased, the turn-over in the CRe spectrum moves to higher energies and the right (higher energy) side of the ``bump'' steepens since low energy CRe will be able to propagate further into the halo due to the larger wind and the smaller magnetic field while high energy CRe still quickly lose their energy. Conversely, as the wind speed approaches 0\,km\,s$^{-1}$ (not shown), the ``bump'' disappears and the CRe spectrum retains its spectral shape from 0.2\,kpc, highlighting the importance of the wind velocity on the character of the large-scale halo.

\begin{figure*}
	\makebox[0.49\linewidth][c]{ \includegraphics[width=.5\linewidth]{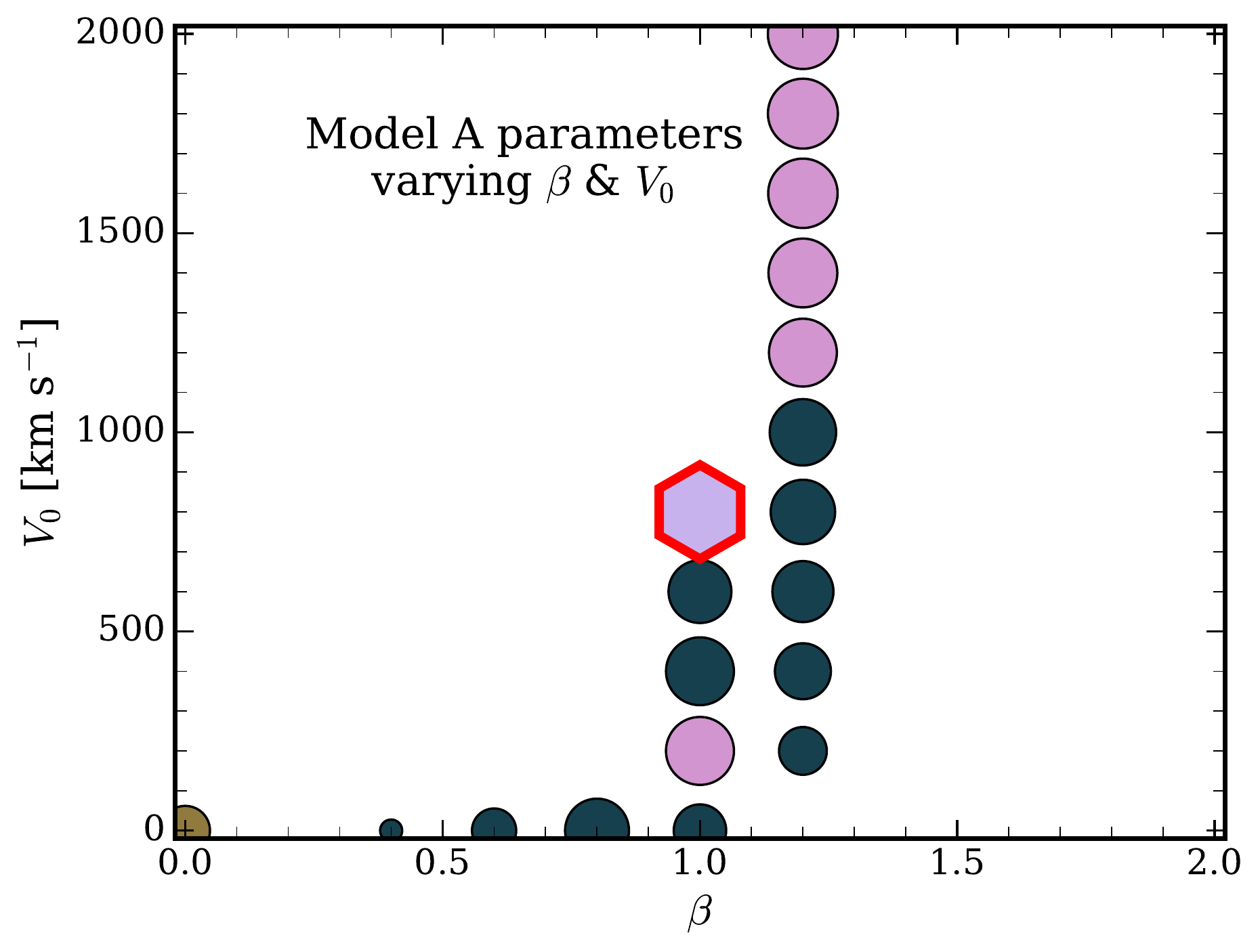} }
	\hspace{0pt}
	\makebox[0.49\linewidth][c]{ \includegraphics[width=.5\linewidth]{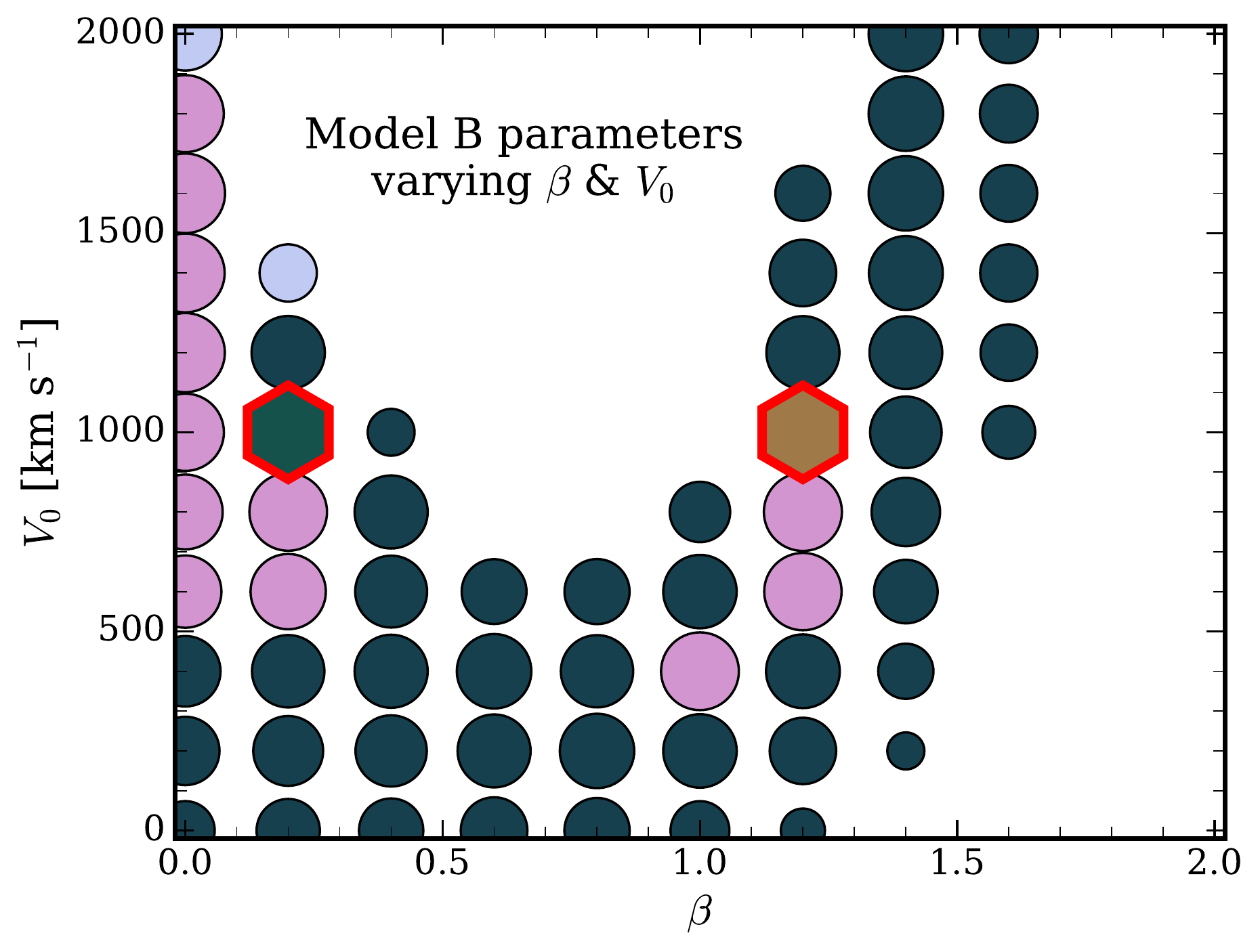} }
	\caption{ The relative $\chi^2_{\mathrm{avg,ext}}$ of the extended radio emission for models similar to Models~A (left) and B (right) in the magnetic field power-law drop-off, $\beta$ and wind velocity, $V_0$, plane. The larger dot, the better the fit. The color denotes the \emph{most constraining} wavelength where 92\,cm, 22\,cm, 6\,cm, and 3\,cm, are denoted by light blue, dark blue, brown, and pink, respectively. The same relation for models similar to Model~B$'$ is nearly identical to the right panel because Model~B and B$'$ have the same magnetic field strength. The red-outlined hexagons denote the positions of models A~(pink), B~(brown), and B$'$~(dark green).
	}
	\label{bv_fit}
\end{figure*}

\begin{figure*}
	\makebox[0.49\linewidth][c]{ \includegraphics[width=.5\linewidth]{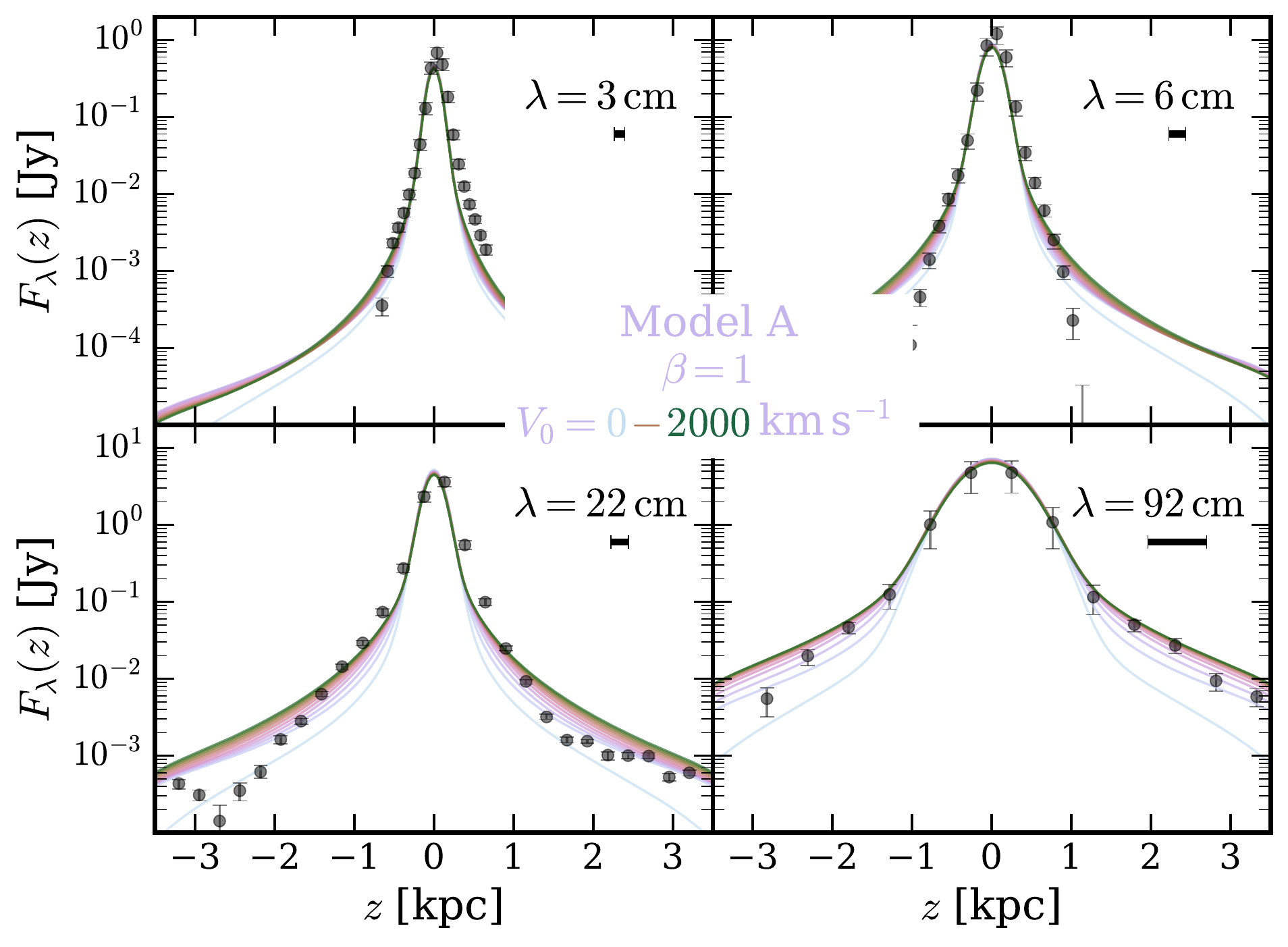} }
	\hspace{0pt}
	\makebox[0.49\linewidth][c]{ \includegraphics[width=.5\linewidth]{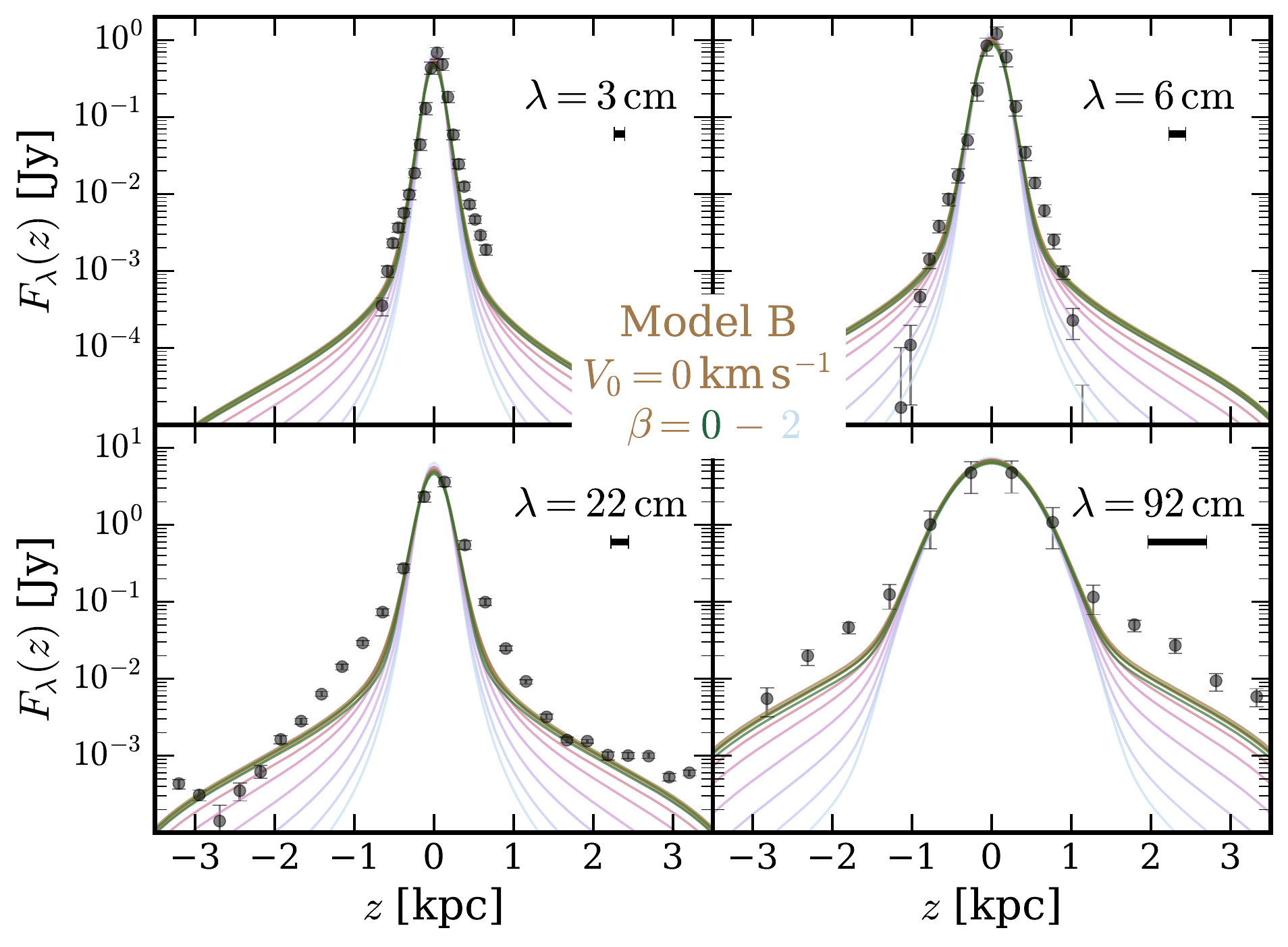} }
	\caption{ Extended synchrotron emission for models from a cross-section of the $\beta{-}V_0$ plane in Figure~\ref{bv_fit}. The left panel shows models with Model~A parameters along with magnetic field drop-offs of $\beta=1$ and letting the wind velocity, $V_0$, vary from 0 (light blue line) to 2000 km\,s$^{-1}$ (dark green line). The right panel shows models with Model~B parameters along with $V_0=0$\,km\,s$^{-1}$ and letting $\beta$ vary from 0 (dark green line) to 2 (light blue line).
	}
	\label{bv_example}
\end{figure*}

\subsubsection{A Halo Magnetic Field$-$Wind Relation}
\label{section:results:extended:beta_wind}

After escaping the central region of the galaxy, CRe need to propagate several kpc to create the large radio halo seen at 22 and 92\,cm. The two most important parameters in our models that affect CRe propagation and the synchrotron emissivity in the halo are the magnetic field power-law drop off --- $B\propto z^{-\beta}$ outside the core, along the minor axis --- and the asymptotic wind velocity, $V_0$. $V_0$ determines the local advection timescale, whereas $\beta$ determines the local magnetic field strength in the halo, the lifetime of CRe due to synchrotron losses, and the relative importance of IC and advection. Since the density drops quickly enough, ionization and bremsstrahlung losses rapidly become weak, and diffusive transport quickly becomes sub-dominant to advection as the physical scale grows, the overall halo  synchrotron emissivity is controlled by $\beta$ and $V_0$.

To explore how these parameters affect the properties of the synchrotron halo, we computed a large suite of models with different $\beta$ and $V_0$, but with core properties identical to Models A and B described throughout this paper (Table~\ref{parameters}). Figure~\ref{bv_fit} shows the relative \mbox{$\chi^2_{\mathrm{avg,ext}} = \sum \chi^2_{\lambda} / N_{\lambda}$} in the $\beta{-}V_0$ plane where $\chi^2_{\lambda}$ and $N_\lambda$ are the $\chi^2$ and the number of data points at a radio wavelength $\lambda$.  We again used a modified $\chi^2$ to approximately weight the observations from all wavelengths evenly. The left and right panels present constraints on variants of Model A and B, respectively. Larger dot sizes indicate a better fit to the data, while the color denotes the \emph{most constraining} wavelength: light blue, dark blue, brown, and pink denote 92, 22, 6, and 3\,cm, respectively. We note that we display a wide range of $\chi^2_{\mathrm{avg,ext}}$ values so that the overall behavior of the models is well-represented, even though some the combinations of $\beta$ and $V_0$ are not especially good fits to the data (compare with Figure~\ref{bv_example}; see below). With the exception of $\beta$ and $V_0$, all models employ the default values listed in Table~\ref{parameters} for Models A and B. As discussed in Section \ref{section:results:extended:diff_wind}, decreasing the diffusion coefficient relative to its fiducial value for either model requires a larger value of $V_0$, and would thus move the locus of best-fit values vertically in Figure~\ref{bv_fit}.

For both sets of models, Figure \ref{bv_fit} shows that the 22\,cm observations (dark blue) dominate the constraints at high-$\beta$ ($\gtrsim1$) and low-$V_0$ ($\lesssim1000$\,km/s). For larger values of $\beta$ the modeled 22\,cm halo becomes too small to reproduce the data. Meanwhile, the 3\,cm (pink) observations tend to dominate the constraints at high $V_0$ in Model~A variants and at low-$\beta$ and high-$V_0$ in Model~B variants. In these regions the modeled emission of the inner halo begins to be overproduced. For Model~B, the 92\,cm data (light blue) dominates the constraints for two points at low $\beta\leq0.2$ and large $V_0$. There, the low-frequency halo emission is overproduced because the weak fall-off in the magnetic energy density and the rapid advection combine to produce a bright, spatially-extended halo.

The red-outlined hexagon in the left panel shows the actual values for the fiducial Model~A and denotes a $\chi^2_{\mathrm{avg,ext}}$ minimum in the $\beta-V_0$ plane. Indeed, this was how the final parameters for Model~A were chosen. As discussed in Section \ref{section:model:observations:fitting}, we first explored the $B_0{-}n_0$ degeneracy using only the integrated emission (Figure \ref{chi_SNR_plot}), and then chose representative models for presentation and discussion (among them, Model~A), with final parameters determined using their detailed halo properties (Figure \ref{bv_fit}).

The two red-outlined hexagons in the right panel show the parameters for the two best-fitting models with the same core parameters as Model~B. These two $\chi^2_{\mathrm{avg,ext}}$ minima have the same $V_0=1000$\,km/s, but very different $\beta= 0.2$ and $\beta=1.2$. The higher $\beta$ model is identical to fiducial Model~B (Table \ref{parameters}). Like Model~A, this is how the final parameters for Model~B were chosen. The low-$\beta$ model is very similar to Model B$'$, but has somewhat different core properties. Whereas the models shown here have core properties identical to Model~B, for the final version of Model B$'$ we adopted a different value for the core density $n_0$ (Table \ref{parameters}) in order to better reproduce the integrated gamma-ray emission. If we instead reproduce the right panel of Figure~\ref{bv_fit} starting with the same core parameters of Model B$'$, we find qualitatively identical results. 

Multiple factors contribute to the degeneracy in the $\beta{-}V_0$ plane. First, as $V_0$ increases, CRe are advected out to larger distances in the halo. If $\beta$ is increased above ${\sim}1.2$ or ${\sim}1.5$ for Model~A and B, respectively, the magnetic field strength decreases too quickly along the minor axis to create the observed extended halo at 22 and 92\,cm. For lower values of $\beta$, the magnetic field remains stronger out to larger distances, which increases the amount of synchrotron emission in the halo. However, if we decrease $\beta$ past some critical value ($\beta\simeq0.8$ for Model~B and $\beta\simeq1.0$ for Model~A), the CRe energy decreases too much to create the extended halo because of the strong synchrotron cooling near the starburst and resulting faint large-scale halo. This problem can be partially mitigated if the core magnetic field $B_0$ is larger and if $\beta$ becomes sufficiently small. Model~B has a larger value of $B_0$ than Model~A, this is why for Model B (right panel) we find some solutions at low $\beta$ and for a range of velocities.

To better demonstrate the behavior of our models in the $\beta{-}V_0$ plane, we present the extended synchrotron emission of an array of models in Figure~\ref{bv_example}. In the left panel, we show models similar to Model~A with $\beta=1$ and let the velocity vary from $0$ (light blue lines) to $2000$\,km\,s$^{-1}$ (dark green lines) in increments of $200$\,km\,s$^{-1}$. We see that even for the $V_0=0$\,km\,s$^{-1}$ model (light blue lines) we over-produce emission at the most extended data points at 6\,cm at $\pm1$\,kpc, while simultaneously under-producing emission in the 22 and 92\,cm halos on kpc scales. As $V_0$ is increased, we continue to over-produce the 3\,cm and 6\,cm data beyond $\pm1$\,kpc, but we also begin to over-produce the 22\,cm halo on $1-2$\,kpc scales, and eventually the 92\,cm data on $\pm3$\,kpc scales. Thus, while the dots in the left panel of Figure~\ref{bv_fit} indicate that for $\beta\simeq1-1.2$, a broad range of $V_0$ may describe the data, the best overall fit occurs for \mbox{$V_0\simeq800$\,km s$^{-1}$}.

The right panel shows models similar to Model~B with \mbox{$V_0=0$\,km\,s$^{-1}$} and letting $\beta$ vary from 0 (dark green lines) to 2 (light blue lines) in increments of 0.2. All models again overproduce the 6\,cm data on scales larger than $\pm1$\,kpc. Simultaneously, all models under-predict the 22\,cm and 92\,cm data. Thus, even though the dots in the right panel of Figure~\ref{bv_fit} indicate that a broad range of low-$V_0$ models may fit the data, the $\chi^2_{\mathrm{avg,ext}}$ minima (the locations of the red-outlined hexagons) all require \mbox{$V_0\simeq1000$\,km s$^{-1}$}.

These sets of constraints on Models A and B (and B$'$) all strongly suggest that advective transport of cosmic rays at \mbox{${\sim}1000$\,km\,s$^{-1}$} is required to produce the large scale halo at both 22 and 92\,cm. Values of the wind speed in the range of the values of $V_0$ we require were argued for in M82 on the basis of X-ray observations by \cite{2009ApJ...697.2030S}. However, as we have emphasized, all our models fail to reproduce the sharp truncation in the 3\,cm and 6\,cm data reported by \cite{2013A&A...555A..23A}. Either there is a diffuse, large-scale, and low surface brightness flux at 3 and 6\,cm missing from the current interferometric measurements of the magnitude predicted by our models, or steady-state cosmic-ray propagation solutions with the physics employed here simply cannot capture the dynamics of the system.

\section{Discussion}
\label{section:discussion}

In Section~\ref{section:results}, we constrained the CR population, gas density, and magnetic field strength as a function of distance along the minor axis. We determined the starburst core gas density, magnetic field strength, and CR energy density by comparing our models to the integrated gamma-ray and radio emission. We subsequently used the observations of the large radio halo to constrain our models along the minor axis. In this section, we discuss the implications of our models,  and the  constraints we can place on the possibility that CRs (Section~\ref{section:discussion:cr_acceleration}), the ISRF (Section~\ref{section:discussion:isrf_acceleration}), or the magnetic field (Section~\ref{section:discussion:magnetic_acceleration}) may be dynamically important in driving the observed galactic wind. 

In the left panel of Figure~\ref{acceleration}, we present our modeled energy densities of the CRp (dark-filled region), the CRe (dotted-outlined region), and the magnetic field (dashed-dotted line) for Model~A (lavender), Model~B (brown), and Model~B$'$ (green). The light-filled regions in each color denote the impact of changing the power-law slope for the magnetic field along the minor axis, $\beta$, by $\pm 0.1$ from the model parameters (See Section~\ref{section:model:galprop:mag_gas}; Table \ref{parameters}). The assumed ISRF energy density we use in our modeling (See Section~\ref{section:model:galprop:isrf}) is denoted by the dashed black line. 

\subsection{Gas Acceleration from Cosmic Rays} 
\label{section:discussion:cr_acceleration}

\begin{figure*}
    \makebox[0.49\linewidth][c]{ \includegraphics[width=0.50\linewidth]{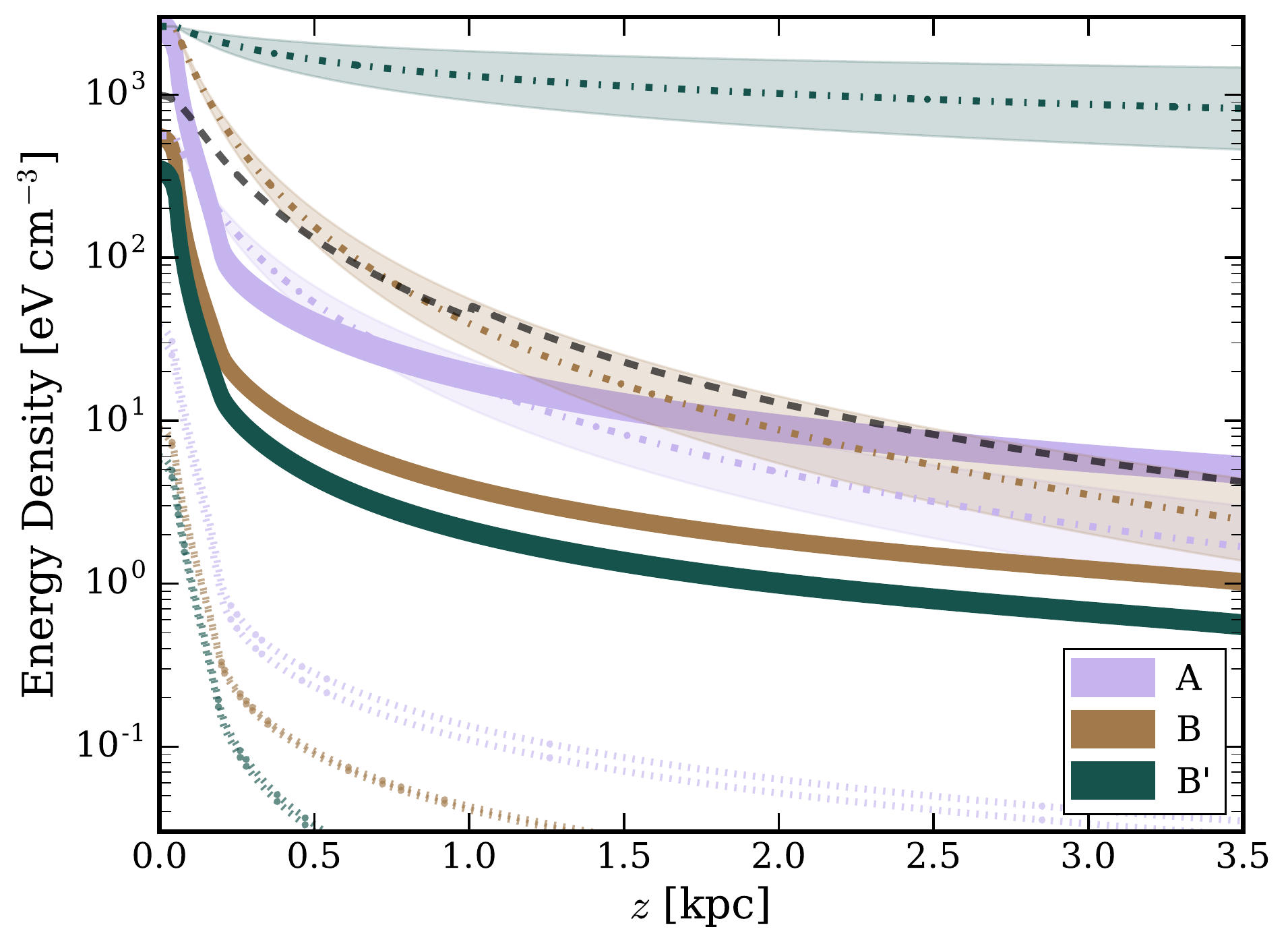} }
	\makebox[0.49\linewidth][c]{ \includegraphics[width=0.50\linewidth]{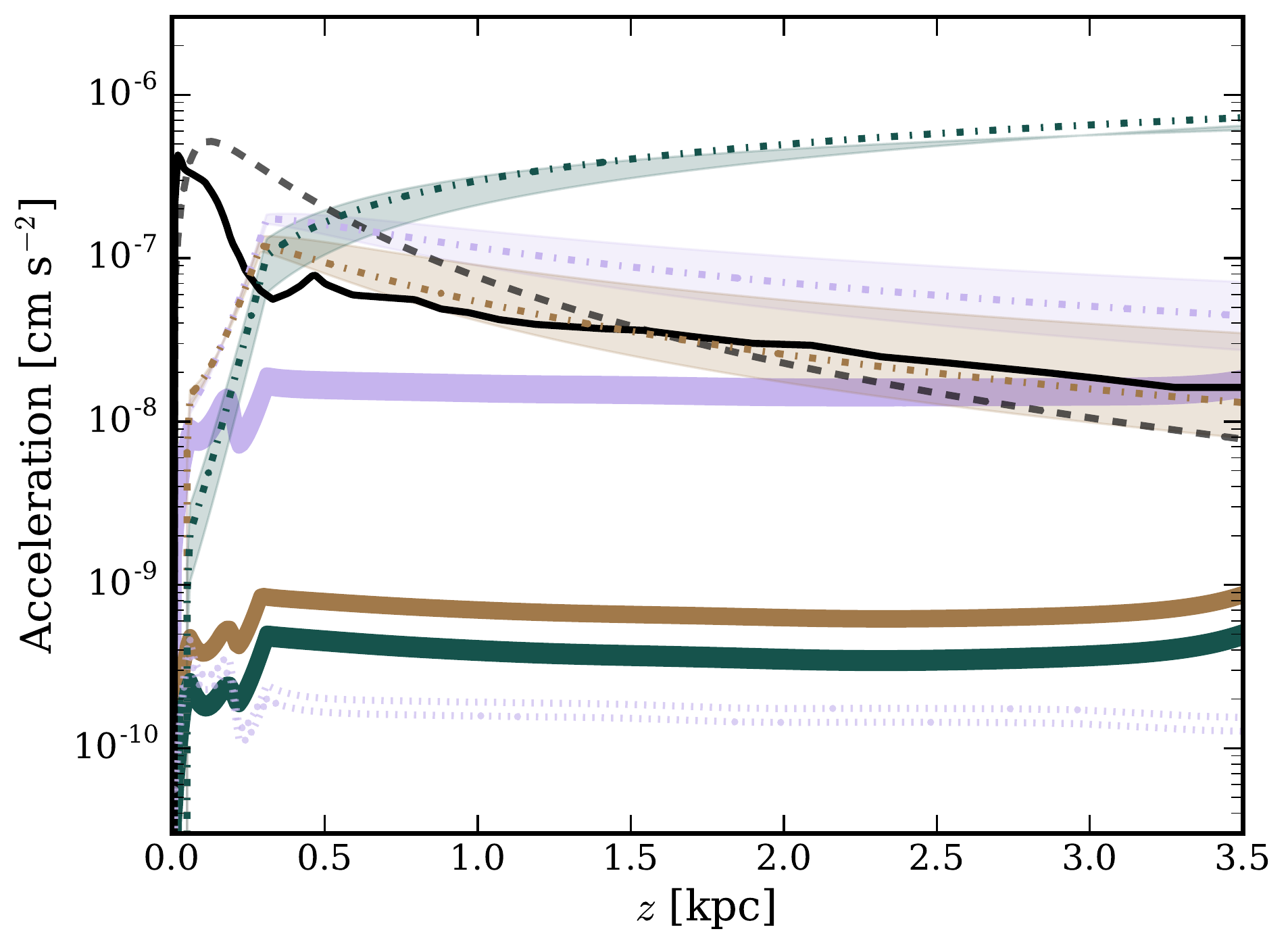} }
	\caption{ Energy densities (left panel) and ``acceleration imparted to the gas'' (right panel) for Model~A (lavender), Model~B (brown), and Model~B$'$ (green). The dark, filled lines denote CRp, the dotted outlined regions denote CRe, the light filled regions denote a range of magnetic field models, and the dot-dashed lines denote the magnetic field model. The shaded region of the magnetic field parameter space depicts the impact of the assumed power-law decay of the magnetic field energy density for models where $\beta$ is shifted by $\pm0.1$ from the standard model parameters. The dashed lines denote the ISRF. The solid black line is the acceleration due gravity based on data from \citet{2012ApJ...757...24G}. }
	\label{acceleration}
\end{figure*}

As CRs propagate through M82 and its halo, they interact with the gas through the magnetic field. We do not model the dynamical interaction, but for simplicity we assume CRs directly interact with the gas through the CR pressure gradient. We do not solve the fluid-dynamic equations for the wind, but we can calculate the acceleration that can be induced by the CR pressure gradient in our models.  

As part of the model outputs of, {\tt GALPROP} returns the isotropic spectrum of each CR species, or $K^2 I_K$ where $K$ is the kinetic energy and $I_K$ is the number intensity (number per area per time per solid angle) per kinetic energy. The total pressure is then \mbox{$P_\mathrm{CR} = ({4\pi}/{3}) \int dK \,I_K {\beta E}/{c}$}. A similar expression is used for the energy density.

The left panel of Figure~\ref{acceleration} shows the energy density in each component, for each model. In the right panel of Figure~\ref{acceleration}, we use the steady-state CR distribution and spectra to constrain the acceleration the CRs might impart to the gas of M82 by calculating ${\nabla P}/{\rho}$ and comparing it to the acceleration due to gravity (denoted as the black solid line) calculated by \citet{2012ApJ...757...24G}. The density profiles used are those constrained from our models. The acceleration due to CRp (CRe) is denoted by the dark-filled (dotted-outlined) region, while the colors --- lavender, brown, and green --- denote Model A, B, or B$'$, respectively. The dashed lines indicate the acceleration of gas due to radiation pressure from the ISRF under simple assumptions (discussed in Section~\ref{section:discussion:isrf_acceleration}). The dot-dashed lines and light-filled regions are due to the magnetic field and are discussed in Section~\ref{section:discussion:magnetic_acceleration}.

We see that CRp acceleration of Model~A is on order $10^{-8}$\,cm\,s$^{-2}$ near 0.5\,kpc, while Model~B+B' have a CRp acceleration on order $10^{-9.5}$\,cm\,s$^{-2}$. The implied acceleration of the gas from CRs in Model~A is ${\sim}1$ order of magnitude below the approximate gravitational acceleration (black dashed), indicating that even in our model with the largest CR energy density and the smallest gas density, CRs are highly \emph{dynamically subdominant} near the disk of the galaxy. However, for distances larger than $\sim$3\,kpc from the disk, CRp may become dynamically relevant for Model~A. In the case of Model~B+B$'$, the CRp acceleration is much smaller. 

The differences in the CR energy densities and ${\nabla P}/{\rho}$ between Models A and B+B$'$ are largely due to the differing CR energy (left panel) and gas densities. Model~B+B$'$ have a smaller total CR energy density within the galaxy and a larger gas density overall compared to Model~A, which yields an overall  factor of ${\sim}20$ difference in the acceleration of the models and a factor of ${\sim}5$ in the central CR energy density. As shown in Figure~\ref{chi_SNR_plot}, Model~A is chosen to be representative of models with the smallest allowable gas density. Thus there is not much room to decrease the gas density further and remain consistent with the gamma-ray and radio observations. However, we note that our models in general require a rapid decrease in the energy density below 0.2\,kpc to fit the spectral steepening of the radio halo discussed in Section~\ref{section:results:extended:cr}. If the gas density could be decreased by a factor of a several on the scale of 0.5\,kpc, then the acceleration due to CRp would increase by the same factor. However, such a small gas density would be in tension with observations by \citet{2015ApJ...814...83L}.  

Overall, our {\tt GALPROP} models show that in a starburst galaxy like M82, the CRs from the starburst are dynamically weak with respect to gravity on the scale of the starburst, and out into the halo along the minor axis. These results strongly constrain models for CR-driven winds, by limiting the overall CR pressure gradient and the associated gas acceleration. As anticipated by earlier one-zone models \citep{LTQ,Lacki2011}, the central CR energy density is set by hadronic losses, such that we expect the central energy density to be of order $\sim\dot{E}_{\rm CR}\tau_{\rm had}/{\rm Volume}$, where $\dot{E}_{\rm CR}$ is the total CR energy injection rate. Scaling for the parameters of Model A (Tables \ref{parameters} \& \ref{outputs}), this simple estimate predicts a core CR energy density \mbox{${\sim}5000$\,eV\,cm$^{-3}$}, a factor of ${\sim}2$ higher than the results of our models, and with a $n_0^{-1}$ dependence that accounts for the decrease in the core energy density between Models A, B, and B$'$. 

The fact that the CRp acceleration reaches the gravitational acceleration on scales of $3$\,kpc in Model A is qualitatively similar to what would be expected for the critical sonic point in a CR-driven wind in the streaming limit \citep{1975ApJ...196..107I,Mao_Ostriker}. However, we note that in such models the gas velocity would be expected to be of order the local CRp sound speed and gravitational circular velocity at the critical point.  A globally consistent solution indicating a CRp-driven wind from our modeling would require that the profiles of these quantites match predictions from models like those of \cite{1975ApJ...196..107I}. This will be the subject of a future effort.

\subsection{Gas Acceleration from the ISRF}
\label{section:discussion:isrf_acceleration}
Radiation pressure on dust grains has been suggested as a mechanism for driving large-scale galactic winds (e.g., \citealt{MQT,wibking}). A detailed assessment for M82 on the basis of the dust-scattered UV halo was done by \cite{coker}.

For reference, in the left panel of Figure~\ref{acceleration}, the energy density of the ISRF is denoted by the black dashed line (See Section~\ref{section:model:galprop:isrf}). In the right panel, we plot the acceleration imparted to our gas models under simple assumptions, and with no attenuation. The acceleration is calculated by integration of the vertical radiation flux from our ISRF source distribution (See Tables~\ref{distributions}~\&~\ref{parameters}) and using the formula $\vec a = \kappa  \vec F_\mathrm{ISRF}/c$ from \citet{rybicki}, where $F_\mathrm{ISRF}$ is the energy flux and $\kappa$ is the absorption opacity per unit gram of gas. We assume that all starlight is unattenuated, and thus take \mbox{$\kappa= 1000$\,cm$^2$\,g$^{-1}$} as representative of the flux-mean dust opacity for a young stellar population (\citealt{2001ApJ...554..778L}).

While the nominal value of the acceleration imparted to gas is greater than the acceleration of gravity below ${\sim}1.5$\,kpc, we have ignored absorption in determining the flux. The work of \cite{coker} implies that the attenuation between the core and the halo corresponds to a UV optical depth of ${\sim}3$ for a foreground dusty screen geometry, which indicates that the acceleration used here outside the core of M82 should be decreased by a factor of ${\sim}20$ and that the acceleration due to starlight on kpc scales is dynamically weak compared to gravity (\citealt{coker}; see also \citealt{socrates}).

\subsection{Gas Acceleration from the Magnetic Field} 
\label{section:discussion:magnetic_acceleration}
From our {\tt GALPROP} models, we determine the strength of the magnetic field along the M82 minor axis. In Section~\ref{section:results:emission:mag_gas} and in Figure~\ref{chi_SNR_plot}, we show that the magnetic field in the starburst core must be $\gtrsim 150$~$\mu$G to replicate the integrated emission. Using measurements of the extended radio halo, in Section~\ref{section:results:extended:beta_wind}, we demonstrate that the power-law drop-off of the magnetic field must have an index $\beta < 1.6$. Due to the rapid decrease in the CR energy density outside the core (See left panel of Figure~\ref{acceleration}), the energy density of CRe further into the halo must be low (as seen in Section~\ref{section:results:extended:cr}), implying we need a relatively strong halo magnetic field to reproduce the bright, extended radio observations.

Putting this information together, we find that the gradient in the magnetic field energy density implies a potential acceleration (for our density profiles) that is dynamically comparable to gravity. In Figure~\ref{acceleration}, we show \mbox{${\nabla P}/{\rho} = {\nabla U_\mathrm{mag}}/{\rho}$}, the gas acceleration due to the gradient in the magnetic field energy density. The dotted-dashed lines denote the magnetic fields used by Model~A (lavender), Model~B (brown), and Model~B$'$ (green). The light shaded regions demonstrate the impact of allowing the magnetic field power-law drop-off, $\beta$, to vary by $\pm 0.1$. In the left panel, we see that the magnetic field energy density is greater than 1\,eV\,cm$^{-3}$ throughout our entire halo. In the right panel, we see that for all models, the magnetic pressure gradient divided by the gas density is larger than gravity on scales larger than ${\sim}0.2$\,kpc. At distances above 0.05\,kpc, the acceleration increases rapidly because the magnetic field energy density falls more slowly than the gas density. Overall, having a slower magnetic field drop-off, gives us a larger magnetic field, thus a larger magnetic pressure gradient. In fact, another way to state the result of the right panel of Figure~\ref{acceleration} is in terms of the Alfv\'en speed $V_A=B/\sqrt{4\pi\rho}\propto \sqrt{U_{\rm mag}/\rho}$. In Model A, $V_A\simeq150-200$\,km s$^{-1}$  on ${\sim}0.5-1.0$\,kpc scales with the assumed density profile. Because of the very slow decrease in the magnetic field of Model B' in the halo, the implied value of $V_A$ reaches $>1000$ km s$^{-1}$ for $z>2$\,kpc.

Overall, Figure~\ref{acceleration}, shows that the magnetic pressure gradient is greater than the CR pressure gradient in our models and that the magnetic pressure could in principle be important in driving the large scale galactic outflow of M82. However, note that we have ignored the magnetic tension force, which would change the total magnetic force depending on the global field topology. Indeed, the strong magnetic field could in principle trap material near M82 and inhibit an outflow, perhaps as seen in the HI wind dynamics reported by \cite{2018ApJ...856...61M}. Much more work is required to understand the dynamical importance of the magnetic field for wind driving.

\section{Conclusions}
\label{section:conclusion}

We use a large suite of two-dimensional axisymmetric {\tt GALPROP} models to constrain the cosmic ray population of the local starburst and super-wind galaxy M82. Using prescriptions for the gas density distribution, magnetic field distribution, and cosmic-ray energy injection rate we seek models that match both the integrated gamma-ray emission, but also the spatially resolved radio emission, particularly along the minor axis where the galactic wind is well-studied by many dynamical and gas tracers. Although subject to a large parameter space, we are able to draw general conclusions about the cosmic-ray energy density and pressure, magnetic field strength, cosmic-ray diffusion rate and advective transport speed. In particular, using the integrated radio and gamma-ray emission, we are able to constrain the locus of potential gas densities and magnetic fields that the cosmic rays see during propagation through the galaxy. As they enter the wind and halo, we are able to constrain the power-law fall-off in the magnetic field strength along the minor axis, and the advection velocity. We use these models to generate maps of the radio and gamma-ray emission at a wide range of energies/frequencies (Figures \ref{radio_ext}, \ref{all_map}, \ref{lofar_map}). Our models lead to a variety of conclusions about the physics of cosmic rays and their propagation in M82, and predictions for future observations.

Our findings from our modeling can be summarized as follows:
\begin{itemize}
    \item There is a degeneracy between the average magnetic field strength and gas density of the starburst core that is strongly constrained by the spectral index of the radio emission, and partially results from the competition between the density-dependent cosmic-ray electron/positron cooling processes, bremsstrahlung and ionization, and the density-independent synchrotron and IC cooling (Figure \ref{chi_SNR_plot}). We provide a quantitative analysis in Section~\ref{section:results:emission:mag_gas} and Appendix~\ref{appendix:spectral}. This result on the magnetic field$-$gas density degeneracy is similar to results found by \citet{2012ApJ...755..106P}. We find a minimum magnetic field strength of 150\,$\mu$G which is similar to the results of \citet{lacki_thompson,yoast-hull}.
    \item For the lowest allowed values of the core gas density and magnetic field strength allowed by the integrated emission, as exemplified by Model A (Figure \ref{chi_SNR_plot}), free-free absorption and emission are important in shaping the radio spectrum at low ($<1$\,GHz) and high frequencies ($>10$\,GHz), respectively (Figure \ref{AB_integrated}; Section~\ref{section:results:emission:mag_gas}).
    \item Due to the relatively high gas densities, the total core cosmic-ray energy density is controlled by strong hadronic losses, implying cosmic-ray pressures that are dynamically weak within the galaxy with respect to the pressure required for hydrostatic equilibrium (Figure \ref{acceleration}; Section~\ref{section:results:emission:energetics}; \ref{section:discussion:cr_acceleration}). This result is in agreement with \cite{LTQ}.
    \item Secondary protons and secondary electron+positrons are non-negligible and need to be taken into account in starburst environments (Figure \ref{b_particle}; Section~\ref{section:results:emission:energetics}). Secondary CRe are important for shaping the integrated radio emission \citep{Rengarajan,Thompson2006,Lacki2011}, but also for providing a source of energetic CRe at the edge of the starburst core that can power the extended synchrotron halo.
    \item The strongly decreasing spectral index as a function of frequency and height above the disk (See Figure \ref{b_spectral}) is due to two factors: (1) Secondary electrons+positrons are no longer produced above a certain height because the hadronic interaction time becomes longer than the advective timescale, (2) The transition from bremsstrahlung+ionization dominated losses for the cosmic-ray electrons/positrons to synchrotron dominated losses, and (3) A strong wind advecting cosmic-ray electrons out of the core (Section~\ref{section:results:extended:cr}).
    \item There is a degeneracy between the cosmic-ray advection velocity, $V_0$, and the rate at which the magnetic field decreases as a function of height along the minor axis, $B\propto r^{-\beta}$ (Figure \ref{bv_fit}). The brightness of the radio halo dictates relatively small values of $\beta$ for realistic values of the cosmic-ray advection speed. For example, in Model~A, which has \mbox{$B_0=150$\,$\mu$G} and \mbox{$n_0=150$\,cm$^{-3}$}, we find a best fit with \mbox{$\beta\simeq1$} and \mbox{$V_0\sim1000$\,km s$^{-1}$} (Figs.~\ref{bv_fit}~\&~\ref{bv_example}, Section~\ref{section:results:extended:beta_wind}). Throughout the parameter space explored, we always find better overall fits to the 22\,cm and 92\,cm data with fast \mbox{$V_0\simeq500-1500$\,km s$^{-1}$} advection, but we are never able to accommodate the steep drop in 3\,cm and 6\,cm flux reported by the observations (Figs.~\ref{bv_fit}, \ref{bv_example}). 
    \item Primary cosmic-ray electrons are strongly calorimetric for all models, whereas secondary electrons and positrons are only partially ($\simeq0.2$) calorimetric for models with lower density and magnetic field (Model A), to more fully calorimetric ($\simeq0.7$ for Model B). Cosmic-ray protons are approximately $50-60$\% calorimetric across the range of models (see Table \ref{outputs}).
    \item Because of strong hadronic losses, cosmic rays have a low core energy density. When we calculate the corresponding pressure, and attempt to estimate the acceleration from the cosmic rays that might be imparted to the gas profiles of our model, we find that the cosmic rays are dynamically weak with respect to gravity (Figure \ref{acceleration}; Section~\ref{section:discussion:cr_acceleration}). Model A has an implied acceleration similar to gravity on ${\sim}3$\,kpc scales.
    \item Because the density drops rapidly outside the core, and because the magnetic field is constrained to fall off fairly slowly (e.g., $\beta\simeq1$ for Model~A, left panel of Figure~\ref{bv_fit}, Table~\ref{parameters}), the implied Alfv\'en velocity increases rapidly along the minor axis (\mbox{$\simeq200$\,km s$^{-1}$} for Model~A on kpc scales). The corresponding magnetic energy density across our range of models is large in the halo, implying that magnetic forces may be dynamically important. The indicated gradient in the magnetic pressure could in principle contribute to driving the large-scale galactic wind, but depends on details of the magnetic field topology that we are unable to constrain (Figure \ref{acceleration}; Section~\ref{section:discussion:magnetic_acceleration}).
\end{itemize}

We have neglected several physical effects that should be examined more fully in future works. Our models evolve to time-steady snapshots and therefore do not capture the physics of a time-variable source of CRs. Myr timescale variations in the star formation history as reported by \cite{Forster_Schreiber2003} for M82 could affect the halo CR energy distribution. We have also neglected CR reacceleration, which could provide a source of CRe in the halo. Finally, we have also ignored the contribution from tertiary CRs.

The primary critical area for future investigation is a self-consistent comparison between dynamical CR wind models and the non-thermal emission of the M82 wind halo.

\section*{Acknowledgments}
We thank Adam Leroy, Paul Martini, Annika Peter, Bj\"{o}rn Adebahr, Paulo Montero-Camacho, and Chris Cappiello for their helpful discussions. TAT thanks Brian Lacki for discussions and collaboration on the physics of cosmic rays. This work is supported by NASA \#80NSSC18K0526 and by NSF \#1516967. TAT acknowledges a Simons Foundation Fellowship and an IBM Einstein Fellowship from the Institute for Advanced Study, Princeton. The computations in this paper were run on the CCAPP condo of the Ruby Cluster at the \citet{OhioSupercomputerCenter1987}. 

\footnotesize{
\bibliographystyle{mnras}
\bibliography{bibtex}
}

\begin{appendices}

\section{Gas Density and Wind Profiles}
\label{appendix:gas_wind}

\begin{figure*}
    \makebox[0.49\linewidth][c]{ \includegraphics[width=0.50\linewidth]{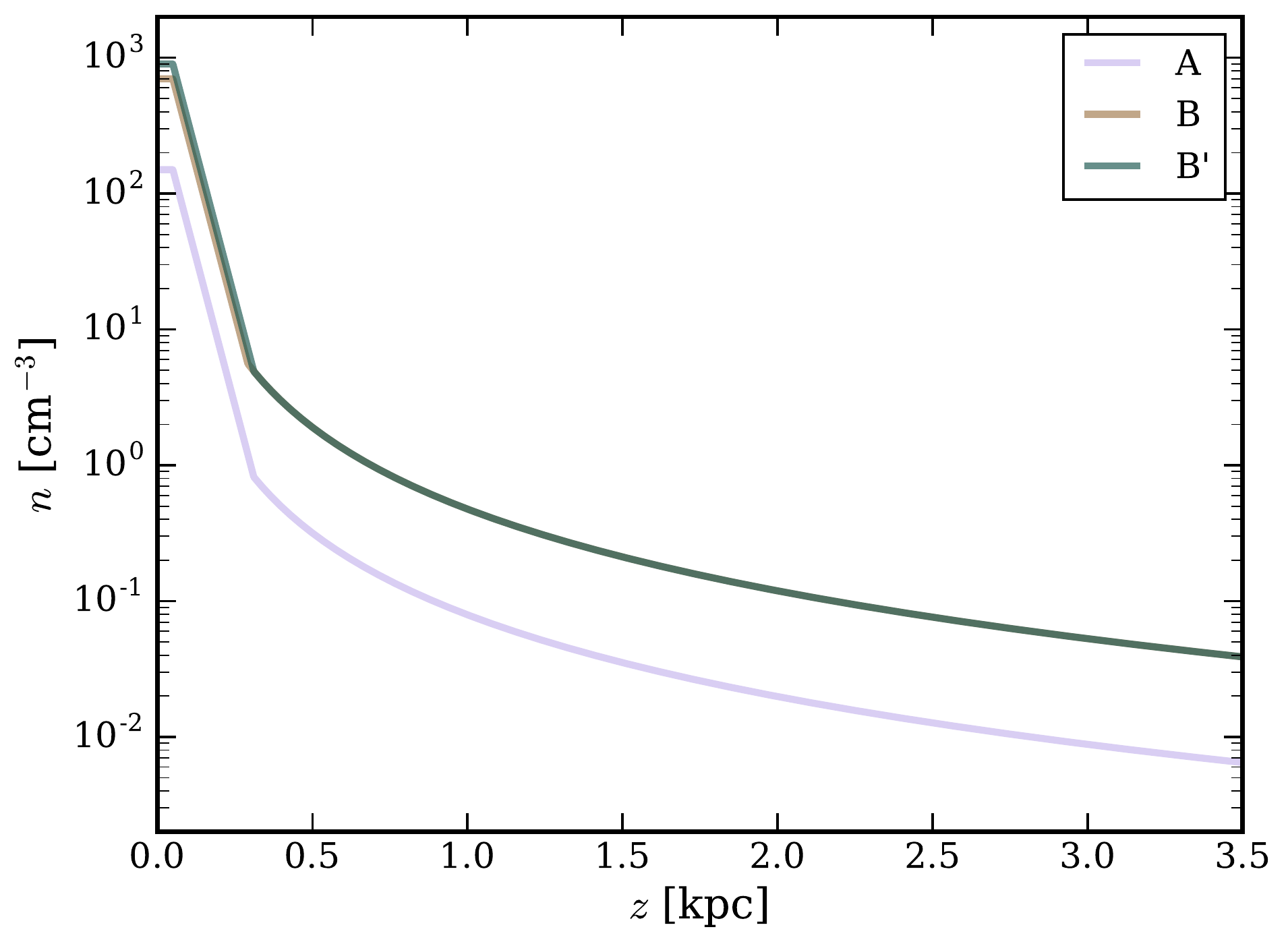} }
	\makebox[0.49\linewidth][c]{ \includegraphics[width=0.50\linewidth]{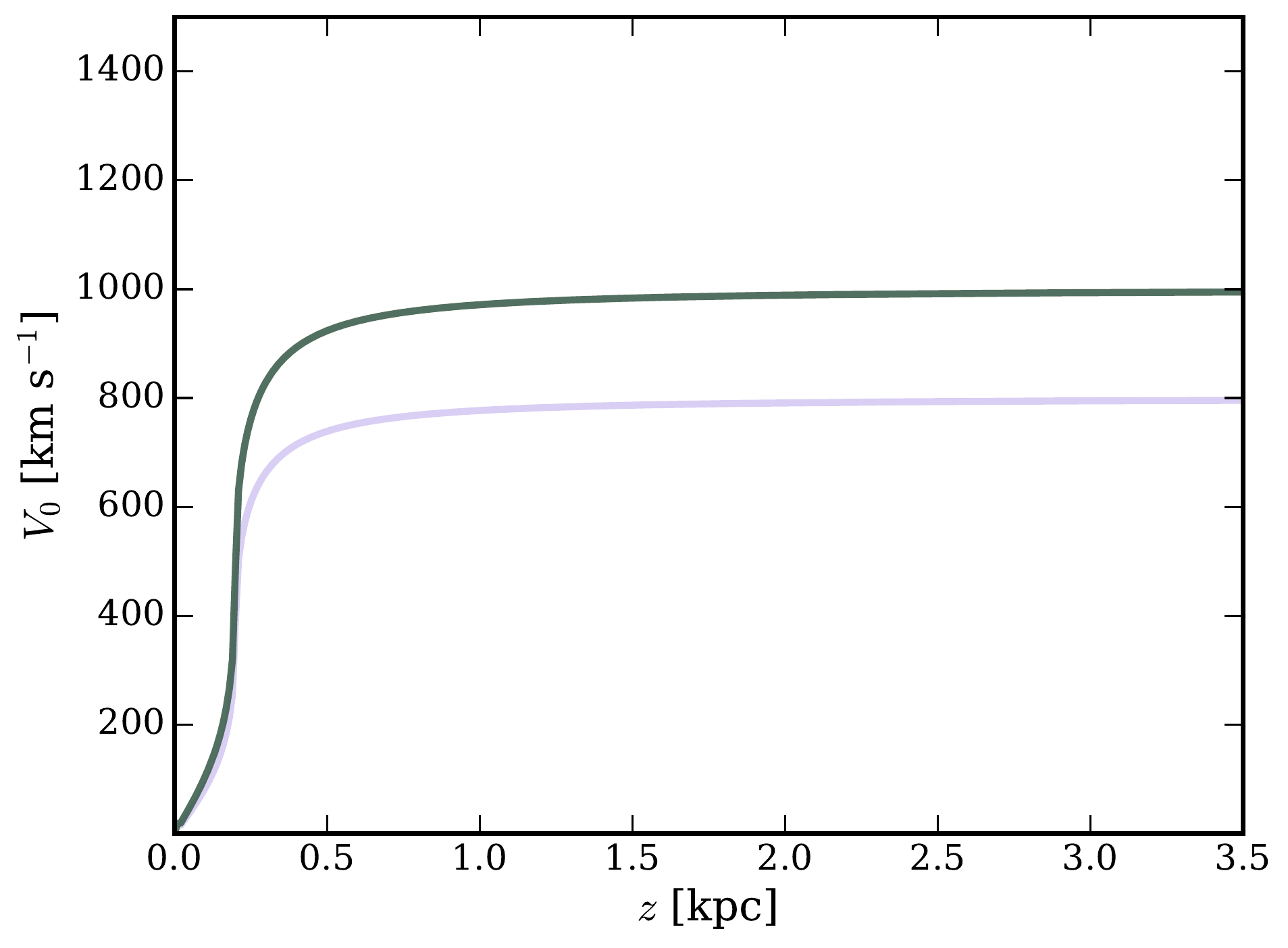} }
	\caption{ The gas density (left) and wind velocity (right) for our three Models A (lavender), B (brown), and B$'$ (green) as a function of distance along the minor axis. Models B and B' have a slight difference in gas density normalization (See Table~\ref{parameters}), but have the same wind component of the gas density and also have the same wind profile. }
	\label{gas_wind}
\end{figure*}

Figure~\ref{gas_wind} shows the gas density (left panel) and advection velocity (right panel) for Models A (lavender), B (brown), and B$'$ (green) as a function of the distance along the minor axis (the lines for B and B$'$ are overlapping). 

The functional form of the gas density is given in Table~\ref{distributions}. The density profile is constant in the core until a height of 0.05\,kpc. Beyond this height, the density decreases exponentially until the wind component of the gas density takes over at $\sim$0.4\,kpc. The wind component assumes a constant galactic mass-loss rate being driven by a spherically-symmetric outflow. We note that Model~A has a different overall normalization for the wind component of the gas density ($\dot{M}/V_n$; See Table~\ref{parameters}), which is important in our modeling. Specifically, if we decrease the wind density normalization, then we can fit the 6\,cm halo (and spectral steepening discussed in Section~\ref{section:results:extended:radio_halo}) much better at 1\,kpc, but at the cost of underproducing the 22\,cm halo at $z>2$\,kpc. Similarly, if we increase the wind density normalization, the models overproduce emission at all wavelengths, but especially at 22 and 6\,cm. As discussed in Section \ref{section:model:galprop:propagation}, we note that the cosmic-ray advection speed, $V_0$, need not be equal to $V_n$ in the normalization of the wind density parameter $\dot{M}/V_n$ in Table \ref{parameters}.

We used a numerical model for the shape of the wind velocity profile and changed the normalization between models, as in the right panel of Figure~\ref{gas_wind}, motivated by energy-driven thermal wind models \citep{Chevalier_Clegg}. The wind velocity reaches half its maximum velocity at 0.2\,kpc \citep{2009ApJ...697.2030S}. We assume the wind velocity is always spherically radial except in the disk where we only consider the vertical component of the wind. We implement this by multiplying the cylindrically radial component of the wind by ($1-\exp |z|/z_\mathrm{core}$), emulating a disk with a 100\,pc height. One might expect that ignoring this factor would significantly change our results since then we would have a wind blowing CRs out of the edge of the core in the disk (at 0.2\,kpc) since the spherical wind velocity reaches half its maximum at a spherical radius of 0.2\,kpc. However, ignoring the additional factor only decreases the final CR energy densities in the core by ${\sim}10$\%. This is because the wind dominates outside $\pm z_\mathrm{core}$ and diffusion from these regions, above and below the disk, keep the disks energy density from drastically changing. 

\section{Analytic Spectral Index Analysis}
\label{appendix:spectral}

\begin{figure*}
	\centering
	\makebox[0.49\linewidth][c]{ \includegraphics[width=.51\linewidth]{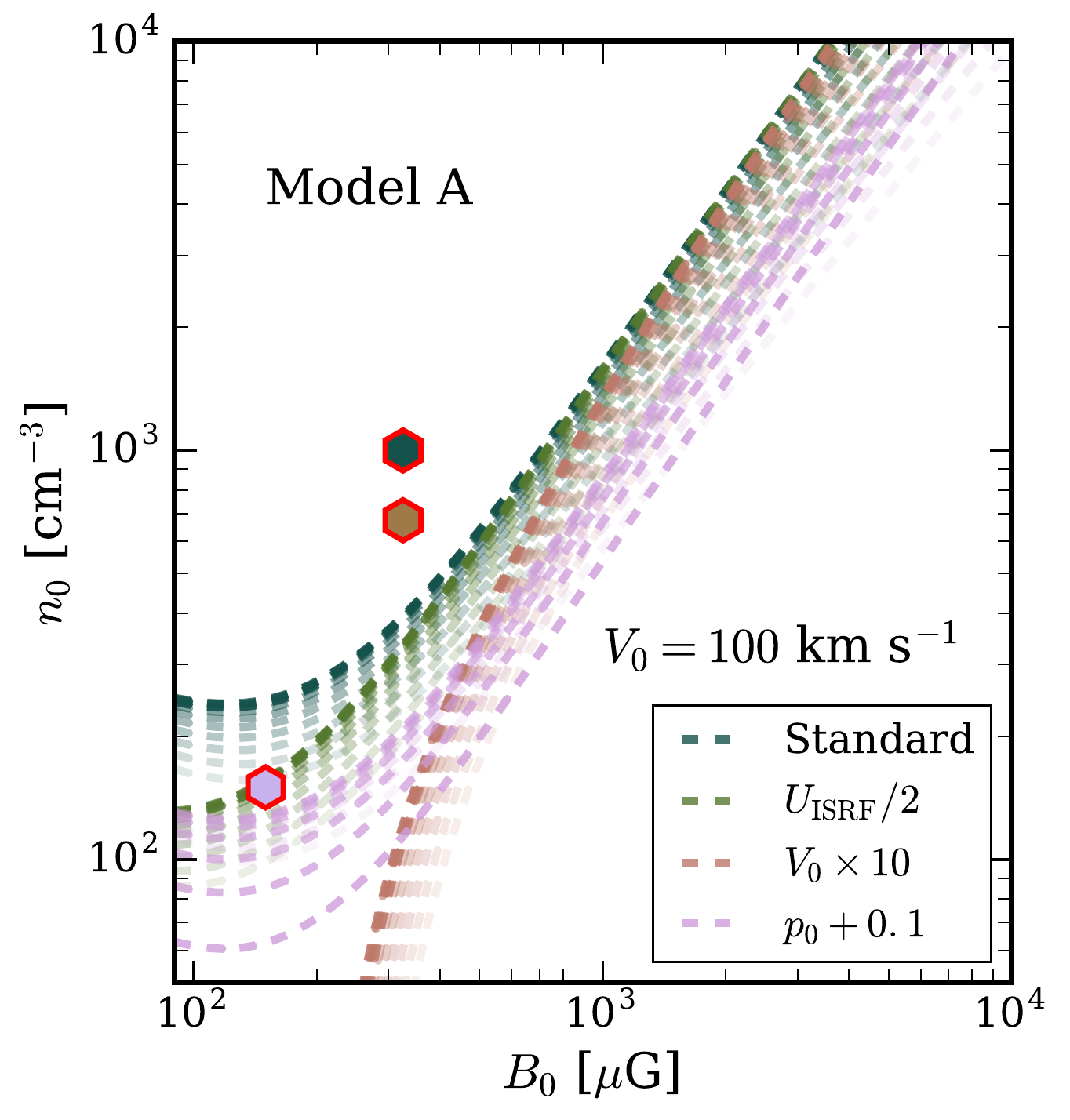} }
	\hspace{0pt}
	\makebox[0.49\linewidth][c]{ \includegraphics[width=.51\linewidth]{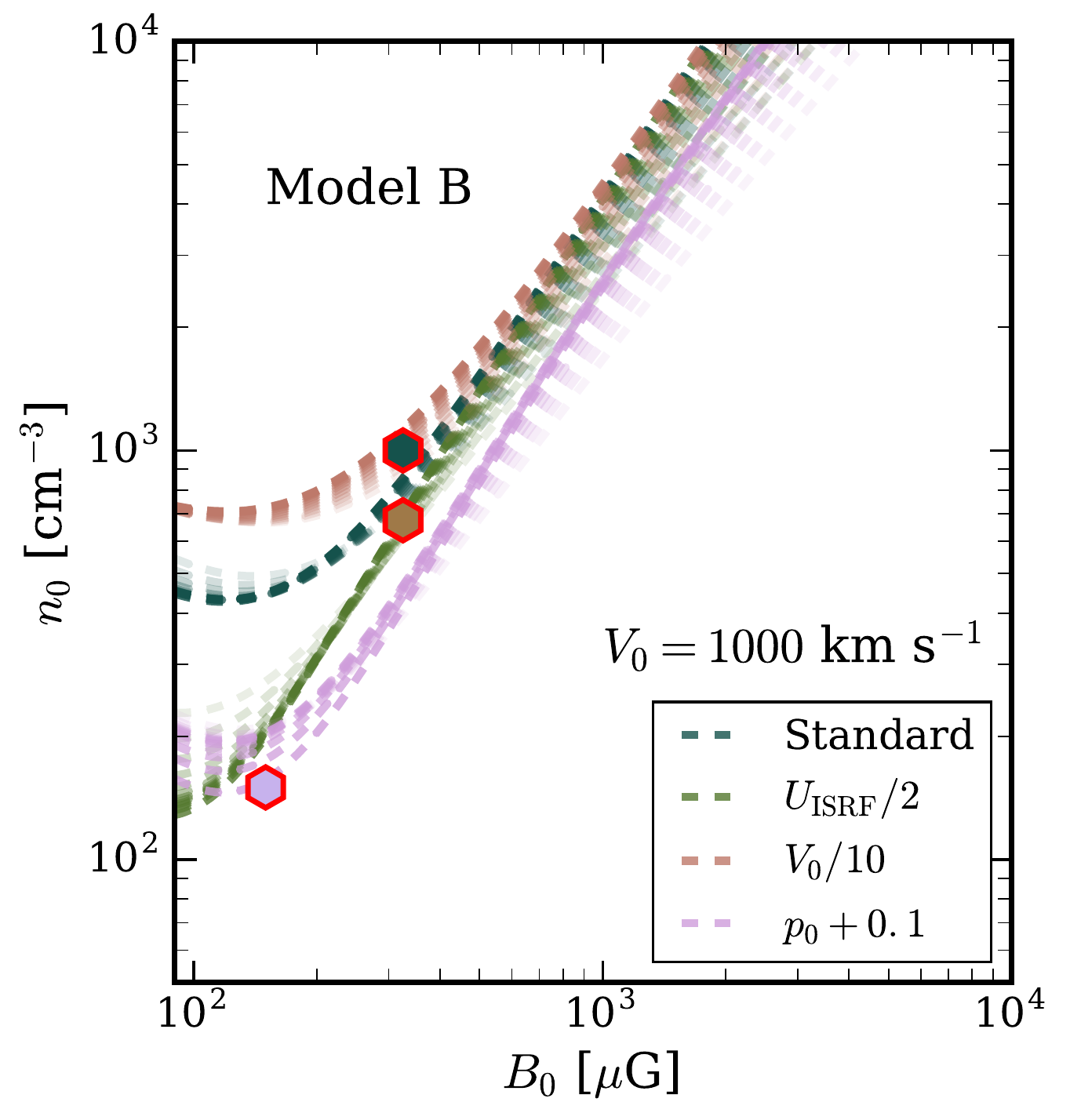} }
	\caption{ Analytic $B_0{-}n_0$ relations when parameters are changed. We choose $\alpha$ from the unabsorbed synchrotron spectra, as seen in Figure~\ref{AB_integrated}, for models A (left plot) and B (right plot). Each color denotes a single set of $U_\mathrm{ISRF}$, $V_0$, and $p_0$. The different lines for each color correspond to choosing a different frequency, $\nu_1$, with its corresponding CRe spectral index, $p$. We choose 12 logarithmically-spaced frequencies between 1\,GHz and 50\,GHz, where the brighter lines are closer to 50\,GHz and the dimmer lines are closer to 1\,GHz. Dark green lines denote our standard model assuming $U_\mathrm{ISRF}=1000$\,ev\,cm$^{-3}$, $p_0=-2.2$, and $V_0=100$ (1000)\,km\,s$^{-1}$ for Model~A~(B). Light green, brown, and pink lines denote the standard model with $U_\mathrm{ISRF}/2$, $V_0\times$ or $/ 10$, and $p_0+0.1$, respectively. The lavender, brown, and green hexagons denote the positions of Models~A~B and B$'$. See Figure~\ref{chi_SNR_plot} for our numerical constraints. }
	\label{chi_analytic}
\end{figure*}

Figure \ref{chi_SNR_plot} presents $\chi_\mathrm{avg,int}$ (defined in Section~\ref{section:results:emission:mag_gas}) contours in the plane of the central magnetic field strength and volumetric gas density, $B_0$ and $n_0$ (see Tables~\ref{distributions} and \ref{parameters}). The degeneracy between these two parameters stems from their importance in the non-thermal gamma-ray and radio CR emission. The magnetic field and gas density control the normalization of the radio and gamma-ray spectra, along with the radio spectral index.

In this section, we use the observed synchrotron spectral index to understand the physics of the degeneracy between $B_0$ vs. $n_0$. A zeroth order approximation for the magnetic field-gas density relation assumes that the synchrotron and bremsstrahlung cooling time scales are approximately equal to each other (see \citet{2013ASSP...34..283T}). In short, for galaxies to have shallow radio spectral indices at $\sim$GHz frequencies, bremsstrahlung or ionization cooling must be comparable to synchrotron and IC cooling to keep the far-infrared radio correlation consistent over many orders of magnitude in magnetic field strength and gas density \cite{LTQ}. We might thus expect a similar relation for a fixed (shallow) radio spectral index for M82, while varying $n_0$ and $B_0$. Setting the bremsstrahlung and synchrotron cooling timescales equal for CRe emitting at GHz frequencies (see eqs.~\ref{tau_bremss}$-$\ref{tau_wind}), we find an expected correlation of the form
\begin{equation}
    n_{100} \sim 0.4 \nu_1 B_{100}^{-\frac23},
\end{equation}
which roughly agrees with the results of Figure \ref{chi_SNR_plot}. However, this estimate ignores other losses such as IC, ionization, and advective losses due to a wind.

To take into account more the effects, we solve the steady-state solution of the position-independent energy-loss equation for CRe in the starburst core given by:
\begin{equation} 
\frac{\partial}{\partial t} \psi = \frac{d}{d E} \left( \dot E \psi \right) + q_{\mathrm{CRe}} \label{model_a1}
\end{equation} 
where $\psi$ is the CRe number density per energy, $E$ is the energy, \mbox{$-\dot E \sim \sum_i {E}/{\tau_i}$}  is the total CRe energy loss rate with $i=$ \{bremss, synch, IC, ion, wind\}, and \mbox{$q_\mathrm{CRe} = q_0 E^{p_0-1}$} is the CRe source spectrum. The values of $\tau_i$ are given in equations \ref{tau_bremss}$-$\ref{tau_wind}. We ignore diffusion in the above expression, as our analysis shows that diffusive losses are subdominant with respect to advective losses thoughout the bulk of the starburst core volume. We solve Equation~\ref{model_a1} assuming $\dot E<0$, ${\partial \psi}/{\partial t} = 0$, and that particle number must be conserved. We find 
\begin{equation}
\psi (E) = \frac{1}{-\dot E} \int_{E}^{E_\mathrm{max}} dE \; q_\mathrm{CRe}.
\end{equation} 

Within a small enough energy bin, we can approximate $E \psi(E) \propto E^p$ and solve for $p$:
\begin{equation}
    p = \frac{d \log E\psi(E)}{d\log E} = 1 - \frac{E}{\dot E} \left( \frac{d \dot E}{dE}- \frac{q_\mathrm{CRe}}{\psi}  \right).\label{p_equation}
\end{equation}
Substituting for $\psi$ and $q_\mathrm{CRe}$, and using the expressions for \mbox{$-\dot E = \dot E_\mathrm{ion} +\dot E_\mathrm{bremss} +\dot E_\mathrm{synch} +\dot E_\mathrm{IC} +\dot E_\mathrm{wind}$},  Equation~\ref{p_equation} then depends on $p$, $p_0$, $n$, $B$, $U_\mathrm{rad}$, $V$, $z$, $E$, and $E_\mathrm{max}$. Using Equation~\ref{e_nu_b}, we eliminate the energy dependence of the resulting expression in favor of the synchrotron frequency $\nu_\mathrm{crit}$ and $B$. If we assume standard values for $p$, $p_0$, $U_\mathrm{ISRF}$, $\nu$, $V$, $z$, and $E_\mathrm{max}$, we can solve Equation~\ref{p_equation} for $n$ in terms of $B$.  Doing so gives us
\begin{equation}
    \begin{split}
        n_{100} =\,& 52.9 \frac{1-F}{X} \left(\frac{\nu_1}{B_{100}}\right) \left(0.24\, B_{100}^2 + U_{1000} \right) \\
        & - 6.2\, \frac{F}{X} \left( \frac{\nu_1}{B_{100}} \right)^{\frac12} \frac{V_{100}}{z_{50}} ,
    \end{split}
\end{equation}
where
\begin{align}
F =\,& {p_0} \left[ 1 - \left( 0.6\, \left(\frac{\nu_1}{B_{100}}\right)^{-\frac12} E_{\mathrm{max},1} \right)^{p_0} \right]^{-1} -p ,\\
X =\,&  \frac{1+F}{1-0.75f_\mathrm{ion}} + 19.2\, F\, \left(\frac{\nu_1}{B_{100}}\right)^{\frac12} ,
\end{align}
and $E_{\mathrm{max},1}~=~{E_\mathrm{max}}\,/\,{1\,\mathrm{GeV}}$. We note that if we assume the \mbox{$E_{\mathrm{max},1} \gg 1$}, then \mbox{$F \approx p_0 - p$} which is ${>}0$ in our case. We assume a CRe injection spectrum with $p_0=-2.2$, the same as the assumed CRp injection spectrum. If there is no free-free emission or absorption, then $p = 2\alpha-1$, where $\alpha$ is the observed radio spectral index. If free-free emission and absorption are important, as it is in our case, then the radio spectral index we observe is not directly related to the synchrotron spectral index from the CRe.

For frequencies above $\sim$1\,GHz, we can ignore the effect of free-free absorption, thus only free-free emission has an effect on the inferred value of $p$. We note that $p$ may change as a function of frequency because bremsstrahlung and ionization cooling dominate for CRe radiating at GHz frequencies, whereas  synchrotron cooling dominates for CRe radiating at higher frequencies. If we assume the radio flux can be approximated as the sum of two power-laws at each frequency --- one from synchrotron with an unknown spectral index, and one from free-free emission with a spectral index of $-0.1$ \citep{1992ARA&A..30..575C} --- then
\begin{equation}
\alpha_\mathrm{measured} = \frac{ \frac{p+1}{2} - 0.1 \left( \frac{\nu_=}{\nu} \right)^{\frac{p+1}{2}+0.1}}{1+ \left( \frac{\nu_=}{\nu}\right)^{\frac{p+1}{2}+0.1}}
\end{equation}
where $\alpha_\mathrm{measured}$ is the measured radio spectral index, $\nu_=$ is the frequency at which the flux of synchrotron and free-free are equal and $\nu$ is the frequency of interest. From here, we solve for $p$ given $\alpha_\mathrm{measured}$. From the data, \mbox{$\nu_= \gtrsim 10$ GHz}. Then for $\alpha_\mathrm{measured}=-0.7$ at $\nu\approx2.5$\,GHz, and letting $\nu_=$ range from \mbox{10 --- $\infty$~GHz}, we find a range for $p$ of $-2.6$ to $-2.4$. For illustrative purposes, we plot this relation with $p=-2.4$ and $-2.6$, $\nu_1=2.5$, and $V_{100} = 10$ on Figure~\ref{chi_SNR_plot} with ionization fractions, $f_\mathrm{ion}$, of 0 and 1 as dotted and dashed lines, respectively. The upper dashed/dotted line denotes $p=-2.4$ and the lower dashed/dotted line denotes $p=-2.6$. 

As discussed in Section~\ref{section:results:emission:mag_gas}, the positive degeneracy between $B_0$ and $n_0$ results from: 1) the complex interactions between the energy-dependant cooling timescales for CRe at different observed synchrotron frequencies and 2) the relative normalization between the radio and gamma-ray spectra that is complicated by the large secondary population created from CRp hadronic interactions. In Figure~\ref{chi_SNR_plot}, we see that our analytic results coincide well with our {\tt GALPROP} models. We note that our analytic results only take into account the constraints from the observed spectral index and not the relative gamma-ray and radio normalizations, while our numerical {\tt GALPROP} models take both factors into account. Our work is consistent with previous results given by \citet{2012ApJ...755..106P}. 

Figure~\ref{chi_analytic} illustrates the effects of changing our assumed parameters on the $B_0$ vs. $n_0$ relation. We use the unabsorbed synchrotron spectra from Figure~\ref{AB_integrated} to determine $\nu_1$ and $p$. The left (right) plot uses the synchrotron spectrum from Model~A~(B). We do not show Model~B' due to its similarity to Model~B. The different lines for each color denote one of 12 logarithmically spaced values for $\nu$ between 1\,GHz and 50\,GHz with the brighter lines being closer to 50\,GHz and the dimmer lines being closer to 1\,GHz. Each color denotes a set of $U_\mathrm{ISRF}$, $V_0$, and $p_0$. Our standard model is shown in dark green, while the other colors denote changing one value from our standard model. The standard model has $U_\mathrm{ISRF}=1000$\,eV\,cm$^{-3}$, $p_0=-2.2$, and $V_0=100$~(1000)~km\,s$^{-1}$ for Model~A~(B). An unabsorbed synchrotron spectrum like Model~B requires a much higher wind velocity than Model~A.

If we decrease $U_\mathrm{ISRF}$ to 500\,eV\,cm$^{-3}$ (light green lines), then the models require a smaller gas density for magnetic fields below $\sim$300\,$\mu$G due to the smaller IC+synchrotron cooling. If we make the injected CRe spectrum harder, $p_0 \rightarrow -2.1$ (pink lines), then we require a stronger magnetic field or smaller gas density to steepen the synchrotron spectrum. For Model~A, the brown lines denote changing $V_0$ from 100 to 1000\,km\,s$^{-1}$ and we see this has a large effect below a few hundred $\mu$G. This drop occurs because both hadronic and wind losses do not directly affect the spectral index. Thus, if the wind is very strong, we no longer require as high a gas density to get the same shallow synchrotron spectrum (ignoring the overall radio flux normalization). For Model~B, the brown lines denote changing $V_0$ from 1000 to 100\,km\,s$^{-1}$ and we see the effect is not as important as in Model~A since Model~B has stronger hadronic+wind cooling. Due to the simplicity of our model, choosing different frequencies (different lines for each color) will give us different results, especially when the gas density and magnetic field are small.

\begin{figure*}
	\centering
	\makebox[0.49\linewidth][c]{ \includegraphics[width=.5\linewidth]{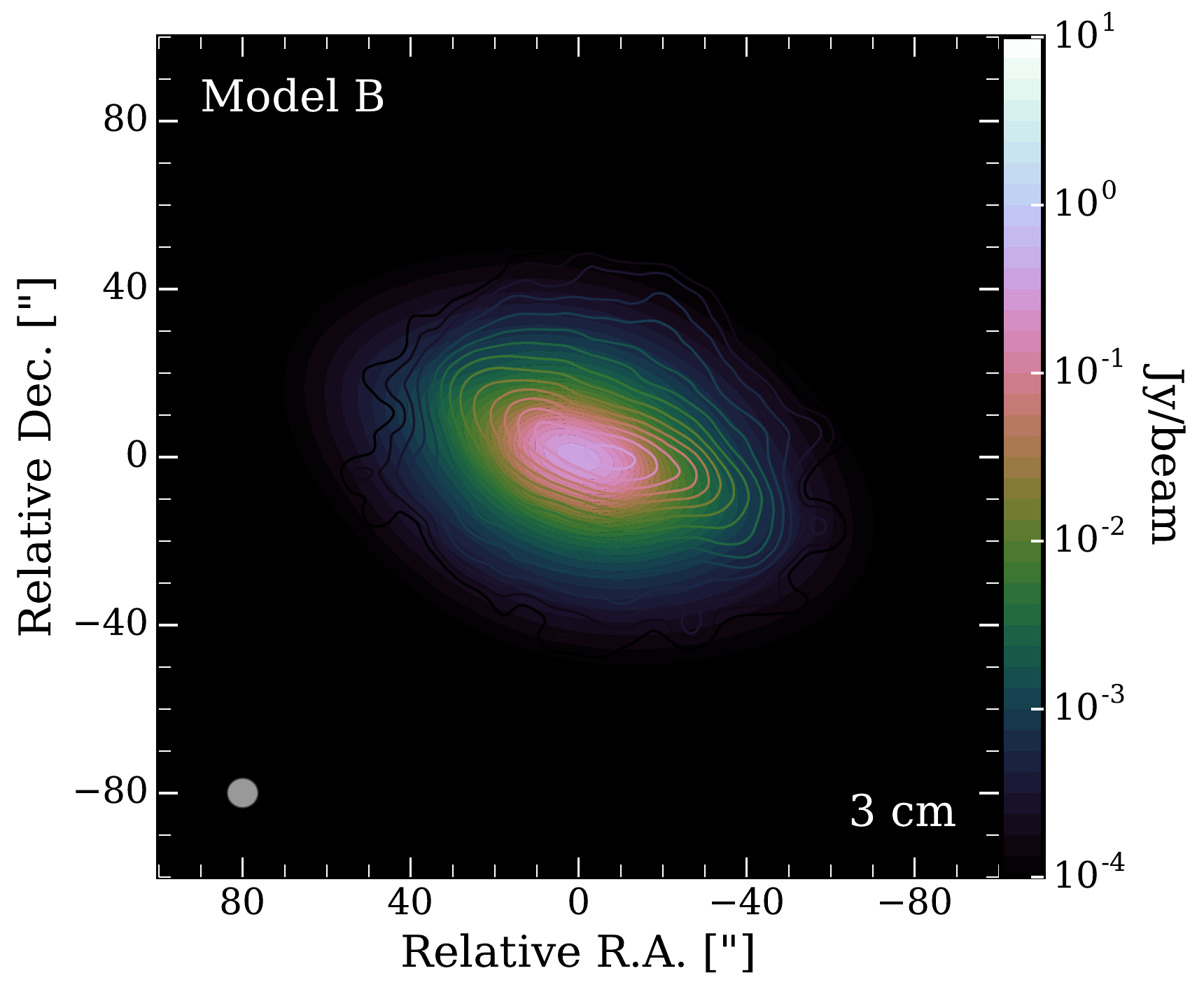} }
	\hspace{0pt}
	\makebox[0.49\linewidth][c]{ \includegraphics[width=.5\linewidth]{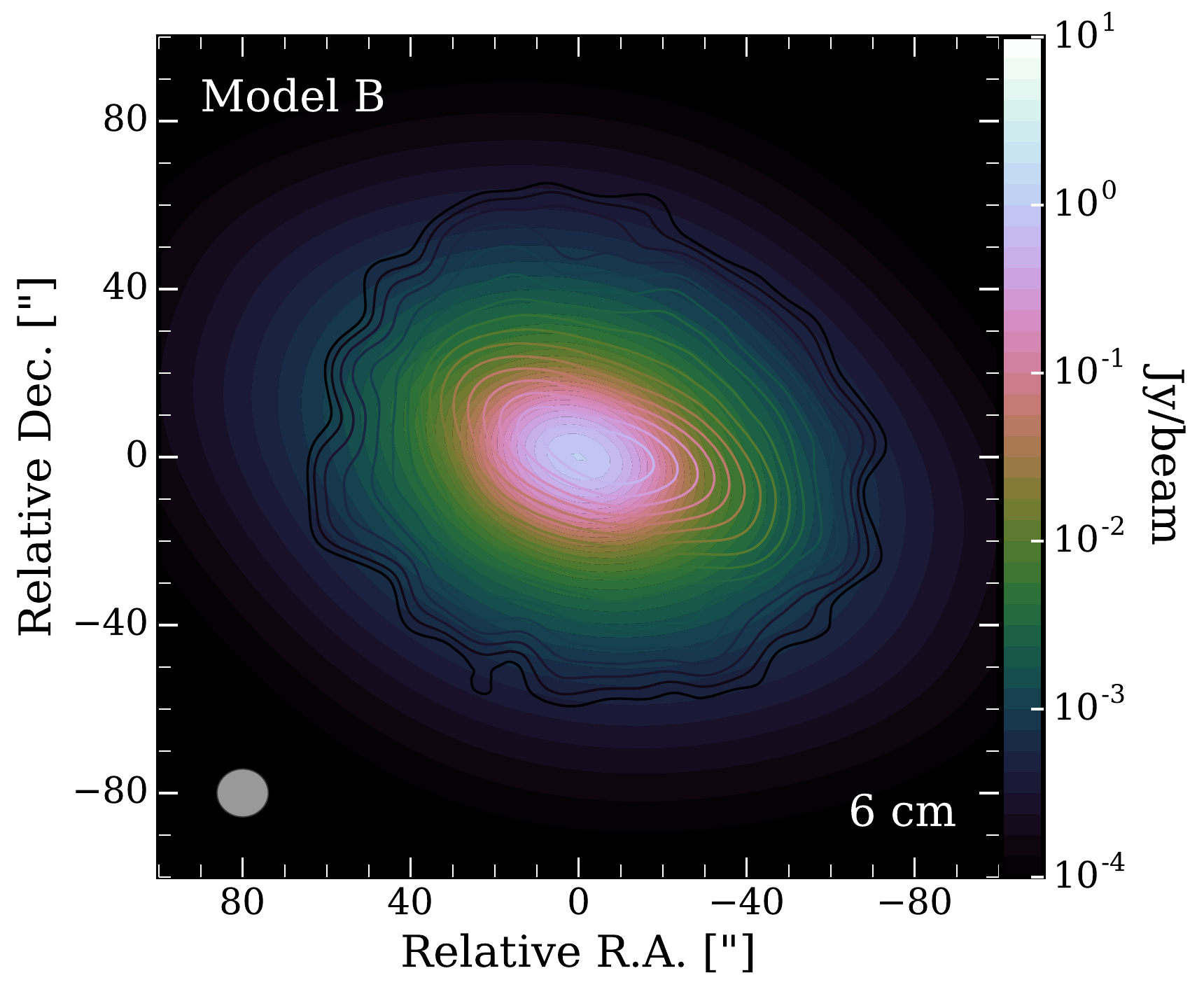} } \\
	\makebox[0.49\linewidth][c]{ \includegraphics[width=.5\linewidth]{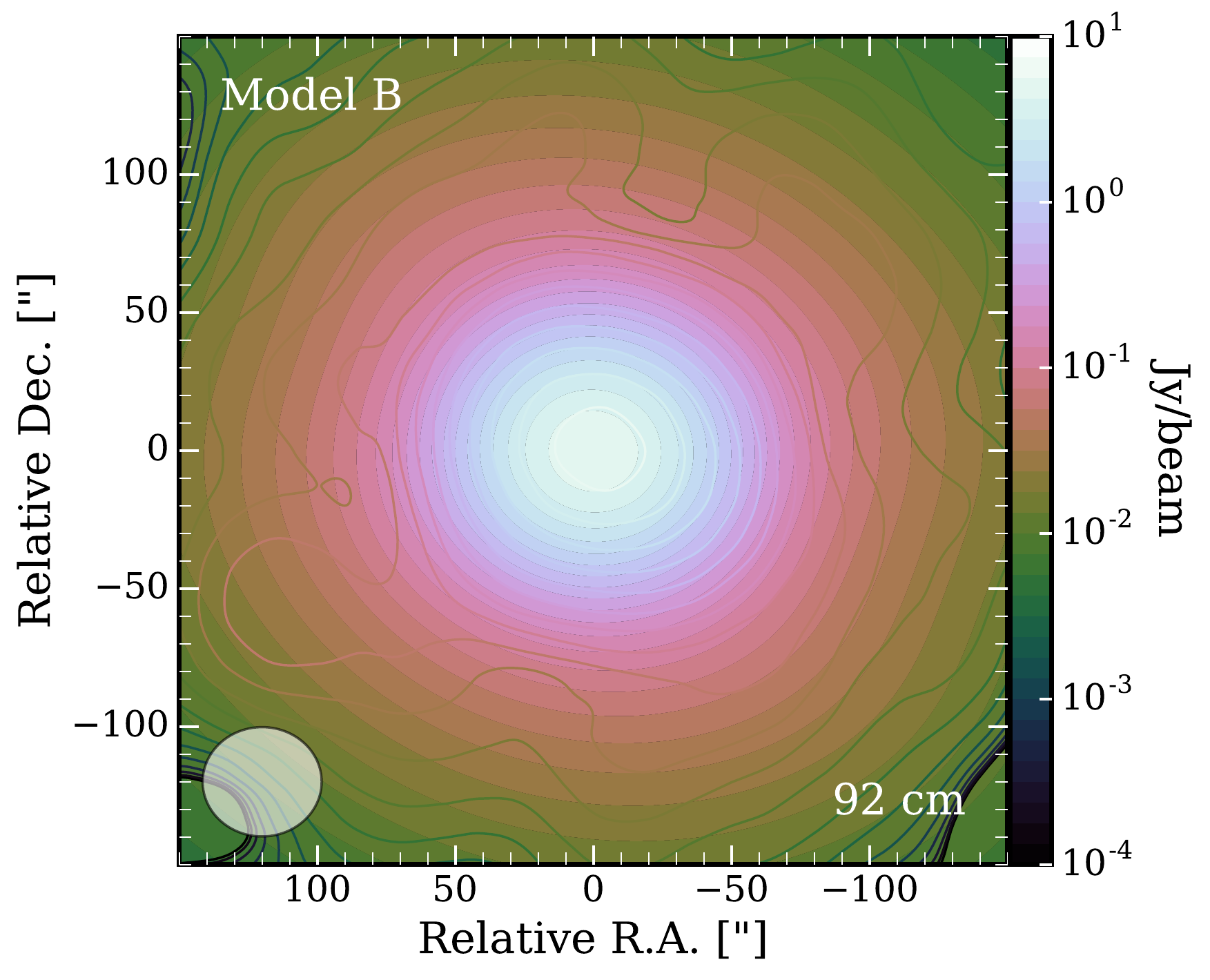} }
	\hspace{0pt}
	\makebox[0.49\linewidth][c]{ \includegraphics[width=.5\linewidth]{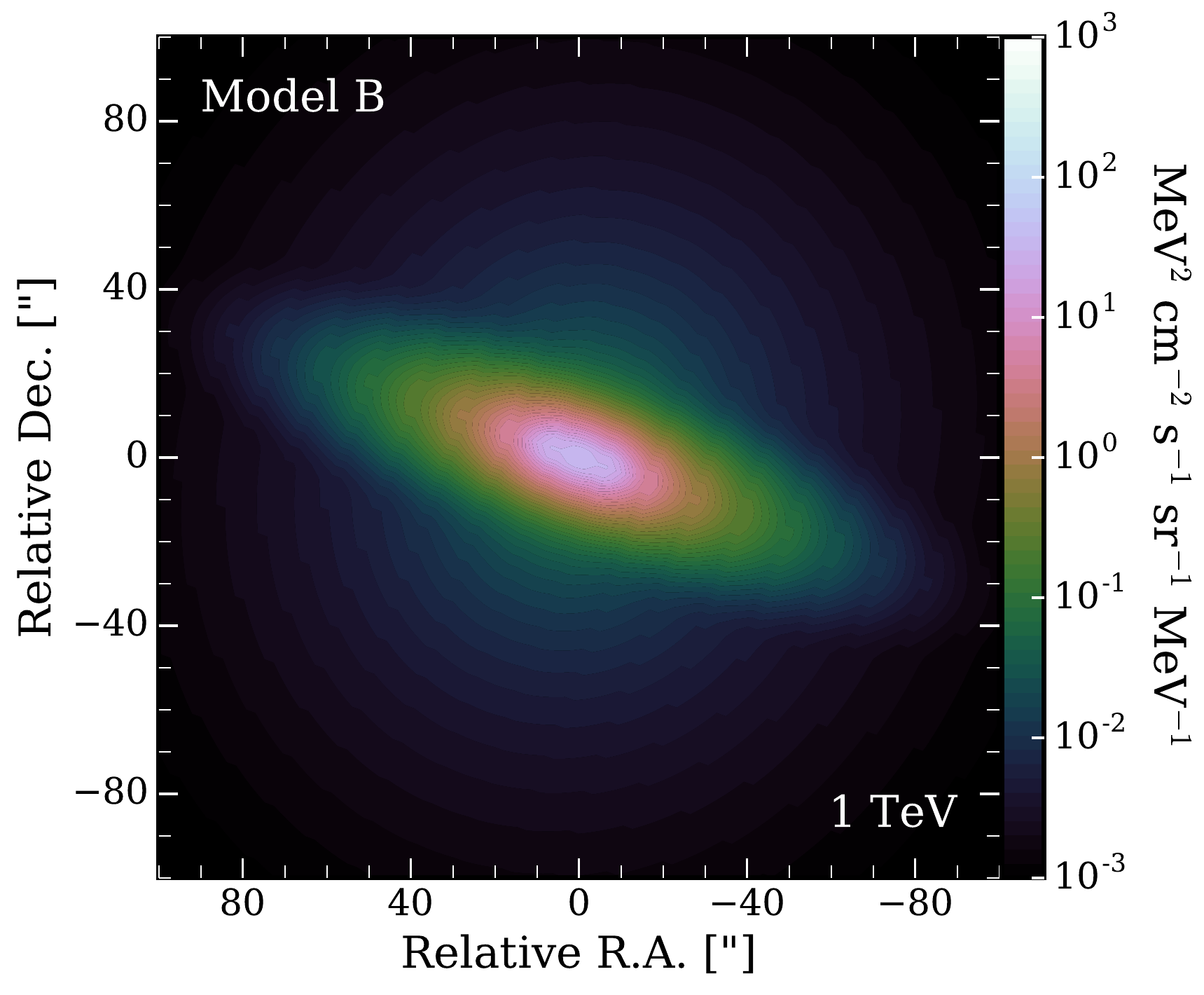} }
	\caption{ Projected images of Model~B at radio wavelengths of 3\,cm (upper-left), 6\,cm (upper-right), and 92\,cm (lower-left) and gamma-ray energy of 1\,TeV (lower-right). The radio maps are compared to colored contours from \citet{2013A&A...555A..23A}. For the radio images, the beam sizes of 7.6"$\times$7.3", 12.5"$\times$11.7", 43.1"$\times$39.6" are shown in the bottom-left corner of each plot for 3, 6, and 92\,cm, respectively. We note that the 92\,cm panel has a larger field of view than the other maps. }
	\label{all_map}
\end{figure*}

\begin{figure*}
	\centering
	\makebox[0.49\linewidth][c]{ \includegraphics[width=.5\linewidth]{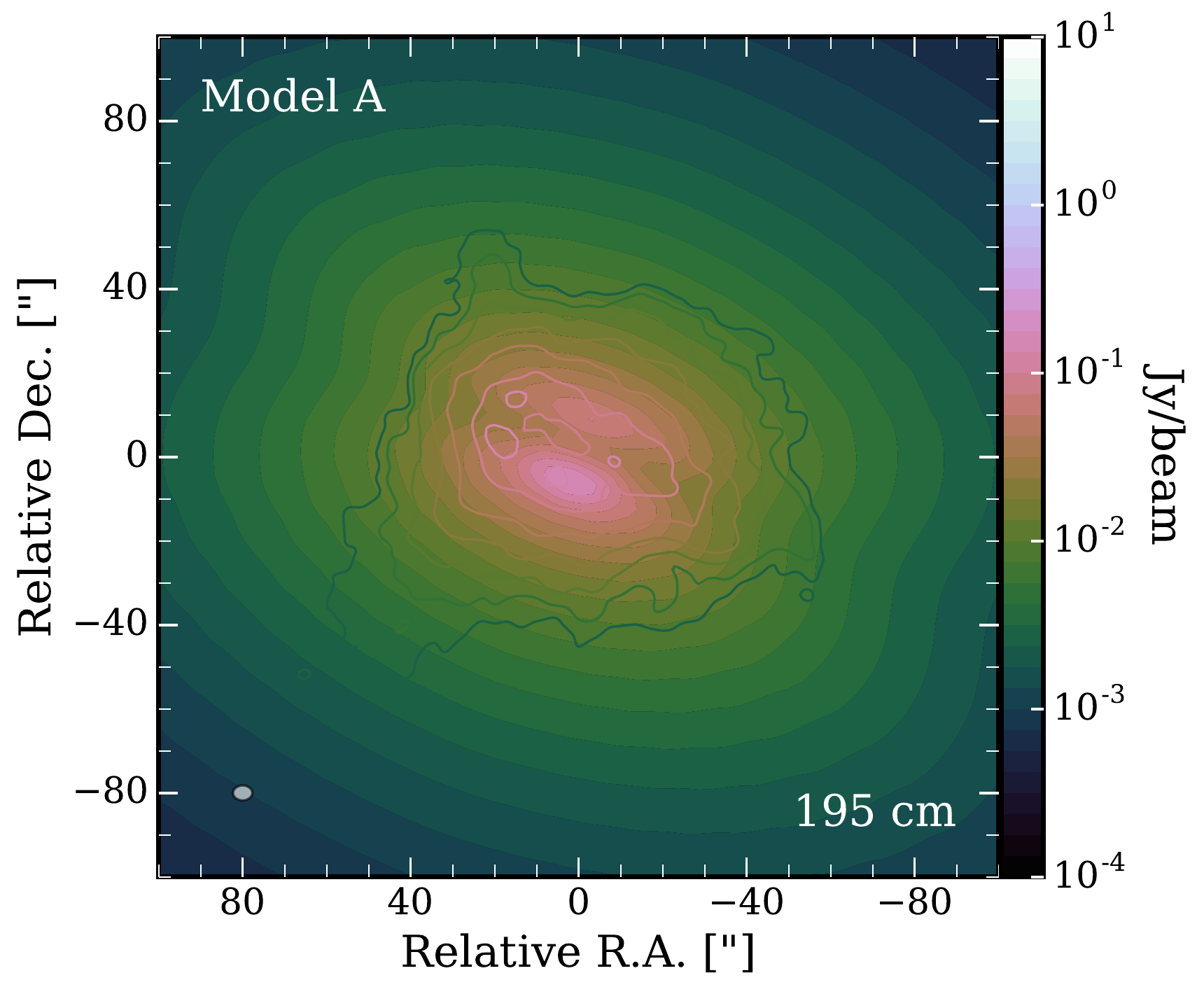} }
	\hspace{0pt}
	\makebox[0.49\linewidth][c]{ \includegraphics[width=.5\linewidth]{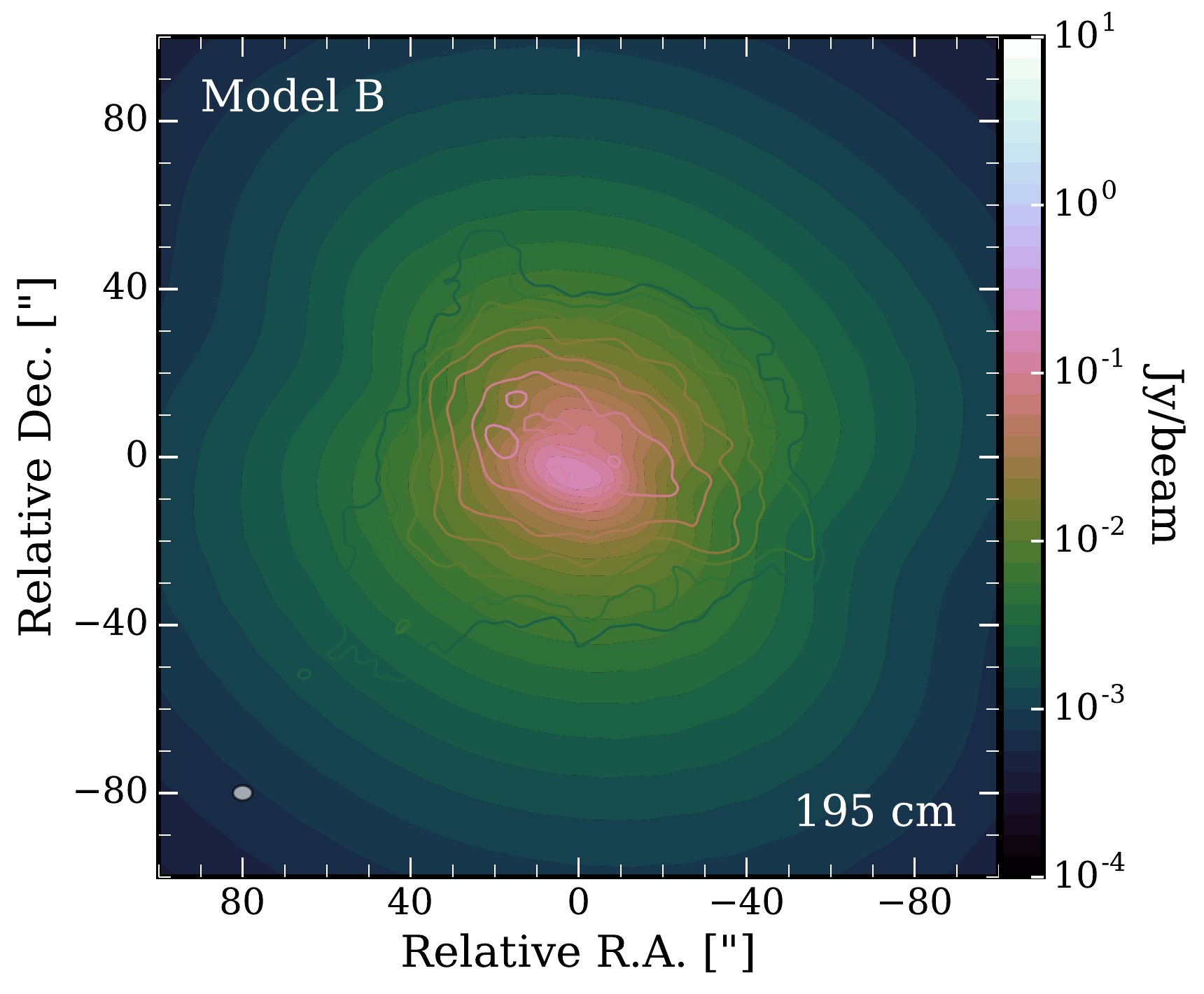} } \\
	\makebox[0.49\linewidth][c]{ \includegraphics[width=.5\linewidth]{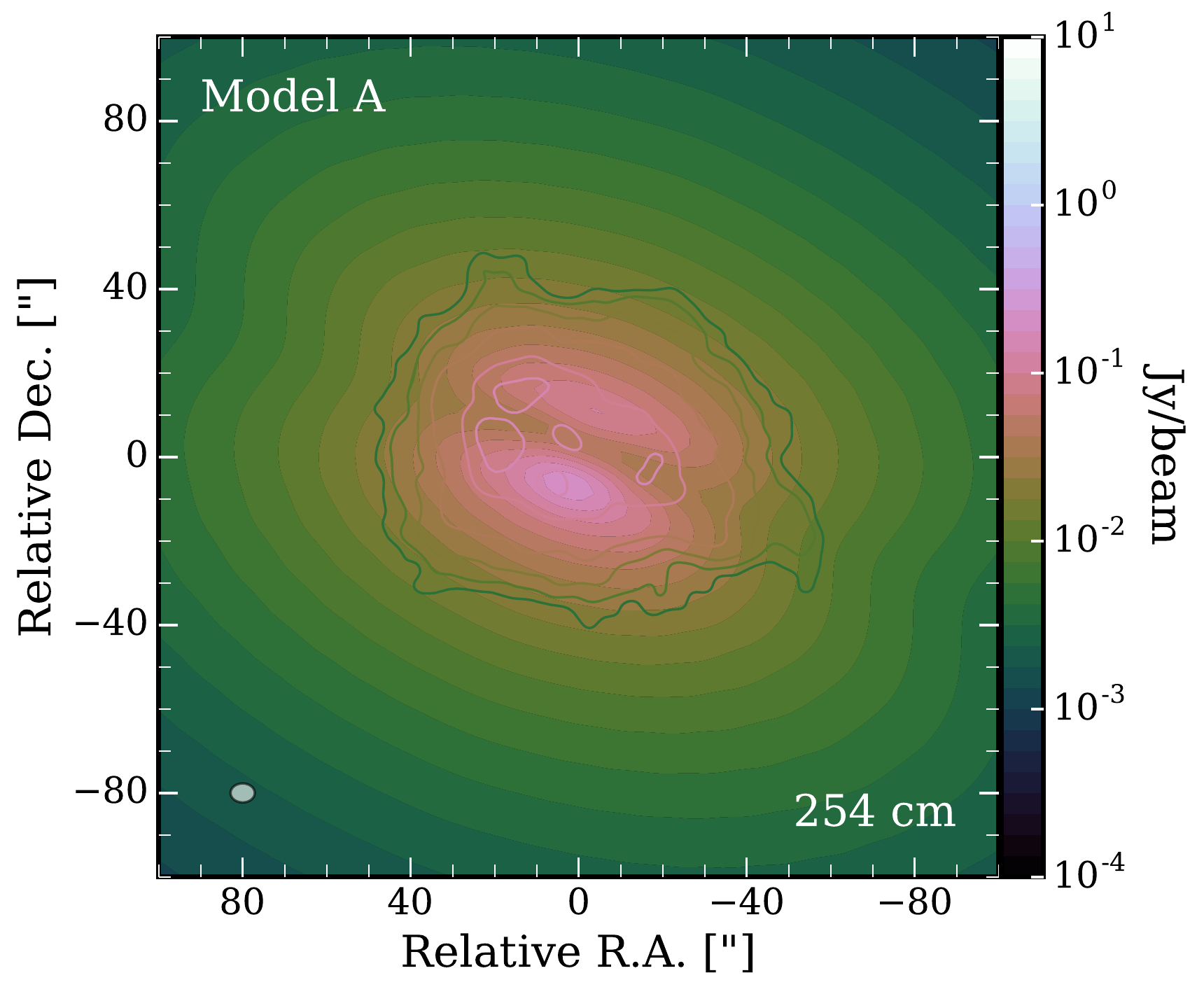} }
	\hspace{0pt}
	\makebox[0.49\linewidth][c]{ \includegraphics[width=.5\linewidth]{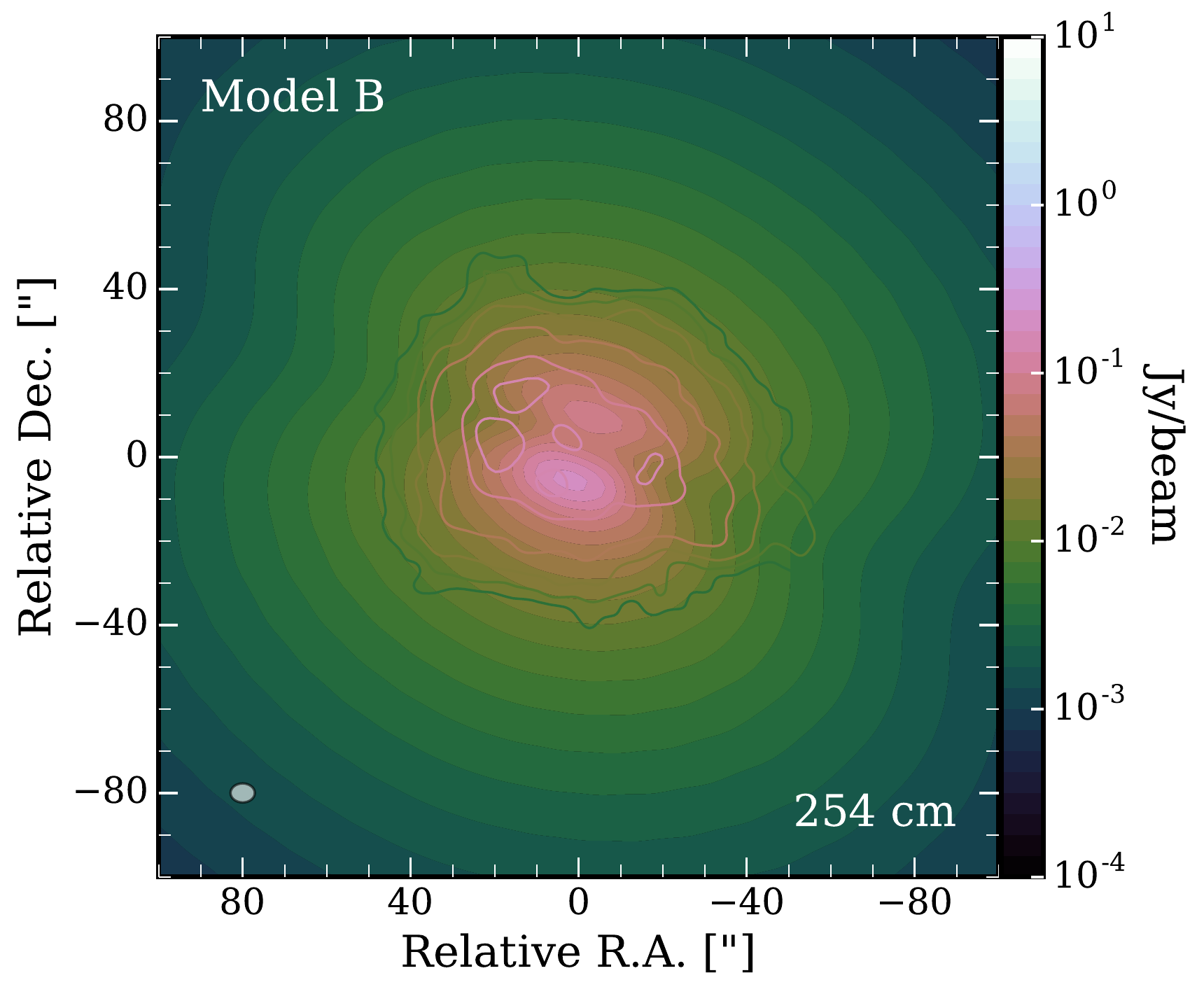} }
	\caption{ Projected images of Model~A (left column) and Model~B (right column) at 195\,cm (154\,MHz) (top row) and 254\,cm (118\,MHz) (bottom row). The radio maps are compared to colored contours from LOFAR \citep{2015A&A...574A.114V}. Our models are convolved with beam sizes of 5.79"$\times$4.52" and 4.66"$\times$3.56" for 195\,cm and 254\,cm, respectively. The beam is shown in the bottom left corner of each plot. These two wavelengths are the two lowest frequency radio data points in Figure~\ref{AB_integrated}.  }
	\label{lofar_map}
\end{figure*}

\section{Effects of Core Shape}
\label{appendix:core_shape}

In order to understand how our results depend on the assumed core density and magnetic field distributions, we tried several variants of the fiducial functional forms given in Table \ref{distributions}.

We assumed the core was a cylinder with $R$ and $z$ dependencies of the form 
\begin{equation}
    B(R,z) = B_0 \exp \left(-\frac{R-R_\mathrm{core}}{R_\mathrm{scale}} \right) \left( \frac{z-z_\mathrm{core}+z_\mathrm{scale}}{z_\mathrm{scale}} \right)^{-\beta},
    \label{bcyl}
\end{equation}
where $R$ is the cylindrical radius. Using this form for $B(R,z)$, we found that in order to obtain extended radio emission consistent with the data, the model required a large value of  $R_\mathrm{scale}>0.5$\,kpc and a small value of $\beta<1$, implying a wide, tall, and highly-magnetic halo. Ultimately, this conclusion follows from the fact that equation (\ref{bcyl}) gives a rapid fall-off in $B$ as a function of $R$, such that the smoothness of the observed radio halo surface brightness as a function of $R$ away from the minor axis, required large $R_{\rm scale}$ and small $\beta$. 

We also explored several models where the core itself was vertically stratified, instead of being characterized by constant density and magnetic field throughout. For example, in one series of models, we used functional forms for the $z$-dependence of the density and magnetic field that continuously exponentially declined from $z=0$\,kpc with the scale lengths of the core (e.g., $\propto\exp(-z/z_{\rm core})$. Our resulting {\it central} gas densities and magnetic fields needed to be much larger in order to match the data, since the volume-\emph{average} gas density and magnetic field of the core needs to be comparable to our final results with our fiducial constant core distribution. That is, we found that our central gas densities and magnetic field values needed to be a factor $\sim$4 times larger to fit the integrated and extended emission. In order to more directly compare with previous one-zone models, and to make our analytic and numerical results consistent, we decided to hold our distributions constant within the core (See Table~\ref{distributions}).

\section{Emission Maps}
\label{appendix:emission_maps}

We expand on Figure~\ref{model_image} and present more emission maps along with data to show how our models qualitatively differ from observations as a function of wavelength in the radio band. The quantitative analysis at 3, 6, 22, and 92\,cm is presented in Section~\ref{section:results:extended}. 

In Figure~\ref{all_map}, we show the projected emission maps of Model~B at 3\,cm (upper-left), 6\,cm (upper-right), and 92\,cm (lower-left), each of which has been convolved with the beam-size of the data, which is shown in the lower left corner of each plot. We also display the gamma-ray emission map at 1\,TeV for comparison with the 1\,GeV map in Figure~\ref{model_image}.

At 3\,cm, we see the asymmetry between the emission above and below the disk as is visible in the data in Figure~\ref{radio_ext}. The biggest apparent difference between our model and the data is the extended emission along the major axis in the disk. Our models slightly overproduce emission in the core, but underproduce emission in the disk. This could alleviate this discrepancy in our models by having CR sources in the disk or increasing the gas density and magnetic field in the disk by changing $R_\mathrm{scale}$ (See Table~\ref{distributions}).

At 6\,cm, we have the same problem replicating the emission in the disk along the major axis. We, again, slightly overproduce emission in the core and have less emission in the disk. Another apparent difference between our models and the data is that we overproduce emission at the edge of the resolved radio halo. All models have this behavior, though Model~A fits the best along the minor axis and both models~A and B have the same fit along the major axis. Model~B$'$ has the worst fit along the major axis.

At 92\,cm, we see our models match the observations pretty well in the core and the inner halo. The data only differs when we are further than $\sim$120" from the core where there may be some feature in the wind that we do not take into account.

At a gamma-ray energy of 1\,TeV, we have the exact same morphology as the 1\,GeV map in Figure~\ref{model_image}. The only difference between the two energies is the normalization, which can be seen in the integrated emission (see Figure~\ref{AB_integrated}). We see the bright core and disk, along with a faint gamma-ray halo. As noted in Section~\ref{section:results:emission}, Model~A has a brighter halo due to the larger CRp energy density in the core and the halo (See Figure~\ref{acceleration}).

In Figure~\ref{AB_integrated}, the two lowest frequency data points in the radio are from the LOFAR experiment \citep{2015A&A...574A.114V}. In Figure~\ref{lofar_map}, we show the projected images of our models~A (left column) and B (right column) at wavelengths (frequencies) of 195\,cm (154\,MHz) and 254\,cm (118\,MHz). The 195\,cm maps are the top row and 254\,cm maps are the bottom row. We do not include these in our analysis due to the importance of free-free emission in replicating the data. In both models, we see an asymmetric image with less emission coming from the very center of the core than right above/below it. This is consistent with the data. Qualitatively, Model~A reproduces the structure of the data in the core better than Model~B. However, Model~A has a much larger halo that overestimates the data more significantly than Model~B. Model~B, over-produces emission in the core and under-produces emission in the disk along the major axis. From Figure~\ref{AB_integrated}, Model~A over-predicts the integrated emission at these wavelengths. We see that we can put better constraints on the free-free structure because of the very small beam size of these measurements.

\end{appendices}

\bsp	
\label{lastpage}
\end{document}